%% file: main.tex
\documentclass[aps, prx, twocolumn, floatfix, superscriptaddress]{revtex4-2}
\usepackage{amsmath,amssymb,amsfonts,amsbsy}
\usepackage{graphicx}
\usepackage{subfig}
\usepackage{dcolumn}
\usepackage{bm}
\usepackage{multirow}
\usepackage{mathtools}
\usepackage{physics}
\usepackage{array}
\usepackage{color}
\usepackage[normalem]{ulem}
\usepackage[per-mode=symbol]{siunitx}
\usepackage{upgreek}
\usepackage{esvect}
\usepackage{booktabs}
\usepackage[utf8]{inputenc}
\usepackage[T1]{fontenc}
\usepackage[pagewise]{lineno}
\usepackage[labelfont=bf]{caption}
\usepackage[figurename=Fig.]{caption}
\usepackage{etoolbox}
\DeclareSIUnit \belm {Bm}
\usepackage{units}
\usepackage{braket}
\usepackage{caption}
\usepackage[hidelinks,bookmarks=false]{hyperref} 
\usepackage[capitalise,nameinlink]{cleveref} 

\def\@setaltaffiliation{\vspace{-\baselineskip}\def\altaffiliation##1{\@par##1\@addpunct.}\altaffiliationes}
\def\@setaltaffiliation{\vspace{-\baselineskip}\def\altaffiliation##1{\@par##1\@addpunct.}\altaffiliationes}

\let\oldequation\align
\let\oldendequation\endalign

\renewenvironment{align}
  {\linenomathNonumbers\oldequation}
  {\oldendequation\endlinenomath}

\begin{document}

\title{Computing with spin qubits at the surface code error threshold}

\author{Xiao~Xue}
\author{Maximilian~Russ}
    \affiliation{QuTech and Kavli Institute of Nanoscience, Delft University of Technology, Lorentzweg 1, 2628 CJ Delft, Netherlands}

\author{Nodar~Samkharadze}
\affiliation{QuTech and Netherlands Organization for Applied Scientific Research (TNO), Stieltjesweg 1, 2628 CK Delft, Netherlands}

\author{Brennan~Undseth}
\affiliation{QuTech and Kavli Institute of Nanoscience, Delft University of Technology, Lorentzweg 1, 2628 CJ Delft, Netherlands}

\author{Amir~Sammak}
\affiliation{QuTech and Netherlands Organization for Applied Scientific Research (TNO), Stieltjesweg 1, 2628 CK Delft, Netherlands}

\author{Giordano~Scappucci}
\author{Lieven~M.~K.~Vandersypen}
\affiliation{QuTech and Kavli Institute of Nanoscience, Delft University of Technology, Lorentzweg 1, 2628 CJ Delft, Netherlands}

\date{\today}

\input{00_Abstract}

\maketitle
\input{01_Introduction}

\input{02_Device}

\input{03_GST_and_1Q}

\input{04_Hamiltonian_engineering_and_2Q}
\input{05_H2_VQE}
\input{06_Conclusion}
\section*{Methods}
\input{Measurement_setup}
\input{Gate_calibration}
\input{Theoretical_model}
\input{Fitting}
\input{Numerical_simulations}
\input{Optimal_pulse_shapes}
\input{gate_set_tomography}

\input{VQE}
\vfill

\input{Extended_figures}
\clearpage

\bibliography{references}

\end{document}

%% file: 00_Abstract.tex
\begin{abstract}

High-fidelity control of quantum bits is paramount for the reliable execution of quantum algorithms and for achieving fault-tolerance, the ability to correct errors faster than they occur~\cite{lidar2013quantum}. The central requirement for fault-tolerance is expressed in terms of an error threshold. Whereas the actual threshold depends on many details, a common target is the $\sim 1\%$ error threshold of the well-known surface code~\cite{fowler2012surface}. Reaching two-qubit gate fidelities above 99\% has been a long-standing major goal for semiconductor spin qubits. These qubits are well positioned for scaling as they can leverage advanced semiconductor technology~\cite{zwerver2021qubits}. Here we report a spin-based quantum processor in silicon with single- and two-qubit gate fidelities all above 99.5\%, extracted from gate set tomography. The average single-qubit gate fidelities remain above 99\% when including crosstalk and idling errors on the neighboring qubit. Utilizing this high-fidelity gate set, we execute the demanding task of calculating molecular ground state energies using a variational quantum eigensolver algorithm~\cite{mcardle2020quantum}. Now that the 99\% barrier for the two-qubit gate fidelity has been surpassed, semiconductor qubits have gained credibility as a leading platform, not only for scaling but also for high-fidelity control. 
\end{abstract}

%% file: 01_Introduction.tex
Quantum computation involves the execution of a large number of elementary operations that take a qubit register through the steps of a quantum algorithm~\cite{nielsen2002quantum}. A major challenge is to implement these operations with sufficient accuracy to arrive at a reliable outcome, even in the presence of decoherence and other error sources. The higher the accuracy, or fidelity, of the operations, the higher the likelihood that near-term applications for quantum computers come in reach~\cite{preskill2018quantum}. Furthermore, for most presently known algorithms, the number of operations that must be concatenated will unavoidably lead to excessive accumulation of errors, and these errors must be removed using quantum error correction~\cite{lidar2013quantum}. Correcting quantum errors faster than they occur is possible when the error probability per operation is below a threshold, the fault-tolerance threshold. For the widely considered surface code, for instance, the fault-tolerance threshold is between 0.6\% and 1\%, under certain assumptions, albeit at the cost of a large redundancy in the number of physical qubits~\cite{fowler2012surface}.\par

Among all the candidate platforms, electron spins in semiconductor quantum dots have advantages for their long coherence times~\cite{veldhorst2015two}, small footprint~\cite{zajac2016scalable}, the potential for scaling up~\cite{vandersypen2017interfacing, li2018crossbar}, and the compatibility with advanced semiconductor manufacturing technology~\cite{zwerver2021qubits}. Single-qubit operations of spin qubits in quantum dots achieve fidelities of 99.9\%~\cite{yoneda2018quantum,yang2019silicon,hendrickx_four-qubit_2021}, but the two-qubit gate fidelities reported vary from 92\% to 98\%~\cite{xue2019benchmarking,huang2019fidelity}. This has limited the two-qubit Bell state fidelities to 94\%~\cite{takeda2020quantum} and quantum algorithms implemented with spin qubits gave only coarsely accurate outcomes~\cite{watson2018programmable,xue2020cmos}.  Pushing the two-qubit gate fidelity well beyond 99\% requires not only low charge noise levels and the elimination of nuclear spins by isotopic enrichment, but also careful Hamiltonian engineering.\par

In this paper, using a precisely engineered two-qubit interaction Hamiltonian, we report the demonstration of single- and two-qubit gates with fidelities above 99.5\%. We use gate set tomography not only to characterize the gates and to quantify the fidelity, but also to improve on the gate calibration. The high-fidelity gates allow us to compute the dissociation energy of molecular hydrogen with a variational quantum eigensolver algorithm, reaching an accuracy for the dissociation energy of around 20 milliHartree, limited by readout errors.

%% file: 02_Device.tex

\begin{figure*}[htbp] 
\center{\includegraphics[width=1\linewidth]{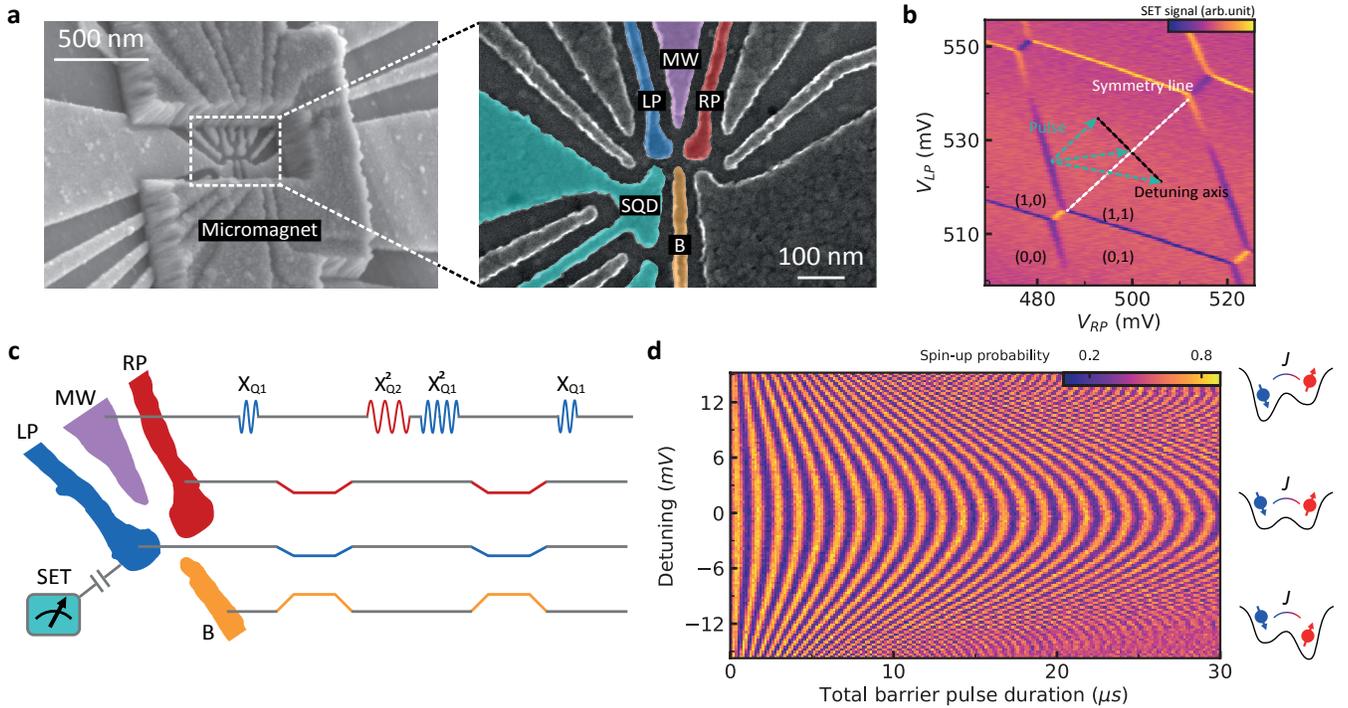}}
\caption{\textbf{a.} Scanning electron microscope images of the silicon quantum processor showing the quantum dot gate pattern and the micromagnet on top. \textbf{b.} Control paths for determining the symmetry operation point in the charge stability diagram. ($M$,$N$) represent the number of electrons in the dots underneath the tip of LP and RP respectively. \textbf{c.} Pulse sequence schematic of a decoupled controlled-phase operation interleaved in a Ramsey interference sequence on Q1. \textbf{d.} Spin-up probability of Q1 after the Ramsey sequence in \textbf{c}, as a function of the detuning in the double dot potential and the total duration of the barrier voltage pulses.}
\label{fig:device}
\end{figure*}

We use a gate-defined double quantum dot in an isotopically enriched $^{28}$Si/SiGe heterostructure (Fig.~\ref{fig:device}a)~\cite{xue2020cmos}, with each dot occupied by one single electron (see Methods). The spin states of the electrons serve as qubits. The spin states are measured with the help of a sensing quantum dot (SQD), which is capacitively coupled to the qubit dots~\cite{elzerman2004single}. A micromagnet on top of the device provides a magnetic field gradient enabling electric-dipole spin resonance (EDSR)~\cite{pioro2008electrically}, and separates the resonance frequencies of the qubits in the presence of an external magnetic field ($\sim$\unit[320]{mT}) to \unit[11.993]{GHz} (Q1) and \unit[11.890]{GHz} (Q2). Single-qubit $X$ and $Y$ gates are implemented by frequency-multiplexed microwave signals applied to gate MW, and virtual $Z$ gates are implemented by a phase update of the reference frame~\cite{vandersypen2005nmr}. The plunger gates (LP and RP) control the chemical potentials of the quantum dots.\par

The native two-qubit gate for spin qubits utilizes the exchange interaction~\cite{loss1998quantum, petta2005coherent}, originating from the wave-function overlap of electrons in neighbouring dots. This selectively shifts the energy of the anti-parallel spin states and thus allows for an electrically pulsed adiabatic \textsc{cphase} gate~\cite{meunier2011efficient,veldhorst2015two,watson2018programmable}. The barrier gate (B) controls the tunnel coupling between the dots, allowing to precisely tune the exchange coupling from < \unit[100]{kHz} to \unit[20]{MHz}. In order to minimize the sensitivity to charge noise, we activate the exchange coupling while avoiding a tilt in the double dot potential~\cite{martins2016noise,reed2016reduced,zhang_2017_tilt} (Fig.~\ref{fig:device}a). 
This symmetric condition can be determined accurately by decoupled adiabatic exchange pulses inside a Ramsey sequence (Fig.~\ref{fig:device}c-d). 
The tunnel barrier is controlled by simultaneously pulsing gate B and compensating on LP and RP to avoid shifts of the electrochemical potentials~\cite{martins2016noise}. The detuning between quantum dots is controlled by additional offsets on the LP and RP pulses in opposite directions. As the decoupling pulses remove additional single-qubit phase accumulation from electron movement in the magnetic field gradient, the spin-up probability of Q1 results in a symmetric Chevron pattern, with the symmetry point at the center (Fig.~\ref{fig:device}e).\par

%% file: 03_GST_and_1Q.tex
\begin{figure}[htbp] 
\center{\includegraphics[width=1\linewidth]{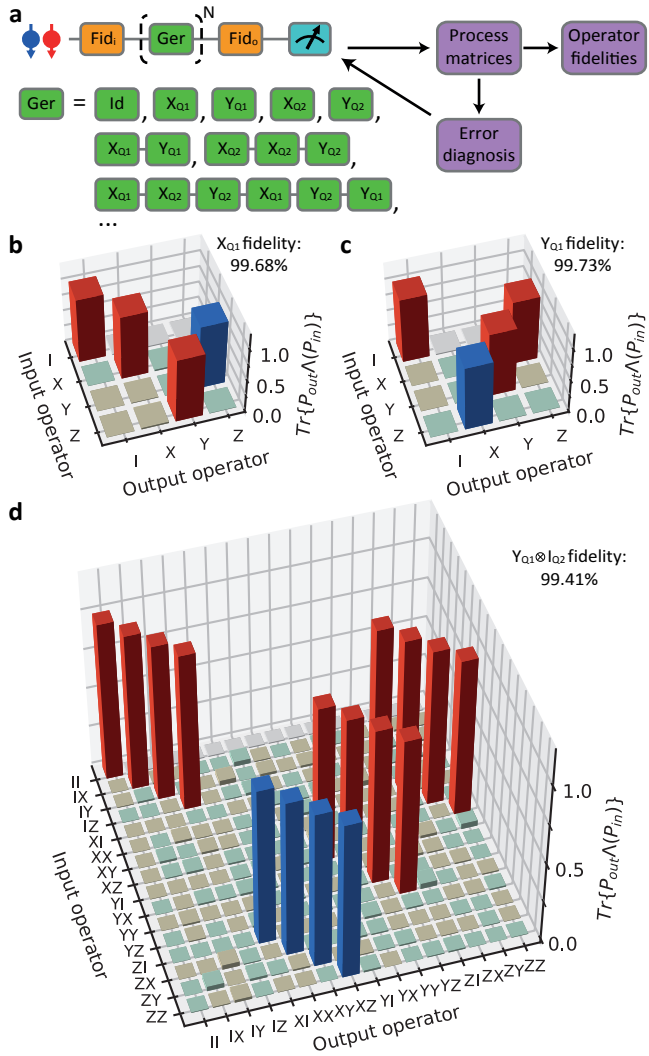}}
\caption{\textbf{a.} Workflow of the GST experiment. Colored blocks show the input and output fiducial sequences ($Fid_i$ and $Fid_o$, orange) and the germ sequences (green). A few examples of single-qubit germ sequences are listed. The outcome is used to adjust pulse parameters in the next run. \textbf{b}-\textbf{c.} PTMs of $X_{Q1}$ and $Y_{Q1}$ in the subspace of Q1. The red (blue) bars are theoretically +1 (-1), and are measured to be positive (negative). The brown (green) bars are theoretically 0 (0) but measured to be positive (negative). \textbf{d.} Experimentally measured PTM of $Y_{Q1}\otimes{I_{Q2}}$ in the complete two-qubit space. The color code is the same as in \textbf{b}-\textbf{c}.}
\label{fig:1QGST}
\end{figure}
\begin{figure}[htbp] 
\center{\includegraphics[width=1\linewidth]{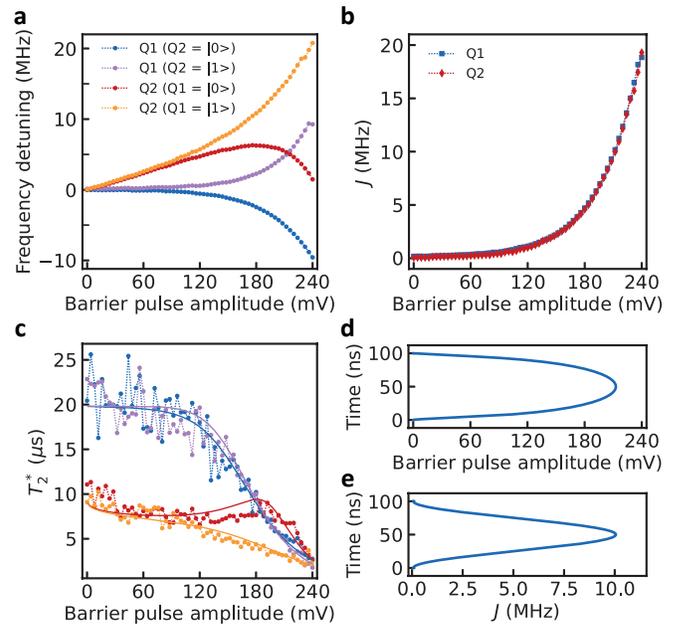}}
\caption{\textbf{a.} Frequency detuning of each qubit conditional on the state of the other qubit as a function of barrier pulse amplitude. The horizontal axis shows the real voltage applied on gate B. \textbf{b.} Exchange strength as a function of barrier pulse amplitude. The data is extracted directly from \textbf{a}. \textbf{c.} $T_2^*$ of each qubit conditional on the state of the other qubit as a function of barrier pulse amplitude (same color code as in \textbf{a}). Each data point is averaged for about 8 minutes. Fitting the $T_2^*$ values to a quasistatic noise model (solid lines, see Methods), the low-frequency amplitudes of the fluctuations are estimated as $\delta f_{Q1}=\unit[11]{kHz}$, $\delta f_{Q2}=\unit[24]{kHz}$, $\delta v_B=\unit[0.4]{m V}$. \textbf{d.} Shape of the barrier pulse, designed to achieve a high-fidelity \textsc{cphase} gate. \textbf{e.} The cosine-shaped $J$ envelope seen by the qubits during the pulse shown in \textbf{d}.}
\label{fig:pulse_engineer}
\end{figure}

Among the various quantum benchmarking techniques, quantum process tomography (QPT) is designed to reconstruct all details in a target process~\cite{nielsen2002quantum}. However, due to the susceptibility of QPT to state preparation and measurement (SPAM) errors, self-consistent benchmarking techniques such as gate set tomography (GST)~\cite{blume2017demonstration} and alternative techniques such as randomized benchmarking (RB)~\cite{magesan2012characterizing} have been developed. Different than RB, GST inherits the advantage of QPT in that it reports the detailed process, which allows us to isolate Hamiltonian errors from stochastic errors and to correct for such errors in the control signals (Extended Data Fig.~\ref{fig:GST_optimization}). We benchmark the fidelities of a universal gate set using gate set tomography~\cite{blume2017demonstration, dehollain2016optimization} (Fig.~\ref{fig:1QGST}a). The gate set we choose contains an idle gate ($I$), sequentially operated single-qubit ${\pi}/2$ rotations about the $\hat{x}$ and $\hat{y}$ axes for each qubit ($X_{Q1}$, $Y_{Q1}$, $X_{Q2}$, and $Y_{Q2}$), and a two-qubit controlled-phase (\textsc{cphase}) gate. A total of 36 fiducial sequences containing $\{null, (X_{Qi})^{n=1,2,3}, {Y_{Qj}}^{n=1,3}\}$ on each qubit, where $null$ unlike the idle gate has no waiting time, are used to tomographically measure the two-qubit state. These fiducials are interleaved by germ sequences and their powers up to a sequence depth of 16. Germs are designed to amplify different types of gate errors in the gate set, such that SPAM errors can be isolated. GST allows using a maximum-likelihood estimator to compute completely positive and trace preserving (CPTP)~\cite{nielsen2002quantum} process matrices for each element of the gate set~\cite{greenbaum2015introduction}. The gate fidelity can be calculated by comparing the measured process using the Pauli transfer matrix (PTM) $\mathcal{M}_\text{exp}$, with the ideal PTM $\mathcal{M}_\text{ideal}$, $F_\text{gate} = (\text{Tr}(\mathcal{M}_\text{exp}^{-1}\mathcal{M}_\text{ideal})+d)/[d(d+1)]$, where $d$ is the dimension of the Hilbert space. These process matrices provide a detailed error diagnosis of the gate set allowing for efficient feedback calibration~(Fig.~\ref{fig:1QGST}a)~\cite{kelly2014optimal}. Analyzing the error generator $\mathcal{L}=\log(\mathcal{M}_\text{exp}\mathcal{M}_\text{ideal}^{-1})$ provides easy access to information. For example, coherent Hamiltonian errors can be isolated from incoherent stochastic errors, and single-qubit errors can be isolated from each other and from two-qubit errors~\cite{blume2021taxonomy}.\par

Figs.~\ref{fig:1QGST}b-c show the reduced PTMs of $X_{Q1}$ and $Y_{Q1}$ operations in the Q1 subspace, and Fig.~\ref{fig:1QGST}d shows the full PTM of $Y_{Q1}$ in two-qubit space ($Y_{Q1}\otimes{I_{Q2}}$) containing additional errors from decoherence and crosstalk on Q2 while operating Q1 (see Extended Data Fig.~\ref{fig:All_PTM} and~\ref{fig:Subspace_PTM} for other PTMs), and from unintentional entanglement due to a residual exchange interaction. The average single-qubit gate fidelity is 99.72\% in the single-qubit subspace ($X_{Q1}$: 99.68\%; $Y_{Q1}$: 99.73\%; $X_{Q2}$: 99.61\%; $Y_{Q2}$: 99.87\%, see the Extended Data Figures for all error bars). 
A metric that is rarely reported is the single-qubit gate fidelity in the full two-qubit space, here 99.16\% on average (see Methods and Extended Data Fig.~\ref{fig:All_PTM}). 
These results highlight that single-qubit benchmarking is not sufficient to identify all errors occurring during single-qubit operations. 
The elimination of idling errors and crosstalk from the microwave drive will be a crucial step in improving the quality of the single-qubit operations further.

%% file: 04_Hamiltonian_engineering_and_2Q.tex
\begin{figure*}[htbp] 
\center{\includegraphics[width=1\linewidth]{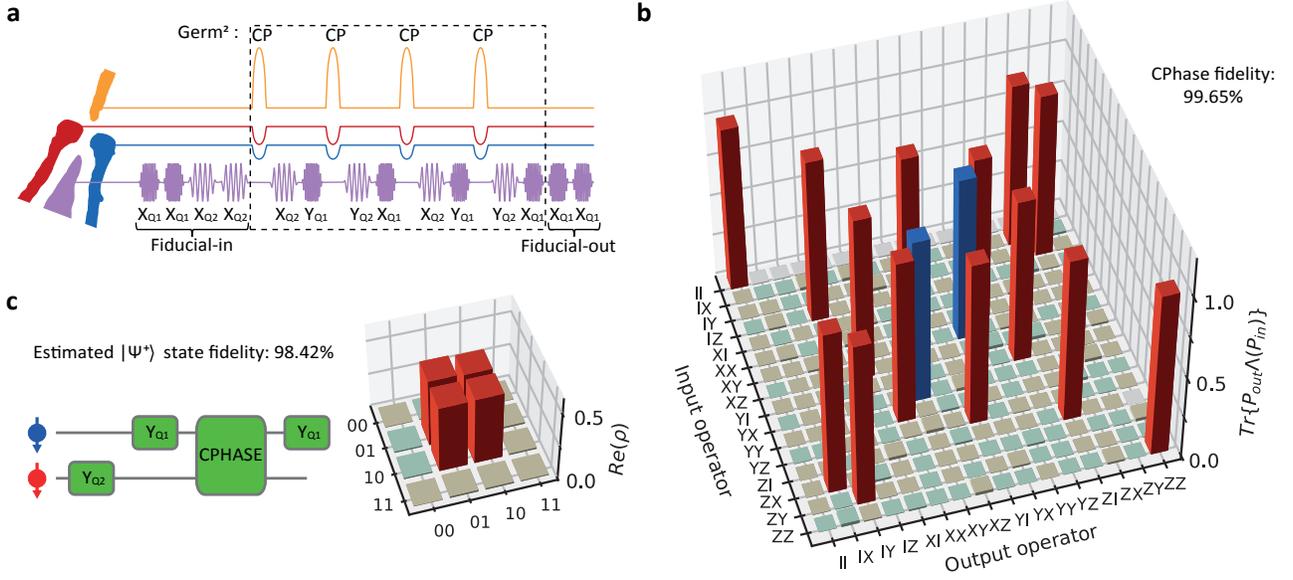}}
\caption{\textbf{a.} A sequence of pulses generated by the AWGs in an example GST sequence. The purple waveforms show the in-phase component of $X/Y$ gates. The \textsc{cphase} gate (shown as CP) is indicated by the orange pulse on gate B with the blue and red compensation pulses on gate LP and gate RP. \textbf{b.} Experimentally determined PTM of a \textsc{cphase} gate. The color code is the same as in Fig.~\ref{fig:1QGST}. \textbf{c} Left is the quantum circuit used to reconstruct the Bell state $\ket{\Psi^+}=(\ket{01} + \ket{10})/\sqrt{2}$ based on the corresponding PTMs. Right is the real part of the reconstructed density matrix of the $\ket{\Psi^+}$ state. The color code is the same as in Fig.\ref{fig:1QGST} except that red (blue) bars here are theoretically +0.5 (-0.5).  
}
\label{fig:2QGST} 
\end{figure*}

For a high-fidelity adiabatic \textsc{cphase} gate, precise control of the exchange coupling, $J$, between the two qubits is required. Specifically, in order to avoid unintended state transitions due to non-adiabatic dynamics, we must be able to carefully shape the envelope of $J$. $J$ is characterized over a wide range using a Ramsey sequence interleaved by a virtual barrier pulse with incremental amplitude $v_B$. Fig.~\ref{fig:pulse_engineer}a shows the measured frequency shift of each qubit as functions of the barrier pulse amplitude and the state of the other qubit. The exchange interaction is modeled to be exponentially dependent on the barrier pulse amplitude $J(v_B)\propto e^{\alpha v_B}$~\cite{cerfontaine2020high, pan_resonant_2020}, while the micromagnet-induced single-qubit frequency shifts follow a linear relationship. By fitting the measured data sets simultaneously to theoretical models (see Methods), $J$ can be extracted very precisely as the difference between the two conditional frequencies of each qubit~\cite{zajac2018resonantly, watson2018programmable} (Fig.~\ref{fig:pulse_engineer}b). 
The barrier pulse $v_B\propto \log(A_{v_B}(1-\cos(2\pi t/t_\text{gate}))/2)$ (Fig.~\ref{fig:pulse_engineer}d) compensates the exponential dependence such that $J\propto (1-\cos(2\pi t/t_\text{gate}))$ follows a cosine window function, which ensures good adiabaticity~\cite{martinis_fast_2014} (Fig.~\ref{fig:pulse_engineer}e). In addition, the virtual gates are calibrated such that the symmetric operation point is maintained for each barrier setting, minimizing the influence of charge noise via the double dot detuning. The most relevant remaining noise sources include charge noise affecting $J$ through fluctuations in the virtual barrier gate $\delta v_B$, and fluctuating qubit frequencies $\delta f_{Q1}, \delta f_{Q2}$ from charge noise entering through artificial spin-orbit coupling from the micromagnet and residual nuclear spin noise coupling through the hyperfine interaction. By analysing the decay of the Ramsey oscillations at each transition frequency, individual dephasing times $T_2^*$ can be extracted, and from there also $\delta v_B$, $\delta f_{Q1}$ and $\delta f_{Q2}$ (Fig.~\ref{fig:pulse_engineer}c). 
\par

Fig.~\ref{fig:2QGST}a shows an example GST pulse sequence that contains twice in a row the germ [$\text{\textsc{cphase}}, X_{Q2}, Y_{Q1}, \text{\textsc{cphase}}, Y_{Q2}, X_{Q1}$]. The PTM of the \textsc{cphase} gate obtained from GST is shown in Fig.~\ref{fig:2QGST}b. Using the detailed information from the error generator to fine tune the calibration parameters, we can achieve a \textsc{cphase} fidelity of $99.65\pm 0.15\%$ (Extended Data Fig.~\ref{fig:Manual_calibration} and~\ref{fig:GST_optimization}). The \textsc{cphase} error generator reveals that at this point incoherent errors dominate. From the obtained PTMs we can numerically estimate Bell state fidelities by multiplications of the PTMs necessary to construct the corresponding state, giving an estimate of 97.75\% - 98.42\% for the four Bell states (Fig.~\ref{fig:2QGST}c and Extended Data Fig.~\ref{fig:Bell_states}).\par

%% file: 05_H2_VQE.tex
\begin{figure*}[htbp] 
\center{\includegraphics[width=1\linewidth]{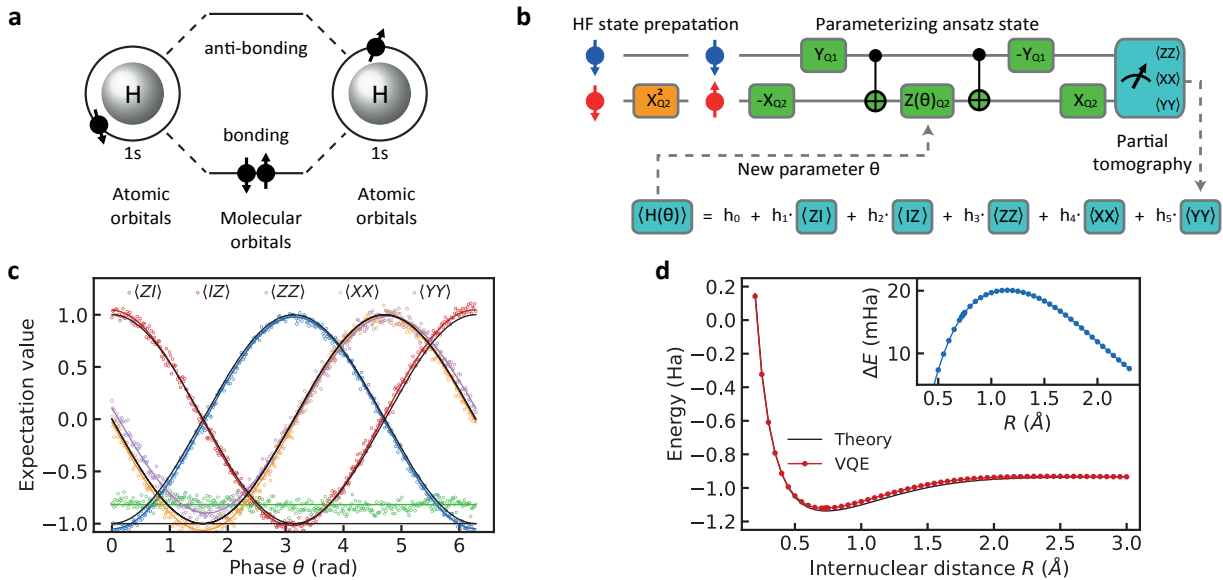}}
\caption{\textbf{a.} Lowest two molecular orbitals of a $H_2$ molecule, formed by the 1$s$ orbitals of two hydrogen atoms. \textbf{b.} The quantum circuit to implement the VQE algorithm for a $H_2$ molecule. The orange block prepares the HF initial state by flipping Q2. The circuit in green blocks creates the parametrized ansatz state. $-X_{Qi}$ and $-Y_{Qj}$ include virtual $Z$ gates. \textsc{CNOT} gates are compiled as [$-Y_{Q2}, \textsc{cphase}, Y_{Q2}$]. To make use of the high-fidelity \textsc{cphase} gate, such compilation is preferred instead of using a single controlled-phase gate with incremental length for creating the parametrized ansatz state. \textbf{c.} Expectation values of the operators in the two-qubit Hamiltonian under BK transformation as a function of $\theta$. Black solid lines show the predicted values. The colored solid lines are sinusoidal fits to the data (and a constant fit for the case of $ZZ$). \textbf{d.} Potential energy of the $H_2$ molecule at varying $R$. The VQE data is normalized to the theoretical energy at large $R$ to directly compare the dissociation energy with the theoretical value. The inset shows the error in the normalized experimental data.} 
\label{fig:H2_VQE}
\end{figure*}

We next test the performance of the high-fidelity gate set in the setting of an actual application. Specifically, we implement a variational quantum eigensolver (VQE) algorithm to compute the ground state energy of molecular hydrogen ($H_2$, Fig.~\ref{fig:H2_VQE}a). In a VQE algorithm, a quantum processor is utilized to implement a classically inefficient subroutine (see Methods and Extended Data Fig.~\ref{fig:VQE_workflow}). The second quantized $H_2$ Hamiltonian can be mapped onto two qubits under the Bravyi-Kitaev (BK) transformation 
$H = h_0II + h_1ZI + h_2IZ + h_3ZZ + h_4XX + h_5YY$.
Here $I$, $X$, $Y$ and $Z$ are Pauli operators, for example $ZI$ is shorthand for $Z \otimes I$, and the coefficients $h_0$-$h_5$ are classically computable functions of the internuclear distance, $R$. Fig.~\ref{fig:H2_VQE}b shows the schematic of the VQE algorithm and its circuit implementation for a $H_2$ molecule. The qubit is initialized in $\ket{01}$, which represents double-occupation of the lowest molecular orbital, corresponding to the Hartree-Fock ground state. A parametrized ansatz state is then prepared by considering single- and double-excitation, which after the BK transformation yields $\ket{\psi(\theta)} = e^{-i{\theta}XY}\ket{01}$, with $\theta$ the parameter to variationally optimize. By performing partial tomography on the ansatz state with an initial guess $\theta_0$, the expectation value of the Hamiltonian for $\ket{\psi(\theta_0)}$ can be calculated. A classical computer can efficiently compute the next guess $\theta_1$ as the new input for the quantum computer. This loop is iterated until the result converges. For a $H_2$ molecule, there is only one parameter $\theta$ to optimize, thus a scan of the entire parameter range of $2\pi$ with finite samples is sufficient to interpolate the smoothly changing measured expectation values. This emulates a real variational algorithm where $\theta$ can be estimated to arbitrary precision by increasing the number of repetitions to suppress statistical fluctuations~\cite{hempel2018quantum}. Fig.~\ref{fig:H2_VQE}c shows the partial tomography result after normalization of the visibility window. The data demonstrates high-quality phase control in the quantum circuits. The deviations in the odd parity expectation values indicate correlations in the readout of the two qubits~\cite{chow2010detecting}. Fig.~\ref{fig:H2_VQE}d shows the energy curves of the $H_2$ molecule from both theory~\cite{mcclean2020openfermion} and the VQE experiment. We observe a minimum energy at around \unit[0.72]{\AA}, and an error of \unit[$\sim20$]{mHa} at the theoretical bond length \unit[0.7414]{\AA}. This accuracy matches the results obtained using superconducting and trapped ion qubits with comparable gate fidelities~\cite{ganzhorn2019gate, hempel2018quantum}.\par

%% file: 06_Conclusion.tex

The two-qubit gate with fidelity above 99.5\% and single-qubit gate fidelities in the two-qubit gate space above 99\% on average, place semiconductor spin qubit logic at the error threshold of the surface code. Recently, a two-qubit operation between nuclear spin qubits in silicon, mediated by an electron spin qubit, has been demonstrated to surpass 99\% fidelity as well, further highlighting that semiconductor spin qubits offer precise two-qubit logic~\cite{madzik2021precision}. Independent studies have shown spin qubit readout with a fidelity above 98\% in only a few $\mu$s~\cite{zheng2019rapid}, with further improvements underway~\cite{schaal2020fast}. With a modest effort in reducing crosstalk errors and in extending the device designs, we are optimistic that the individually demonstrated advantages of semiconductor spin qubits can be combined into a fault-tolerant and highly-integrated quantum computer. The same advances will allow us to implement more sophisticated algorithms in the NISQ era, such as solving energies involving excited states of more complex molecules.

\textbf{Acknowledgements}
We acknowledge fruitful discussions with P. Cerfontaine, C. Bureau-Oxton, M. Madzik, A. Morello, J. Helsen, B. Terhal, M. Veldhorst, and all the members of the spin qubit team, and technical assistance by O. Benningshof, M. Sarsby, R. Schouten and R. Vermeulen. This research was funded by the Dutch Ministry for Economic Affairs through the allowance for Topconsortia for Knowledge and Innovation (TKI) and the Army Research Office (ARO) under grant numbers W911NF-17-1-0274. The views and conclusions contained in this document are those of the authors and should not be interpreted as representing the official policies, either expressed or implied, of the ARO or the US Government. The US Government is authorized to reproduce and distribute reprints for government purposes notwithstanding any copyright notation herein.

\textbf{Author contributions} X.X. performed the experiment with help from N.S. and B.U., M.R developed the theory model and analyzed the data with X.X., N.S. fabricated the quantum dot device, A.S. and G.S. designed and grew the Si/SiGe heterostructure, X.X. and L.M.K.V. conceived the project, L.M.K.V. supervised the project, X.X., M.R. and L.M.K.V. wrote the manuscript with input from all authors.

\textbf{Competing interests}
The authors declare no competing interests.

\textbf{Data availability}
Data supporting this work are available at zenodo, https://doi.org/10.5281/zenodo.5044450.

\textbf{Code availability}
The codes used for data acquisition and processing are from the open source python packages QCoDeS, which is available at https://github.com/QCoDeS/Qcodes, QTT, which is available at https://github.com/QuTech-Delft/qtt, and PycQED, which is available at https://github.com/DiCarloLab-Delft/PycQED{\textunderscore}py3. The codes used for the design and analysis of the gate set tomography experiment are from pyGSTi, which is available at https://github.com/pyGSTio/pyGSTi. And the codes used for the design and analysis of the variational quantum eigensolver experiment are from OpenFermion, which is available at https://github.com/quantumlib/OpenFermion.

%% file: Measurement_setup.tex
\section{Measurement setup}

The measurement setup and device are similar to the one used in Ref.~\cite{xue2020cmos}. We summarize a few key points and all the differences here. The gates LP, RP, and B are connected to arbitrary waveform generators (AWGs, Tektronix 5014C) via coaxial cables. The position in the charge stability diagram of the quantum dots is controlled by voltage pulses applied on LP and RP. Linear combinations of the voltage pulses on B, LP and RP are used to control the exchange coupling between the two qubits at the symmetry point. The compensation coefficients are: $v_{LP}/v_B = -0.081, v_{RP}/v_B = -0.104$. A vector signal generator (VSG, Keysight E8267D) is connected to gate MW and sends frequency-multiplexed microwave bursts (not necessarily time-multiplexed) to implement electric-dipole spin resonance (EDSR). The VSG has two I/Q input channels, receiving I/Q modulation pulses from two channels of an AWG. I/Q modulation is used to control the frequency, phase, and length of the microwave bursts. The current signal of the sensing quantum dot is converted to a voltage signal and recorded by a digitizer card (Spectrum M4i.44), and then converted into 0 or 1 by comparing it to a threshold value.\par

Two differences between the present setup and the one in Ref.~\cite{xue2020cmos} are that 1)  the programmable mechanical switch is configured such that gate MW is always connected to the VSG, and not to the cryo-CMOS control chip; 2) a second AWG of the same model is connected to gate B with its clock synchronized with the first AWG.\par

%% file: Gate_calibration.tex
\section{Gate calibration}

In the gate set used in this work, $\{I$, $X_{Q1}$, $Y_{Q1}$, $X_{Q2}$, $Y_{Q2}$, $\textsc{cphase} \}$, the duration of the $I$ gate and the \textsc{cphase} are set to \SI{100}{\nano\second}, and we calibrate and keep the amplitudes of the single-qubit drives fixed and in the linear response regime. The envelope of the single-qubit gates are shaped following a ``tukey'' window, as it allows adiabatic single-qubit gates with relatively small amplitudes, thus avoids distortion caused by nonlinear response.
The general tukey window of length $t_p$ is given by
\begin{align}
    W(t,r) = \begin{cases} 
      \frac{1}{2}\left[1-\cos\left(\frac{2\pi t }{r t_p}\right)\right] & 0 \leq t \leq \frac{r t_p}{2} \\
      1 & \frac{r t_p}{2} < t\leq t_p -\frac{r t_p}{2} \\
       \frac{1}{2}\left[1-\cos\left(\frac{2\pi (t_p-t) }{r t_p}\right)\right] & t_p -\frac{r t_p}{2}\leq t \leq t_p,
   \end{cases}
   \label{eq:tukey_window}
\end{align}
where $r=0.5$ for our pulses. 
Apart from these fixed parameters, there are 11 free parameters that must be calibrated: single-qubit frequencies $f_{Q1}$ and $f_{Q2}$, burst lengths for single-qubit gates $t_{XY1}$ and $t_{XY2}$, phase shifts caused by single-qubit gates on the addressed qubit itself $\phi_{11}$ and $\phi_{22}$, phase shifts caused by single-qubit gates on the unaddressed ``victim qubit'' $\phi_{12}$ and $\phi_{21}$ ($\phi_{12}$ is the phase shift on Q1 induced by a gate on Q2 and similar for $\phi_{21}$), the peak amplitude of the \textsc{cphase} gate $A_{v_B}$, and phase shifts caused by the gate voltage pulses used for \textsc{cphase} gate on the qubits $\theta_1$ and $\theta_2$ (in addition, we absorb into $\theta_1$ and $\theta_2$ the 90 degree phase shifts needed to transform $\text{diag}(1,i,i,1)$ into $\text{diag}(1,1,1,-1)$).\par
For single-qubit gates, $f_{Q1}$ and $f_{Q2}$ are calibrated by standard Ramsey sequences, which are automatically executed every two hours. The EDSR burst times $t_{XY1}$ and $t_{XY2}$ are initially calibrated by an AllXY calibration protocol~\cite{reed2013entanglement}. The phases $\phi_{11}$, $\phi_{12}$, $\phi_{21}$, and $\phi_{22}$ are initially calibrated by measuring the phase shift of the victim qubit (Q1 for $\phi_{11}$ and $\phi_{21}$; Q2 for $\phi_{22}$ and $\phi_{12}$) in a Ramsey sequence interleaved by a pair of [$X_{Qi}$, $-X_{Qi}$] gates on the addressed qubit (Q1 for $\phi_{11}$ and $\phi_{12}$; Q2 for $\phi_{22}$ and $\phi_{21}$) (Extended Data Fig.~\ref{fig:Manual_calibration}). \par
The optimal pulse design presented in Fig.~\ref{fig:pulse_engineer} gives a rough guidance of the pulse amplitude $A_{v_B}$. In a more precise calibration of the \textsc{cphase} gate, an optional $\pi$-rotation is applied to the control qubit (e.g. Q1) to prepare it into the $\ket{0}$ or $\ket{1}$ state, followed by a Ramsey sequence on the target qubit (Q2) interleaved by an exchange pulse. The amplitude is precisely tuned to bring Q2 completely out of phase (by 180 degree) between the two measurements (Extended Data Fig.~\ref{fig:Manual_calibration} d-e). The phase $\theta_2$ is determined such that the phase of Q2 changes by zero ($\pi$) when Q1 is in the state $\ket{0}$ ($\ket{1}$),  corresponding to the $\textsc{cphase}=\text{diag}(1,1,1,-1)$ in the standard basis. The same measurement is then performed again with the Q2 as the control qubit and Q1 as the target qubit to determine $\theta_1$~\cite{watson2018programmable}.\par
In such a ``conventional'' calibration procedure of the \textsc{cphase} gate, we notice that the two qubits experience different conditional phases (Extended Data Fig.~\ref{fig:Manual_calibration}). We believe that this effect is caused by off-resonant driving from the optional $\pi$-rotation on the control qubit. Similar effects can also affect the calibration of the phase crosstalk from single-qubit gates.\par

This motivates us to use the results from GST as feedback to adjust the gate parameters. The error generators not only describe the total errors of the gates, but also distinguish Hamiltonian errors (coherent errors) from stochastic errors (incoherent errors). We use the information on 7 different Hamiltonian errors ($IX$, $IY$, $XI$, $YI$, $ZI$, $IZ$ and $ZZ$) of each gate, to correct all 11 gate parameters, except $f_{Q1}$ and $f_{Q2}$, for which calibrations using standard Ramsey sequences are sufficient.
For single-qubit gates, $t_{XY1}$ and $t_{XY2}$ are adjusted according to the $IX$, $IY$, $XI$ and $YI$ errors. The phases $\phi_{11}$, $\phi_{12}$, $\phi_{21}$, and $\phi_{22}$ are adjusted according to the $ZI$ and $IZ$ errors. For the \textsc{cphase} gate, $\theta_1$ and $\theta_2$ are adjusted according to the $ZI$ and $IZ$ errors, and $A_{v_B}$ is adjusted according to the $ZZ$ error. The adjusted gates are then used in a new GST loop.

%% file: Theoretical_model.tex
\section{Theoretical model}
In this section we describe the theoretical model used for the fitting, the pulse optimization, and the numerical simulations.
The dynamics of two electron spins in the $(1,1)$ charge configuration can be well-described by an extended Heisenberg model~\cite{loss1998quantum}
\begin{align}
H &=J\,(\vv{S}_1\cdot \vv{S}_2-\frac{1}{4}) +  g\mu_B\vv{B}_1\cdot\vv{S}_1+  g\mu_B\vv{B}_2\cdot\vv{S}_2,
\label{eq:ham_matrix}
\end{align}
with $\vv{S}_j = \hbar(X_j,Y_j,Z_j)^T/2$, where $X_j,Y_j,Z_j$ are the single-qubit Pauli-matrices acting on spin $j=1,2$, $\mu_B$ the Bohr's magneton, $g\approx 2$ the g-factor in silicon, and $\hbar=h/(2\pi)$ the reduced Planck constant. The first and second term describe the interaction of the electron spin in dot 1 and dot 2 with the magnetic fields $\vv{B}_j=(B_{x,j},0,B_{z,j})^T$ originating from the externally applied field and the micromagnet. The transverse components $B_{x,j}$ induce spin-flips, thus, single-qubit gates if modulated resonantly via EDSR.
For later convenience we define the resonance frequencies $hf_{Q1}=g\mu_B B_{z,1}$ and $hf_{Q2}=g\mu_B B_{z,2}$, and the energy difference between the qubits $\Delta E_z= g\mu_B (B_{z,2}- B_{z,1})$. The last term in the Hamiltonian of Eq.~\eqref{eq:ham_matrix} describes the exchange interaction $J$ between the spins in neighboring dots. The exchange interaction originates from the overlap of the wave-functions through virtual tunneling events and is in general a non-linear function of the applied barrier voltage $v_B$. 
We note that $v_B$ determines the compensation pulses on LP and RP for virtual barrier control.
We model $J$ as an exponential function~\cite{cerfontaine2020high,pan_resonant_2020}
\begin{align}
J(v_B) = J_\text{res} e^{2\alpha v_B},
\label{eq:J_sim_expression}
\end{align}
where $J_\text{res}\approx\unit[20-100]{kHz}$ is the residual exchange interaction during idle and single-qubit operations and $\alpha$ the lever arm.
In general the magnetic fields $\vv{B}_j$ depend on the exact position of the electron. We include this in our model $B_{z,j}\rightarrow B_{z,j}(v_B)= B_{z,j}(0) + \beta_j v_B^\gamma$, where $\beta_j$ amounts for the impact of the barrier voltage on the resonance frequency of qubit $j$. The transition energies described in the main text are now given by diagonalizing Hamiltonian from Eq.~\eqref{eq:ham_matrix} and computing the energy difference between the eigenstates corresponding to the computational basis states $\lbrace\ket{00},\ket{01},\ket{10},\ket{11}\rbrace$~\cite{russ_high-fidelity_2018}. We have
\begin{align}
    hf_\text{Q1 (Q2$=\ket{0}$)} &= \mathcal{E}(\ket{10}) - \mathcal{E}(\ket{00}),\label{eq:resonance1} \\
    hf_\text{Q1 (Q2$=\ket{1}$)} &= \mathcal{E}(\ket{11}) - \mathcal{E}(\ket{01}),\label{eq:resonance2} \\
    hf_\text{Q2 (Q1$=\ket{0}$)} &= \mathcal{E}(\ket{01}) - \mathcal{E}(\ket{00}),\label{eq:resonance3} \\
    hf_\text{Q2 (Q1$=\ket{1}$)} &= \mathcal{E}(\ket{11}) - \mathcal{E}(\ket{10}),\label{eq:resonance4}
\end{align}
where $\mathcal{E}(\ket{\xi})$ denotes the eigenenergy of eigenstate $\ket{\xi}$ and $\ket{0}=\ket{\downarrow}$ is defined by the magnetic field direction.\par

In the presence of noise, qubits start to loose information. In silicon, charge noise and nuclear noise are the dominating noise sources. In the absence of two-qubit coupling and correlated charge noise, both qubits decohere largely independently of each other, giving rise to a decoherence time set by the interaction with the nuclear spins and charge noise coupling to the qubit via intrinsic and artificial (via the inhomogeneous magnetic field) spin-orbit interaction. We describe this effect by $f_{Q1}\rightarrow f_{Q1} + \delta f_{Q1}$ and $f_{Q2}\rightarrow f_{Q2} + \delta f_{Q2}$, where $\delta f_{Q1}$ and $\delta f_{Q1}$ are the single-qubit frequency fluctuations. Charge noise additionally can affect both qubits via correlated frequency shifts and the exchange interaction through the barrier voltage, which we model as $v_B\rightarrow v_B + \delta v_B$. In the presence of finite exchange coupling one can define four distinct pure dephasing times, each corresponding to the dephasing of a single qubit with the other qubit in a specific basis state. In a quasistatic approximation the four dephasing times are then given by
\begin{widetext}

\begin{align}
    T_2^\star(\text{Q1 (Q2$=\ket{0}$)}) &= \frac{1}{\sqrt{2}\pi\sqrt{\left[\frac{d\left(hf_\text{Q1 (Q2$=\ket{0}$)})\right)}{d v_B}\right]^2\delta v_B^2
    + \left[\frac{d\left(hf_\text{Q1 (Q2$=\ket{0}$)}\right)}{d hf_{Q1}}\right]^2 \delta f_{Q1}^2
    +\left[\frac{d\left(hf_\text{Q1 (Q2$=\ket{0}$)}\right)}{d hf_{Q2}}\right]^2\delta f_{Q2}^2 }},\label{eq:T2star_1} \\
    T_2^\star(\text{Q1 (Q2$=\ket{1}$)}) &= \frac{1}{\sqrt{2}\pi\sqrt{\left[\frac{d\left(hf_\text{Q1 (Q2$=\ket{1}$)})\right)}{d v_B}\right]^2\delta v_B^2
    + \left[\frac{d\left(hf_\text{Q1 (Q2$=\ket{1}$)}\right)}{d hf_{Q1}}\right]^2 \delta f_{Q1}^2
    +\left[\frac{d\left(hf_\text{Q1 (Q2$=\ket{1}$)}\right)}{d hf_{Q2}}\right]^2\delta f_{Q2}^2 }}, \label{eq:T2star_2} \\
    T_2^\star(\text{Q2 (Q1$=\ket{0}$)}) &= \frac{1}{\sqrt{2}\pi\sqrt{\left[\frac{d\left(hf_\text{Q2 (Q1$=\ket{0}$)})\right)}{d v_B}\right]^2\delta v_B^2
    + \left[\frac{d\left(hf_\text{Q2 (Q1$=\ket{0}$)}\right)}{d hf_{Q1}}\right]^2 \delta f_{Q1}^2
    +\left[\frac{d\left(hf_\text{Q2 (Q1$=\ket{0}$)}\right)}{d hf_{Q2}}\right]^2\delta f_{Q2}^2 }}, \label{eq:T2star_3} \\
    T_2^\star(\text{Q2 (Q1$=\ket{1}$)}) &= \frac{1}{\sqrt{2}\pi\sqrt{\left[\frac{d\left(hf_\text{Q2 (Q1$=\ket{1}$)})\right)}{d v_B}\right]^2\delta v_B^2
    + \left[\frac{d\left(hf_\text{Q2 (Q1$=\ket{1}$)}\right)}{d hf_{Q1}}\right]^2 \delta f_{Q1}^2
    +\left[\frac{d\left(hf_\text{Q2 (Q1$=\ket{1}$)}\right)}{d hf_{Q2}}\right]^2\delta f_{Q2}^2 }}.
    \label{eq:T2star_4} 
\end{align}
\end{widetext}

%% file: Fitting.tex
\section{Fitting qubit frequencies and dephasing times}
The transition energies Eqs.~\eqref{eq:resonance1}-\eqref{eq:resonance4} are fitted simultaneously to the measured results from the Ramsey experiment (see Fig.~\ref{fig:pulse_engineer}a). For the fitting we use the \texttt{NonLinearModelFit} function from the software Mathematica with the least square method. The best fits yield the following parameters $\alpha=\unit[12.1\pm 0.05]{V^{-1}}$, $\beta_1=\unit[-2.91\pm 0.11]{MHz^\gamma/V^\gamma}$, $\beta_2=\unit[67.2\pm 0.63]{MHz^\gamma/V^\gamma}$, and $\gamma=\unit[1.20\pm 0.01]{}$, and $J_\text{res}=\unit[58.8\pm 1.8]{kHz}$.\par

The dephasing times Eqs.~\eqref{eq:T2star_1}-\eqref{eq:T2star_4} are fitted simultaneously to the measured results from the Ramsey experiment (see Fig.~\ref{fig:pulse_engineer}c) using the same method. The best fits yield the following parameters $\delta v_B=\unit[0.40\pm 0.01]{m V}$, $\delta f_{Q1}=\unit[11\pm 0.1]{kHz}$, and $\delta f_{Q2}=\unit[24\pm 0.7]{kHz}$. 

%% file: Numerical_simulations.tex
\section{Numerical simulations}
\label{sec:num_sim}

For all numerical simulations,  we solve the time-dependent Schr{\"o}dinger equation 
\begin{align}
    i\hbar \ket{\psi(t)} = H \ket{\psi(t)}
\end{align}
and iteratively compute the unitary propagator according to
\begin{align}
    U(t+\Delta t) = e^{-\frac{i}{\hbar} H(t+\Delta t)}U(t).
\end{align}
Here, $H(t+\Delta t$ is discretized into $N$ segments of length $\Delta t$ such that $H(t)$ is constant in the time-interval $\left[t,t+\Delta t\right]$.
All simulations are performed in the rotating frame of the external magnetic field $(B_{z,1}+B_{z,2})/2$ and neglecting the counter-rotating terms, making the so-called rotating wave approximation (RWA). This allows us to chose $\Delta t=\unit[10]{ps}$ as a sufficiently small time step. 

For the noise simulations, we included classical fluctuations of $f_{Q1}\rightarrow f_{Q1} + \delta f_{Q1}$, $f_{Q2}\rightarrow f_{Q2} + \delta f_{Q2}$, and $v_B\rightarrow v_B + \delta v_B$. We assume the noise coupling to the resonance frequencies $\delta f_{Q1}$ and $\delta f_{Q2}$ to be quasistatic and assume $1/f$ noise for $v_B$ which we describe by its spectral density $S_{\delta v_B}(\omega)= \delta v_B/\omega$. To compute time traces of the fluctuation we use the approach introduced in Refs.~\cite{yang_achieving_2019,koski_strong_2020} to generate time-correlated time traces. The fluctuations are discretized into $N$ segments with time $\Delta t$ such that $\delta v_B(t)$ is constant in the time interval $\left[t,t+\Delta t\right)$, with the same $\Delta t$ as above. Consequently, fluctuations which are faster than $f_\text{max}=\frac{1}{\Delta t}$ are truncated. 

%% file: Optimal_pulse_shapes.tex
\section{\textsc{cphase} gate}
We realize a universal $\textsc{cphase}=\text{diag}(1,1,1,-1)$ gate by adiabatically pulsing the exchange interaction using a carefully designed pulse shape. Starting from Eq.~\eqref{eq:ham_matrix}, the full dynamics can be projected on the odd-parity space spanned by $\ket{01}$ and $\ket{10}$. The entangling exchange gate is reduced in this subspace to a global phase shift thus the goal is to minimize any dynamics inside the subspace. Introducing a new set of Pauli operators in this subspace
$\sigma_x=\ket{01}\bra{10}+\ket{10}\bra{01}$, $\sigma_y=-i \ket{01}\bra{10}+i\ket{10}\bra{01}$, and $\sigma_z=\ket{01}\bra{01}-\ket{10}\bra{10}$, we find 
\begin{align}
H_{\text{sub}}(t)=\frac{1}{2}\big( - J\left(v_B(t)\right) + \Delta E_z \,\sigma_z+J\left(v_B(t)\right)\, \sigma_x\big).
\label{eq:HheisSub}
\end{align}
In order to investigate the adiabatic behaviour, it is convenient to switch into the adiabatic frame defined by $U_\text{ad}=e^{-\frac{i}{2} \tan^{-1}\left(\frac{J\left(v_B(t)\right)}{\Delta E_z}\right)\sigma_y}$. The Hamiltonian accordingly transforms as
\begin{align}
H_{\text{ad}}&=U_\text{ad}^\dagger(t) H_{\text{sub}}(t) U_\text{ad}(t) -i \hbar U_\text{ad}^\dagger(t)\dot{U}_\text{ad}(t)\label{eq:Ham_ramp_part1}\\
&\approx \frac{1}{2}\bigg(- J\left(v_B(t)\right) +  \Delta E_z\,\sigma_z 
-\frac{\hbar\dot{J}}{\Delta E_z}\sigma_y \bigg),
\label{eq:Ham_ramp_part2}
\end{align}
where the first term is unaffected and describes the global phase accumulation due to the exchange interaction, the second term describes the single-qubit phase accumulations, and the last term $f(t)=\hbar\dot{J}/(2\Delta E_z)$ describes the diabatic deviation proportional to the derivative of the exchange pulse. From Eq.~\eqref{eq:Ham_ramp_part1} to Eq.~\eqref{eq:Ham_ramp_part2} we assumed a constant $\Delta E_z(t)\approx \Delta E_z$, and $J(t)\ll \Delta E_z$. The transition probability from state $\ket{\uparrow\downarrow}$ to $\ket{\downarrow\uparrow}$ using a pulse of length $t_p$ is then given by~\cite{martinis_fast_2014}
\begin{align}
P_{\ket{\uparrow\downarrow}\rightarrow\ket{\downarrow\uparrow}} &\approx \left|\int_{0}^{t_p} f(t) e^{-\frac{i}{\hbar} \Delta E_z t }\, dt \right|^2 \label{eq:ESD_part1} \\
&= S\big(f(t)\big).
\label{eq:ESD}
\end{align}
From the first to the second line we replaced the integral by the (short time-scale) Fourier transform, allowing us to describe the spin-flip error probability by the energy spectral density (ESD) of the input signal $f(t)$. Minimizing such errors is therefore identical to minimizing the ESD of a pulse, a well-known and solved problem from classical signal processing and statistics. Optimal shapes are commonly referred to as window functions $W(t)$ due to their property to restrict the spectral resolution of signals. 
A high-fidelity exchange pulse is consequently given by $J(0)=J(t_p)$ and 
\begin{align}
    \int_0^{t_p}dt J\left(v_B(t)\right)/\hbar=\pi/2,
    \label{eq:cphase_condition}
\end{align}
while setting $J(t)= A_{v_B} W(t) J_\text{res}$~\cite{martinis_fast_2014}, with a scaling factor $A_{v_B}$ that is to be determined. In this work, we have chosen the cosine window
\begin{align}
    W(t)=\frac{1}{2}\left[1-\cos\left(\frac{2\pi t }{t_p}\right)\right],
\end{align}
from signal processing which has a high spectral resolution. The amplitude $A_{v_B}$ follows from condition Eq.~\eqref{eq:cphase_condition}. For a pulse length of $t_p=\unit[100]{ns}$ and a cosine pulse shape we find $A_{v_B} J_\text{res}=\unit[10.06]{MHz}$. Our numerical simulations predict an average gate infidelity $1-F_\text{gate}<10^{-6}$ without noise and $1-F=0.22\times 10^{-3}$ with the inclusion of noise through the fluctuations $\delta f_{Q1}$, $\delta f_{Q2}$, and $\delta v_B$ discussed in section~\ref{sec:num_sim}. 
As explained in the main text, due to the exponential voltage-exchange relation the target pulse shape for J(t) must be converted to a barrier gate pulse, following~\cite{russ_soon_to_be_appear}
\begin{align}
    v_B(t)=\frac{1}{2\alpha}\text{log}\left(A_{v_B}\, W(t)\right).
\end{align}
The numerical simulations with the fitted noise parameters in the simplified model from Section~\ref{sec:num_sim} predict a gate fidelity above 99.97\%. The measured PTMs reveal significantly higher rates of incoherent errors, which we attribute to drifts in the barrier voltage on a timescale much longer than the timescale on which $\delta f_{Q1}$, $\delta f_{Q2}$, and $\delta v_B$ were determined. 
 

%% file: gate_set_tomography.tex
\section{Gate set tomography analysis}
We designed a customized gate set tomography (GST) experiment using the gate set $\{ I$, $X_{Q1}$, $Y_{Q1}$, $X_{Q2}$, $Y_{Q2}$, $\textsc{cphase}\}$, where $I$ is a $\unit[100]{ns}$ idle gate, $X_{Q1}$ ($Y_{Q1}$) and $X_{Q2}$ ($Y_{Q2}$) are single qubit $\pi/2$ gates on qubit 1 and 2 with rotation axis $\hat{x}$ ($\hat{y}$), and $\textsc{cphase}=\text{diag}(1,1,1,-1)$. A classic two-qubit GST experiment consists of a set of germs designed to amplify any error rate in the sequence when repeated,  and a set of 36 fiducials composed by the 11 elementary operations $\{\textit{null}$, $X_{Q1}$, $X_{Q1}X_{Q1}$, $X_{Q1}X_{Q1}X_{Q1}$, $Y_{Q1}$, $Y_{Q1}Y_{Q1}Y_{Q1}$, $X_{Q2}$, $X_{Q2}X_{Q2}$, $X_{Q2}X_{Q2}X_{Q2}$, $Y_{Q2}$, $Y_{Q2}Y_{Q2}Y_{Q2}\}$ required to do quantum process tomography of the germs~\cite{greenbaum2015introduction}. Note, that the $\textit{null}$ gate is the instruction for doing nothing in zero time, different from the idle gate. The germs and fiducials are then compiled into GST sequences such that each sequence consists of two fiducials interleaved by a single germ or power of germs (as illustrated in Fig.~\ref{fig:1QGST}a of the main text)~\cite{nielsen2020gate}. The GST sequences are classified by their germ powers into lengths $L=1,2,4,8,16\cdots$, where a sequence of length $n$ consists of $n$ gates plus the fiducial gates. After the execution of all sequences a maximum-likelihood estimation (MLE) is performed to estimate the process matrices of each gate in the gate set and the SPAM probabilities. We use the open source \texttt{pyGSTi} python package~\cite{nielsen_python_2019,nielsen_probing_2020} to perform the MLE, as well as to design a reduced GST experiment by eliminating redundant circuits, and to provide statistical error bars by computing all involved Hessians. The circuit optimization allows us to perform GST with a maximum sequence length $L_\text{max}=16$ using 1685 different sequences in total.
The \texttt{pyGSTi} package quantifies the Markovian-model violation of the experimental data counting the number of standard deviations exceeding their expectation values under the $\chi^2$ hypothesis~\cite{nielsen_probing_2020}. This model violation is internally translated into a more accessible goodness ratio from $0-5$ with 5 being the best~\cite{nielsen_python_2019}, where we get a 4 out of 5 rating indicating remarkably small deviations from expected results.

From the gate set tomography experiment, we have extracted the Pauli transfer matrix (PTM) $\mathcal{M}_\text{exp}$ describing each gate in our gate set $\{I$,  $X_{Q1}$, $Y_{Q1}$, $X_{Q2}$, $Y_{Q2}$, $\textsc{cphase}\}$. The PTM is isomorphically related to the conventionally used $\chi$-matrix describing a quantum process. A completely positive trace-preserving (CPTP) two-qubit PTM has 240 parameters describing the process. To get insight in the errors of the gates in the experiment, we first compute the error in the PTM given by $E=\mathcal{M}_\text{exp}\mathcal{M}_\text{ideal}^{-1}$, where we have adapted the convention to add the error after the ideal gate. The average gate fidelity is then conveniently given by
\begin{align}
F_\text{gate} = \frac{\text{Tr}(\mathcal{M}_\text{exp}^{-1}\mathcal{M}_\text{ideal})+d}{d(d+1)}.
\label{eq:averageGate}
\end{align}
It is related to the entanglement fidelity via $F_\text{ent}=\frac{d+1}{d}F_\text{gate}$~\cite{white_measuring_2007}, where $d$ is the dimension of the two-qubit Hilbert space. While the PTM $\mathcal{M}$ perfectly describes the errors, it is more intuitive to analyze the corresponding error generator $\mathcal{L}=\log(E)$ of the process~\cite{blume2021taxonomy}. The error generator relates $\mathcal{L}$ to the error PTM $E$ in a similar way as a Hamiltonian relates $H$ to a unitary operation $U=e^{-i H}$. The error generator can be separated into several blocks. A full discussion about the error generator can be found in Ref.~\cite{blume2021taxonomy}. In this work, we have used the error generator to distinguish the dynamics originating from coherent Hamiltonian errors, which can be corrected by adjusting gate parameters (see Extended Data Figure~\ref{fig:GST_optimization}, from noisy/stochastic dynamics which cannot be corrected easily. The coherent errors can be extracted by projecting $\mathcal{L}$ onto the $4\times 4$-dimensional Hamiltonian space $H$. In the Hilbert-Schmidt space, the Hamiltonian projection is given by~\cite{blume2021taxonomy}
\begin{align}
    H_{mn} = -\frac{i}{d^2}\text{Tr}\left[\left(P_m^T\otimes P_n^T\otimes \boldsymbol{1}_d - \boldsymbol{1}_d\otimes P_m\otimes P_n \right)\mathcal{L}_\text{sup}\right],
\end{align}
where $\mathcal{L}_\text{sup}$ is the error generator in Liouville superoperator form, $P_m\in{I,X,Y,Z}$ are the extended Pauli matrices with $m,n=0,1,2,3$, $\boldsymbol{1}_d$ is the $d$-dimensional Identity matrix, and $d=4$ is the dimension of the two-qubit Hilbert space. To improve the calibration of our gate set, we use the information of 7 different Hamiltonian errors ($IX$, $IY$, $XI$, $YI$, $ZI$, $IZ$ and $ZZ$).  To estimate coherent Hamiltonian errors and incoherent stochastic errors, two new metrics are considered~\cite{blume2021taxonomy}; the Jamiolkowski probability 
\begin{align}
    \epsilon_J(\mathcal{L}) = -\text{Tr}(\rho_J(\mathcal{L})\ket{\Psi}\bra{\Psi})),
    \label{eq:Jam_prob}
\end{align}
which describes the amount of incoherent error in the process, and the Jamiolkowski amplitude
\begin{align}
    \theta_J(\mathcal{L}) = ||(1-\ket{\Psi}\bra{\Psi})\rho_J(\mathcal{L})\ket{\Psi}||_2,
    \label{eq:Jam_amp}
\end{align} which approximately describes the amount of coherent Hamiltonian errors (Extended Data Table.~\ref{tab:GST_results}). Here, $\rho_J(\mathcal{L}) = (\mathcal{L}\otimes\boldsymbol{1}_{d^2})[\ket{\Psi}\bra{\Psi}]$ is the Jamiolkowski state and $\ket{\Psi}$ is a maximally entangling four-qubit state which originates from the relation of quantum processes to states in a Hilbert space twice the dimension via the Choi-Jamiolkowski isomorphism~\cite{jamiolkowski_linear_1972}. For small errors, the average gate infidelity can be approximated by~\cite{blume2021taxonomy}
\begin{align}
    1-F_\text{gate}=\frac{d}{d+1}\left[\epsilon_J(\mathcal{L})+\theta_J(\mathcal{L})^2\right].
\end{align}
For a comparison of the performance of the single-qubit gates with previous experiments reporting single-qubit gate fidelities, we compute the fidelities projected to the single-qubit space from the PTMs or the error generators. In Fig.~\ref{fig:1QGST} and Extended Data Fig.~\ref{fig:Subspace_PTM}, single-qubit gate fidelities are estimated by projecting the PTMs onto corresponding subspace. Let $\mathcal{P}_{j}$ be the projector on the subspace of qubit $j$ then the fidelity is given by
\begin{align}
F_\text{sub} = \frac{\text{Tr}(\mathcal{P}_{j}\mathcal{M}_\text{exp}^{-1}\mathcal{P}_{j}\mathcal{M}_\text{ideal})+(d/2)}{(d/2)((d/2)+1)}.
\label{eq:averageGate_sub}
\end{align}
Error bars for the fidelity projected to the subspace are computed using standard error propagation of the confidence intervals of $\mathcal{M}_\text{exp}$ provided by the \texttt{pyGSTi} package.
A more optimistic estimation for the fidelities in the single-qubit subspace is given by projecting the error generators instead of the PTMs.

%% file: VQE.tex
\section{VQE}

We follow the approach of Ref.~\cite{hempel2018quantum} to using the VQE algorithm to compute the ground state of molecular hydrogen, after mapping this state onto the state of two qubits. We include this information here for completeness.
The Hamiltonian of a molecular system in atomic units ($\hbar = 1$) reads 
\begin{align}
    H = &-\sum_{i} \frac{\nabla_{R_i}^2}{2M_i} -\sum_{j} \frac{\nabla_{r_j}^2}{2} - \sum_{i,j} \frac{Q_i}{\abs{R_i - r_j}}\nonumber\\
    &+\sum_{i,j>i} \frac{Q_iQ_j}{\abs{R_i - R_j}} + \sum_{i,j>i} \frac{1}{\abs{r_i - r_j}},
\end{align}
where $R_i$, $M_i$ and $Q_i$ are the position, mass and charge of the $i$-th nuclei, and $r_j$ is the position of the $j$-th electron. The first two sums describe the kinetic energies of the nuclei and electrons, respectively. The last three sums describe the Coulomb repulsion between nuclei and electrons, nuclei and nuclei, and electrons and electrons, respectively. As we are primarily interested in the electronic structure of the molecule, and nuclear masses are a few orders of magnitude larger than the electron masses, the nuclei are treated as static point charges under the Born-Oppenheimer approximation. Consequentially, the electronic Hamiltonian can be simplified to
\begin{align}
    H_e = &-\sum_{i} \frac{\nabla_{r_i}^2}{2} - \sum_{i,j} \frac{Q_i}{\abs{R_i - r_j}} + \sum_{i,j>i} \frac{1}{\abs{r_i - r_j}}.
\end{align}
Switching into the second-quantization representation, described by fermionic creation and annihilation operators, $a_p^{\dagger}$ and $a_q$, acting on a finite basis, the Hamiltonian becomes
\begin{align}
    H_e = \sum_{pq}h_{pq}a_p^{\dagger}a_q + \sum_{pqrs}h_{pqrs}a_p^{\dagger}a_q^{\dagger}a_ra_s.
\end{align}
The anti-symmetry under exchange is retained through the anti-commutation relation of the operators.
The weights of the two sums are given by the integrals 
\begin{align}
    &h_{pq} = \int d\sigma\psi_p^*(\sigma)(\frac{\nabla_{r_i}^2}{2} - \sum_{i} \frac{Q_i}{\abs{R_i - r}})\psi_q^*(\sigma), \\
    &h_{pqrs} = \int d\sigma_1d\sigma_2\frac{\psi_p^*(\sigma_1)\psi_q^*(\sigma_2)\psi_s(\sigma_1)\psi_r(\sigma_2)}{\abs{r_1 - r_2}}.
\end{align}
Such a second-quantized molecular Hamiltonian can be mapped onto qubits using the Jordan-Wigner (JW) or the Bravyi-Kitaev (BK) transformation~\cite{mcardle2020quantum}. The JW transformation directly encodes the occupation number (0 or 1) of the $i$-th spin-orbital into the state ($\ket{0}$ or $\ket{1}$) of the $i$-th qubit. The number of qubits required after JW transformation is thus the same as the number of spin-orbitals that are of interest. The BK transformation, on the other hand, encodes the information in both the occupation number and parities -- whether there is an even or odd occupation in a subset of spin-orbitals.\par

Taking molecular hydrogen in the Hartree-Fock basis as an example, we are interested in investigating the bonding ($\ket{O_1\uparrow}$, $\ket{O_1\downarrow}$) and the anti-bonding orbital state ($\ket{O_2\uparrow}$, $\ket{O_2\downarrow}$). The initial guess of the solution is the Hartree-Fock (HF) state in which both electrons occupy the $\ket{O_1}$ orbital. The JW transformation encodes the HF initial state as $\ket{0011}$, representing $\ket{N_{O_2\downarrow} N_{O_2\uparrow} N_{O_1\downarrow} N_{O_1\uparrow}}$ from left to right, where $N_{O_iS}$ is the occupation of the $O_iS$ spin-orbital with $S=\uparrow,\downarrow$. The BK transformation encodes the HF initial state as $\ket{0001}$, where the first and the third qubit (counting from the right) encode the occupation number of the first and third spin-orbital ($N_{O_1\uparrow} = 1$ and $N_{O_2\uparrow} = 0$), the second qubit encodes the parity of the first two spin-orbitals ($(N_{O_1\uparrow} + N_{O_1\downarrow})$ mod 2 = 0), and the fourth qubit encodes the parity of all four spin-orbitals ($(N_{O_1\uparrow} + N_{O_1\downarrow} + N_{O_2\uparrow} + N_{O_2\downarrow})$ mod 2 = 0). With the standard transformation rules for fermionic creation and annihilation operators, the system Hamiltonian becomes a four-qubit Hamiltonian 
\begin{align}
    H_{JW} = &~g_0I + g_1Z_1 + g_2Z_2 + g_3Z_3 + g_4Z_4\nonumber\\
             &~+ g_5Z_1Z_2 + g_6Z_1Z_3 + g_7Z_1Z_4\nonumber\\ 
             &~+ g_8Z_2Z_3 + g_9Z_2Z_4 + g_{10}Z_3Z_4\nonumber\\
             &~+ g_{11}Y_1X_2X_3X_4 + g_{12}Y_1Y_2X_3X_4\nonumber\\
             &~+ g_{13}X_1X_2Y_3Y_4 + g_{14}X_1Y_2Y_3Y_4,\\
    H_{BK} = &~g_0I + g_1Z_1 + g_2Z_2 + g_3Z_3\nonumber\\
             &~+ g_4Z_1Z_2 + g_5Z_1Z_3 + g_6Z_2Z_4\nonumber\\
             &~+ g_7Z_1Z_2Z_3 + g_8Z1Z_3Z_4 + g_{9}Z_2Z_3Z_4\nonumber\\
             &~+ g_{10}Z_1Z_2Z_3Z_4 + g_{11}X_1Z_2X_3\nonumber\\ 
             &~+ g_{12}Y_1Z_2Y_3 + g_{13}Y_1Z_2Y_3Z_4.
\end{align}
We see that due to the symmetry of the represented system in $H_{BK}$, qubit 2 and qubit 4 are never flipped, allowing us to reduce the dimension of the Hamiltonian to
\begin{align}
    H_{BK}^\text{reduced} = &~h_0I + h_1Z_1 + h_2Z_2 + h_3Z_1Z_2 \nonumber\\
                        &~+ h_4X_0X_1 + h_5Y_0Y_1.
\end{align}
This reduced representation requires only two qubits to simulate the hydrogen molecule. The HF initial state thus becomes $\ket{01}$. We emphasize that such a reduction of the BK Hamiltonian is not a special case for $H_2$ molecule but is connected to symmetry considerations to reduce the complexity of systems, in a scalable way.\par

The variational quantum eigensolver (VQE) is a method to compute the ground state energy of the Hamiltonian.
The total energy can be directly calculated by measuring the expectation value of each Hamiltonian term. This can be done easily by partial quantum state tomography. All the expectation values are then added up with a set of weights ($h_0$ through $h_5$). The weights are only functions of the internuclear separation ($R$) and can be computed efficiently by a classical computer. Here we use the \texttt{OpenFermion} python package to compute these weights~\cite{mcclean2020openfermion}.\par
The main task of the quantum processor is then to encode the molecular spin-orbital state into the qubits. The starting point is the HF initial state, which is believed to largely overlap with the actual ground state. In order to find the actual ground state, the initial state needs to be ``parameterized'' into an ansatz to explore a subspace of all possible states. We apply the unitary coupled cluster (UCC) theory to the parameterized ansatz state, which is widely believed to be a powerful approach and cannot be efficiently executed on a classical computer~\cite{taube2006new}. The UCC operator has a format
\begin{align}
    U_{UCC}(\vv{\theta}) = e^{\sum_n (T_n(\vv{\theta}) - T_n^{\dagger}(\vv{\theta}))},
\end{align}
with 
\begin{align}
    &T_1(\vv{\theta}) = \sum_{m,i} \vv{\theta}_i^m a_m^{\dagger}a_i,\\
    &T_2(\vv{\theta}) = \sum_{m,n,i,j} \vv{\theta}_{i,j}^{m,n} a_m^{\dagger}a_n^{\dagger}a_ia_j
\end{align}
representing single-excitation and double-excitation of the electrons. The indices $i, j$ label the occupied spin-orbitals and $m, n$ are the labels of the unoccupied spin-orbitals. The vector $\vv{\theta}$ is the set of all parameters to optimize. In the case of a $H_2$ molecule, the UCC operator is transformed into a qubit operator as
\begin{align}
    U_{UCC}^{BK}(\vv{\theta}) = e^{-i\theta XY},
\end{align}
where $\theta$ is a single parameter to variationally optimize.

%% file: Extended_figures.tex
\renewcommand{\figurename}{Extended Data Fig.}

\setcounter{figure}{0}

\begin{figure*}[htbp] 
\center{\includegraphics[width=1\linewidth]{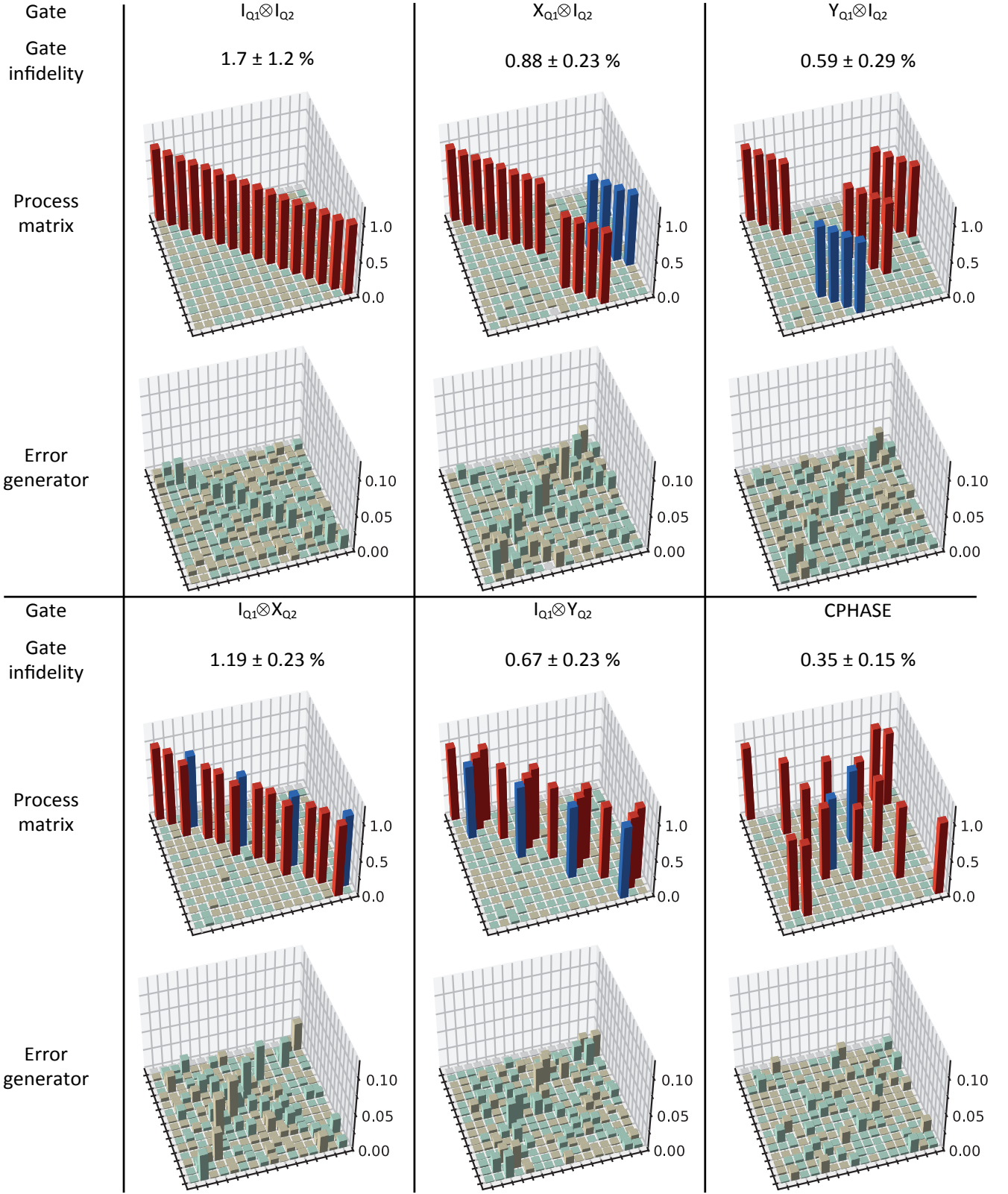}}
\caption{Average gate infidelity, process matrices (PTMs), and error generators of the 6 quantum gates in the chosen gate set. These results are analyzed by the \texttt{pyGSTi} package using maximum-likelihood estimation.} 
\label{fig:All_PTM}
\end{figure*}

\begin{figure*}[htbp] 
\center{\includegraphics[width=1\linewidth]{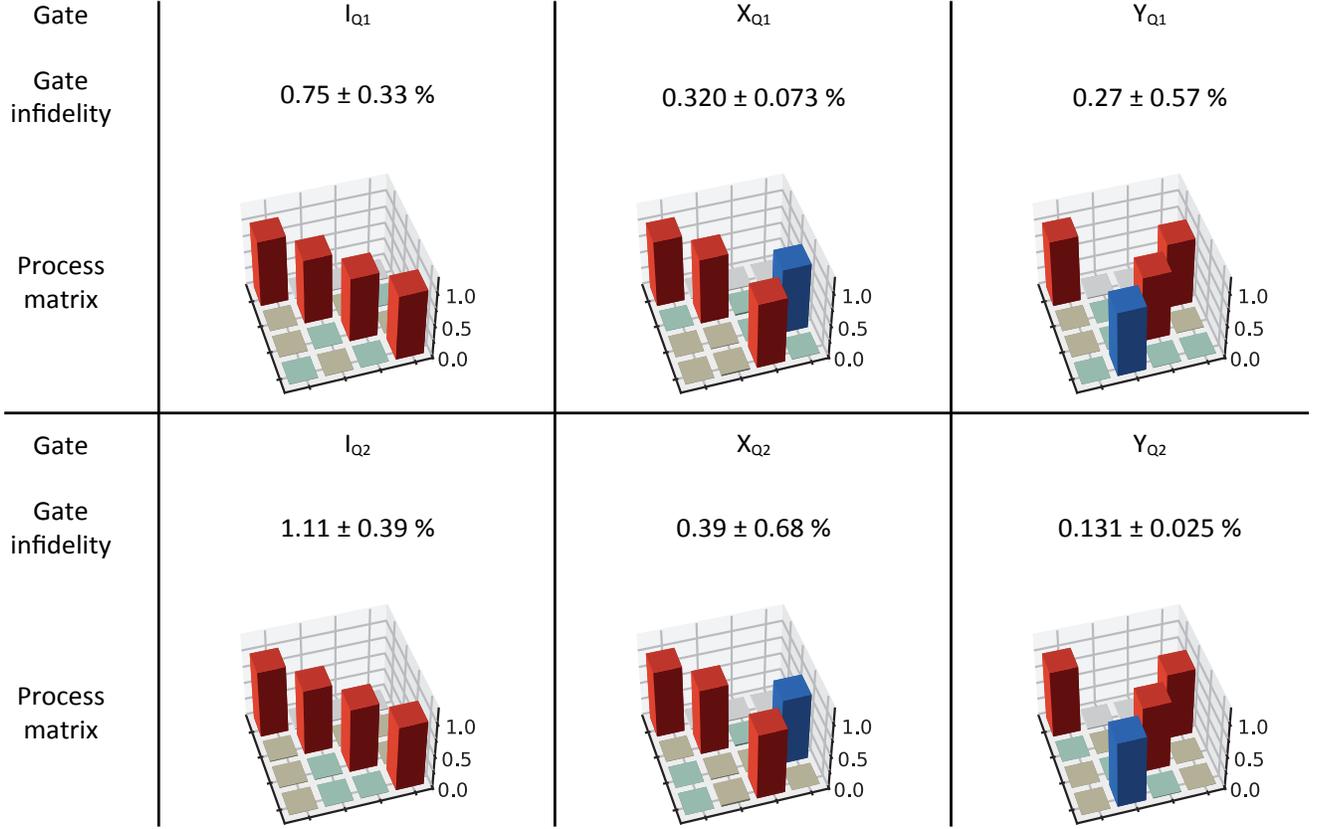}}
\caption{Average gate infidelity and process matrices (PTMs) of the identity gates (idle gates) and single-qubit $X/Y$ gates in the subspace of the individual qubits. The individual PTMs are calculated from the PTMs in the two-qubit space (see Methods).} 
\label{fig:Subspace_PTM}
\end{figure*}


\captionsetup[table]{name=Extended Data Table}
\begin{table*}[htbp]
\begin{tabular}{|c|c|c|c|c|c|c|c|}
\hline
     & $1-F_\text{gate}$ & $1-F_\text{sub}$ & $\epsilon_J$ & $\theta_J$ & $ D$ & $||\cdot||_\diamond$ \\ \hline
$I$    & $ 0.017 \pm 0.012 $ & $\begin{array}{c}
    \text{Q1: }0.0075\pm 0.0033   \\
    \text{Q2: }0.0111\pm 0.0039 
\end{array}$ & $0.021$ & $0.0097$ & 
$0.024 \pm 0.015$ & $0.038 \pm 0.019$ \\ \hline
$X_\text{Q1}$     & $ 0.0088 \pm 0.0023 $ & $0.00320\pm 0.00073$ & $0.010$ & $0.027$ & 
$0.032 \pm 0.022$ & $0.047 \pm 0.035$ \\ \hline
$Y_\text{Q1}$    & $ 0.0059 \pm 0.0029 $ & $0.0027\pm 0.0057$ & $0.0069$ & $0.022$ &
$0.0256 \pm 0.0073$ & $0.034 \pm 0.022$ \\ \hline
$X_\text{Q2}$     & $ 0.0119 \pm 0.0023 $ & $0.0039\pm 0.0068$ & $0.014$ & $0.028$ &
$0.035 \pm 0.030$ & $0.044 \pm 0.041$ \\ \hline
$Y_\text{Q2}$     & $ 0.0067 \pm 0.0023 $ & $0.00131\pm 0.00025$ & $0.0079$ & $0.022$ &
$0.0265 \pm 0.0080$ & $0.034 \pm 0.014$ \\ \hline
\textsc{cphase} & $ 0.0035 \pm 0.0015 $ & $-$ & $0.0042$ & $0.016$ &
$0.018 \pm 0.014$ & $0.023 \pm 0.010$ \\ \hline
\end{tabular}
\caption{Detailed overview of important metrics of the gate set $[ I$, $X_{Q1}$, $Y_{Q1}$, $X_{Q2}$, $Y_{Q2}$, $\textsc{cphase}]$: the average gate fidelity $F_\text{gate}$ (see Eq.~\eqref{eq:averageGate}) and the fidelity reduced to the single-qubit subspace (see Eq.~\eqref{eq:averageGate_sub}), the Jamiolkowski probability $\epsilon_J$ (see Eq.~\eqref{eq:Jam_prob}), Jamiolkowski amplitude $\theta_J$ (see Eq.~\eqref{eq:Jam_amp}), the trace distance $D(\mathcal{M}_\text{ideal},\mathcal{M}_\text{exp})=||\mathcal{M}_\text{ideal}-\mathcal{M}_\text{exp}||_1/2$, and the diamond norm $||\mathcal{M}_\text{ideal},\mathcal{M}_\text{exp}||_\diamond=\max_{\rho}\|(\mathcal{M}_\text{ideal} \otimes \boldsymbol{1}_{d^2})\rho - (\mathcal{M}_\text{exp} \otimes \boldsymbol{1}_{d^2})\rho\|_1/2$.}
\label{tab:GST_results}
\end{table*}

\begin{figure*}[htbp] 
\center{\includegraphics[width=1\linewidth]{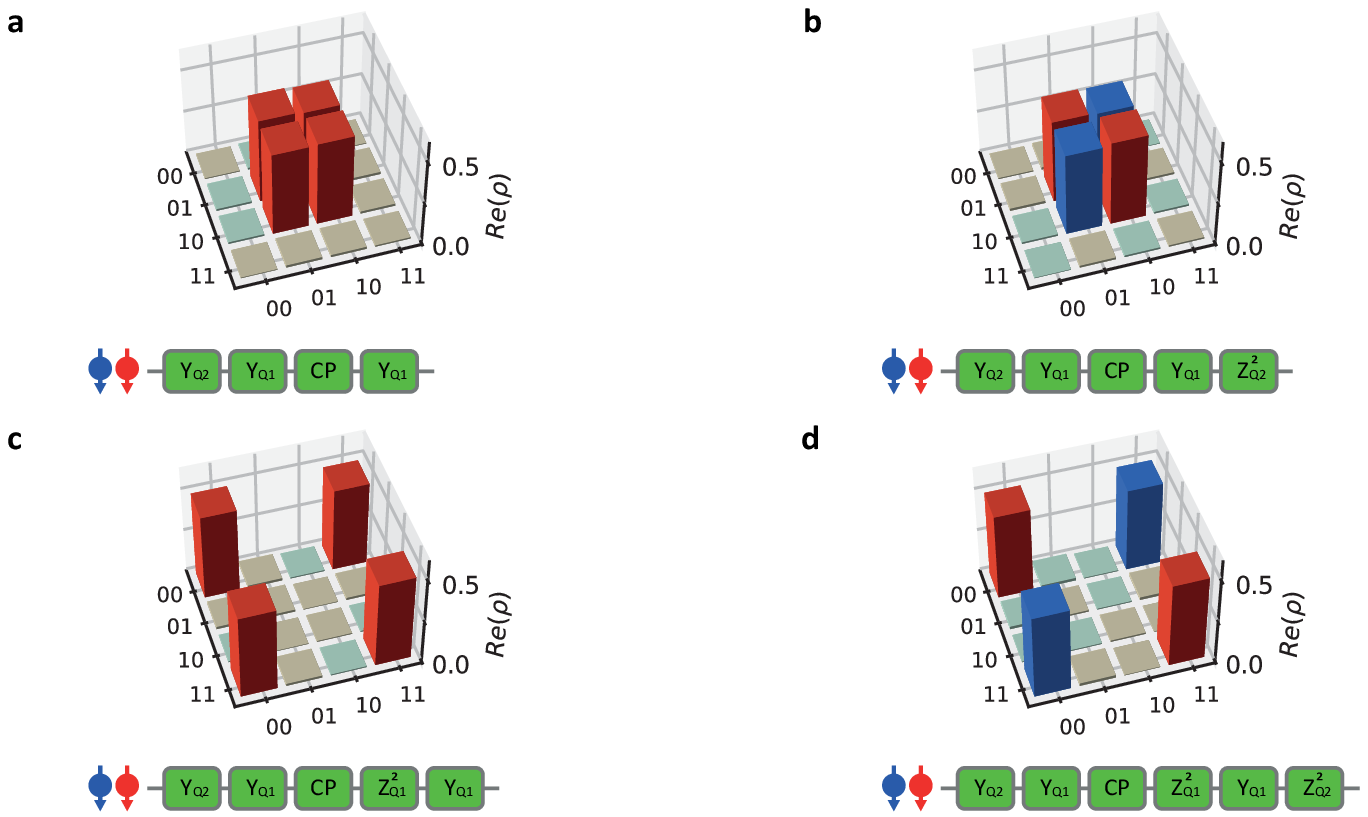}}
\caption{Top panels show the real part of the reconstructed density matrices of the four Bell states
$\ket{\Psi^+}=(\ket{01} + \ket{10})/\sqrt{2}$ (\textbf{a}), $\ket{\Psi^-}=(\ket{01} - \ket{10})/\sqrt{2}$ (\textbf{b}), $\ket{\Phi^+}=(\ket{00} + \ket{11})/\sqrt{2}$ (\textbf{c}), $\ket{\Phi^+}=(\ket{00} - \ket{11})/\sqrt{2}$ (\textbf{d}). The color code is the same as in Fig.~\ref{fig:2QGST}. Bottom panels show the quantum circuit used to reconstruct the Bell states. $Z_{Qi}^2$ is a virtual $\pi$-rotation around the $\hat{z}$ axis on the $i$th qubit executed by a reference frame change. We numerically estimate the state fidelities to be 98.42\% for $\ket{\Psi^+}$ and $\ket{\Psi^-}$ state, and 97.75\% for $\ket{\Phi^+}$ and $\ket{\Phi^-}$ state.
} 
\label{fig:Bell_states}
\end{figure*}

\begin{figure*}[htbp] 
\center{\includegraphics[width=1\linewidth]{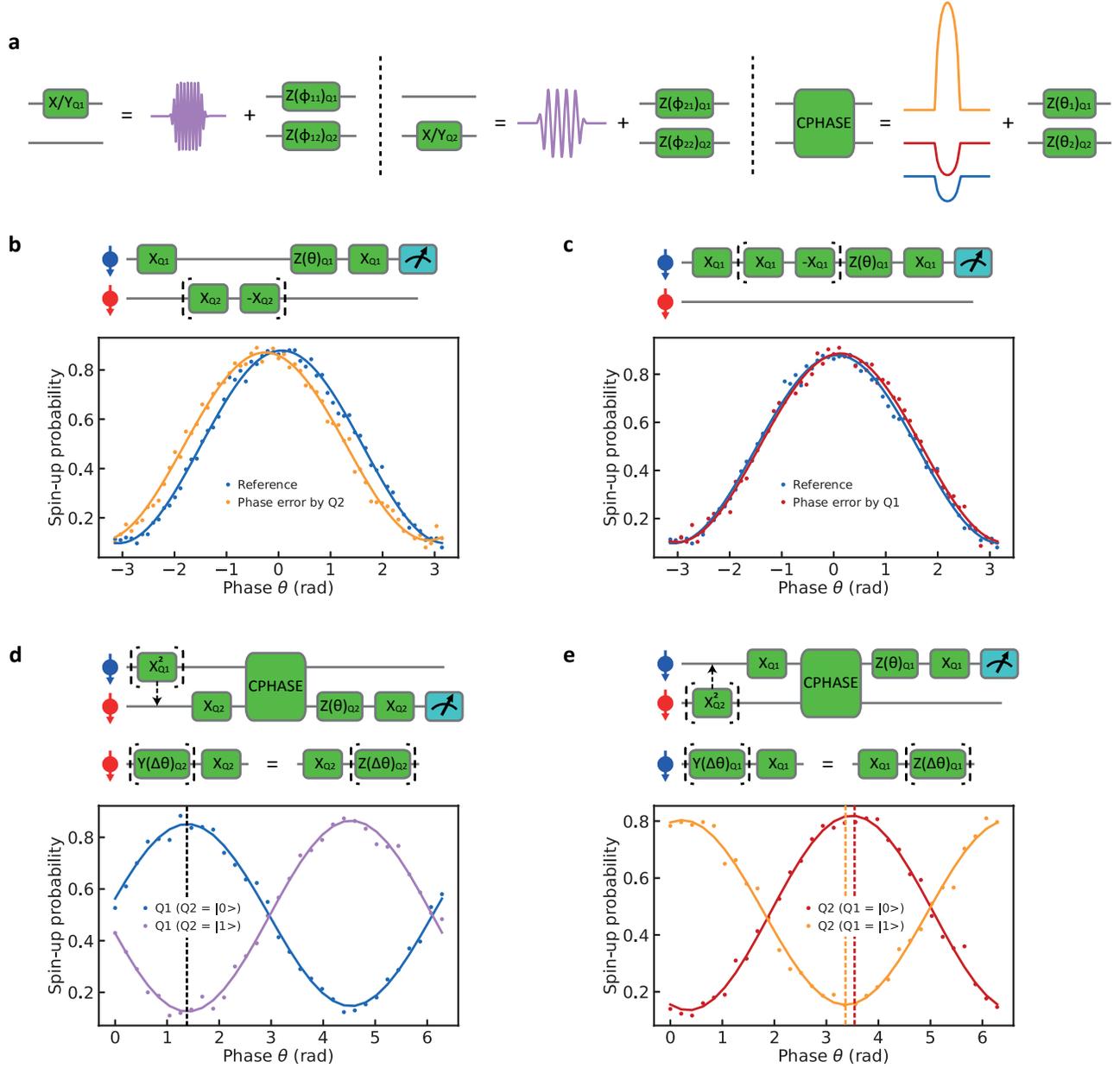}}
\caption{\textbf{a.} Decomposition of single- and two-qubit gates. After each microwave burst for single-qubit rotations, a corresponding phase correction is applied to each qubit. The \textsc{cphase} gate is implemented by a barrier voltage pulse on gate B (orange) and negative compensation pulses on gates LP (blue) and RP (red), with the same shape as the barrier pulse. Single-qubit phase corrections are then applied on each qubit to compensate the frequency detuning induced by electron movement in the magnetic field gradient. \textbf{b}-\textbf{c.} Calibration of phase corrections on Q1 induced by a single-qubit gate applied on Q2 ($\phi_{21}$, \textbf{b}) and on Q1 ($\phi_{11}$, \textbf{c}). A relative phase shift, $2\phi_{21}$ ($2\phi_{11}$), is determined by interleaving the target gate (a $\pi/2$ rotation) and its inverse (a $-\pi/2$ rotation) on Q2 (Q1) in a Ramsey interference sequence. \textbf{d}-\textbf{e.} Calibration of phase corrections on each qubit after the \textsc{cphase} gate, using Q1 (\textbf{d}) and Q2 (\textbf{e}) as the control-qubit respectively. When the amplitude of the barrier pulse is perfectly calibrated, the two curves in each experiment should both be out of phase by 180 degree. However, when the barrier pulse amplitude is calibrated such that one of the two experiments shows a 180 degree phase difference (\textbf{d}), the phase difference in the other calibration experiment always deviates by a few degrees. One possible explanation is that the optional $\pi$ rotation applied to the control-qubit induces a small off-resonance rotation on the other qubit, causing an additional phase on the target qubit to appear in the measurement due to the commutation relation of the Pauli operators.
} 
\label{fig:Manual_calibration}
\end{figure*}


\begin{figure*}[htbp] 
\center{\includegraphics[width=1\linewidth]{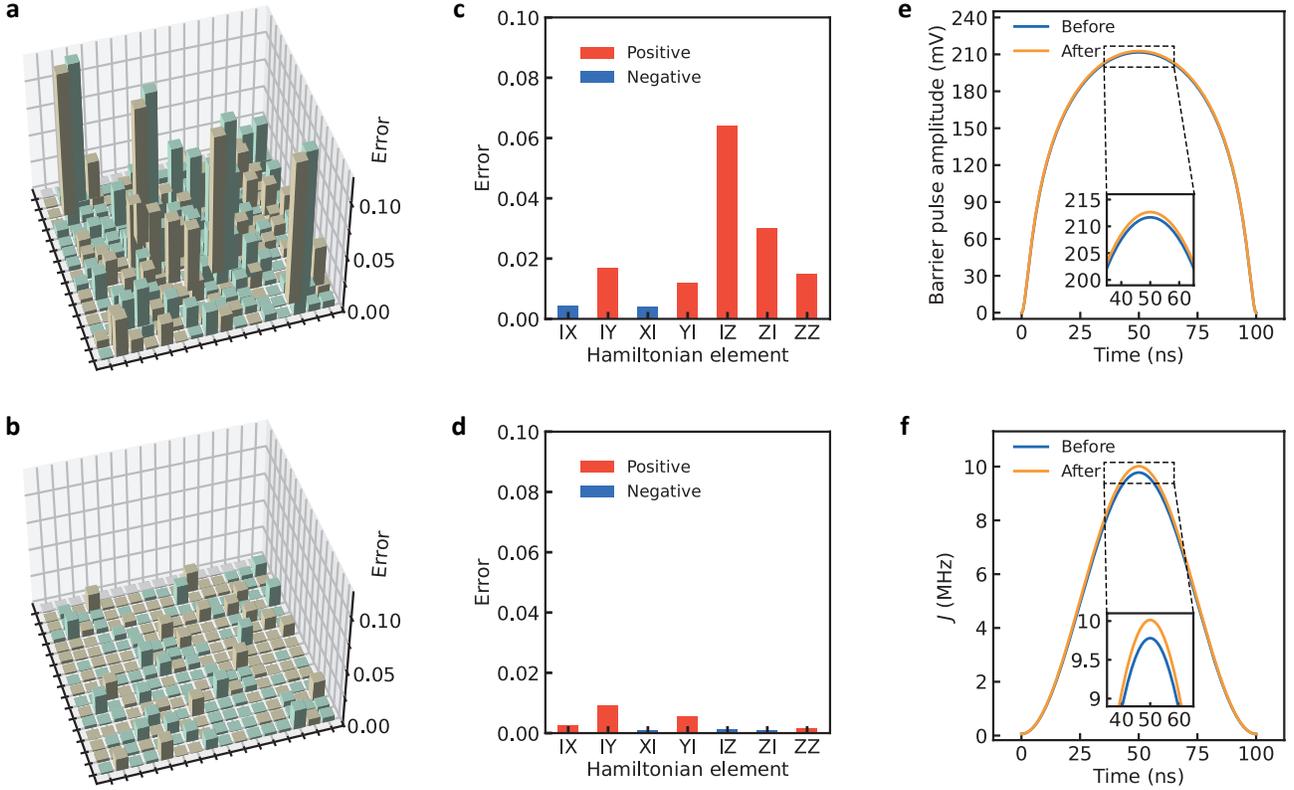}}
\caption{\textbf{a}-\textbf{b.}Full error generators for (\textbf{a}) a \textsc{cphase} gate calibrated by conventional Ramsey sequences and (\textbf{b}) after improving the calibration using the information extracted from \textbf{a}, resulting in fidelities of 97.86\% and 99.65\%, respectively. \textbf{c}-\textbf{d.} Seven Hamiltonian errors ($IX$, $IY$, $XI$, $YI$, $IZ$, $ZI$ and $ZZ$) extracted from the error generators shown in \textbf{a} (\textbf{c}) and \textbf{b} (\textbf{d}). Due to the crosstalk-induced additional phases shown in Extended Data Fig.~\ref{fig:Manual_calibration}, errors $IZ$, $ZI$ and $ZZ$ occur systematically in conventional calibrations. 
\textbf{e}-\textbf{f.} Shapes of the barrier pulses (\textbf{e}) and their corresponding $J$ envelopes (\textbf{f}) for a \textsc{cphase} gate before and after being corrected by GST. Since the Hamiltonian to generate a \textsc{cphase} gate is $H = (II + IZ + ZI - ZZ)/2$, the positive $ZZ$ error shown in \textbf{c} is corrected by increasing the amplitude of the pulse. The $IZ$ and $ZI$ errors are corrected by decreasing the phase shifts $\theta_1$ and $\theta_2$ after the \textsc{cphase} gate. Hamiltonian errors in single-qubit gates are corrected similarly.}
\label{fig:GST_optimization}
\end{figure*}


\begin{figure*}[htbp] 
\center{\includegraphics[width=1\linewidth]{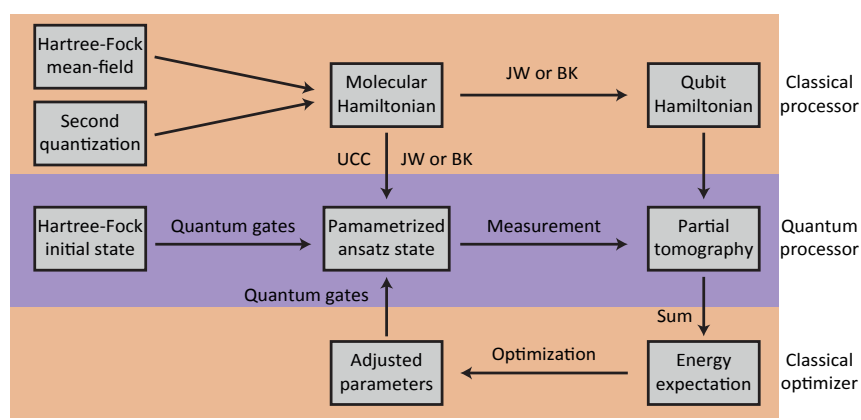}}
\caption{Workflow of the VQE algorithm. The qubit Hamiltonian is typically transformed from the molecular Hamiltonian by JW transformation or BK transformation by a classical processor (see Methods). A HF initial state is encoded into the qubit states according to JW or BK transformation, and then transformed by the quantum processor into a parameterized ansatz state by considering single- and double-excitation in the molecule using unitary coupled cluster (UCC) theory. The expectation value of each individual Hamiltonian term is directly measured by partial state tomography. The expectation of the total energy is then calculated by weighted sum of the individual expectations. The result is fed into a classical optimizer, which suggests a new parameterized ansatz state for the next run. This process is iterated until the expectation of the total energy converges.} 
\label{fig:VQE_workflow}
\end{figure*}



%% file: main.bbl
\begin{thebibliography}{55}%
\makeatletter
\providecommand \@ifxundefined [1]{%
 \@ifx{#1\undefined}
}%
\providecommand \@ifnum [1]{%
 \ifnum #1\expandafter \@firstoftwo
 \else \expandafter \@secondoftwo
 \fi
}%
\providecommand \@ifx [1]{%
 \ifx #1\expandafter \@firstoftwo
 \else \expandafter \@secondoftwo
 \fi
}%
\providecommand \natexlab [1]{#1}%
\providecommand \enquote  [1]{``#1''}%
\providecommand \bibnamefont  [1]{#1}%
\providecommand \bibfnamefont [1]{#1}%
\providecommand \citenamefont [1]{#1}%
\providecommand \href@noop [0]{\@secondoftwo}%
\providecommand \href [0]{\begingroup \@sanitize@url \@href}%
\providecommand \@href[1]{\@@startlink{#1}\@@href}%
\providecommand \@@href[1]{\endgroup#1\@@endlink}%
\providecommand \@sanitize@url [0]{\catcode `\\12\catcode `\$12\catcode
  `\&12\catcode `\#12\catcode `\^12\catcode `\_12\catcode `\%12\relax}%
\providecommand \@@startlink[1]{}%
\providecommand \@@endlink[0]{}%
\providecommand \url  [0]{\begingroup\@sanitize@url \@url }%
\providecommand \@url [1]{\endgroup\@href {#1}{\urlprefix }}%
\providecommand \urlprefix  [0]{URL }%
\providecommand \Eprint [0]{\href }%
\providecommand \doibase [0]{https://doi.org/}%
\providecommand \selectlanguage [0]{\@gobble}%
\providecommand \bibinfo  [0]{\@secondoftwo}%
\providecommand \bibfield  [0]{\@secondoftwo}%
\providecommand \translation [1]{[#1]}%
\providecommand \BibitemOpen [0]{}%
\providecommand \bibitemStop [0]{}%
\providecommand \bibitemNoStop [0]{.\EOS\space}%
\providecommand \EOS [0]{\spacefactor3000\relax}%
\providecommand \BibitemShut  [1]{\csname bibitem#1\endcsname}%
\let\auto@bib@innerbib\@empty
\bibitem [{\citenamefont {Lidar}\ and\ \citenamefont
  {Brun}(2013)}]{lidar2013quantum}%
  \BibitemOpen
  \bibfield  {author} {\bibinfo {author} {\bibfnamefont {D.~A.}\ \bibnamefont
  {Lidar}}\ and\ \bibinfo {author} {\bibfnamefont {T.~A.}\ \bibnamefont
  {Brun}},\ }\href@noop {} {\emph {\bibinfo {title} {Quantum error
  correction}}}\ (\bibinfo  {publisher} {Cambridge university press},\ \bibinfo
  {year} {2013})\BibitemShut {NoStop}%
\bibitem [{\citenamefont {Fowler}\ \emph {et~al.}(2012)\citenamefont {Fowler},
  \citenamefont {Mariantoni}, \citenamefont {Martinis},\ and\ \citenamefont
  {Cleland}}]{fowler2012surface}%
  \BibitemOpen
  \bibfield  {author} {\bibinfo {author} {\bibfnamefont {A.~G.}\ \bibnamefont
  {Fowler}}, \bibinfo {author} {\bibfnamefont {M.}~\bibnamefont {Mariantoni}},
  \bibinfo {author} {\bibfnamefont {J.~M.}\ \bibnamefont {Martinis}},\ and\
  \bibinfo {author} {\bibfnamefont {A.~N.}\ \bibnamefont {Cleland}},\
  }\bibfield  {title} {\bibinfo {title} {Surface codes: {Towards} practical
  large-scale quantum computation},\ }\href
  {https://doi.org/10.1103/PhysRevA.86.032324} {\bibfield  {journal} {\bibinfo
  {journal} {Phys. Rev. A}\ }\textbf {\bibinfo {volume} {86}},\ \bibinfo
  {pages} {032324} (\bibinfo {year} {2012})}\BibitemShut {NoStop}%
\bibitem [{\citenamefont {Zwerver}\ \emph {et~al.}(2021)\citenamefont
  {Zwerver}, \citenamefont {Kr{\"a}henmann}, \citenamefont {Watson},
  \citenamefont {Lampert}, \citenamefont {George}, \citenamefont
  {Pillarisetty}, \citenamefont {Bojarski}, \citenamefont {Amin}, \citenamefont
  {Amitonov}, \citenamefont {Boter}, \citenamefont {Caudillo}, \citenamefont
  {Corras-Serrano}, \citenamefont {Dehollain}, \citenamefont {Droulers},
  \citenamefont {Henry}, \citenamefont {Kotlyar}, \citenamefont {Lodari},
  \citenamefont {Luthi}, \citenamefont {Michalak}, \citenamefont {Mueller},
  \citenamefont {Neyens}, \citenamefont {Roberts}, \citenamefont {Samkharadze},
  \citenamefont {Zheng}, \citenamefont {Zietz}, \citenamefont {Scappucci},
  \citenamefont {Veldhorst}, \citenamefont {Vandersypen},\ and\ \citenamefont
  {Clarke}}]{zwerver2021qubits}%
  \BibitemOpen
  \bibfield  {author} {\bibinfo {author} {\bibfnamefont {A.~M.~J.}\
  \bibnamefont {Zwerver}}, \bibinfo {author} {\bibfnamefont {T.}~\bibnamefont
  {Kr{\"a}henmann}}, \bibinfo {author} {\bibfnamefont {T.~F.}\ \bibnamefont
  {Watson}}, \bibinfo {author} {\bibfnamefont {L.}~\bibnamefont {Lampert}},
  \bibinfo {author} {\bibfnamefont {H.~C.}\ \bibnamefont {George}}, \bibinfo
  {author} {\bibfnamefont {R.}~\bibnamefont {Pillarisetty}}, \bibinfo {author}
  {\bibfnamefont {S.~A.}\ \bibnamefont {Bojarski}}, \bibinfo {author}
  {\bibfnamefont {P.}~\bibnamefont {Amin}}, \bibinfo {author} {\bibfnamefont
  {S.~V.}\ \bibnamefont {Amitonov}}, \bibinfo {author} {\bibfnamefont {J.~M.}\
  \bibnamefont {Boter}}, \bibinfo {author} {\bibfnamefont {R.}~\bibnamefont
  {Caudillo}}, \bibinfo {author} {\bibfnamefont {D.}~\bibnamefont
  {Corras-Serrano}}, \bibinfo {author} {\bibfnamefont {J.~P.}\ \bibnamefont
  {Dehollain}}, \bibinfo {author} {\bibfnamefont {G.}~\bibnamefont {Droulers}},
  \bibinfo {author} {\bibfnamefont {E.~M.}\ \bibnamefont {Henry}}, \bibinfo
  {author} {\bibfnamefont {R.}~\bibnamefont {Kotlyar}}, \bibinfo {author}
  {\bibfnamefont {M.}~\bibnamefont {Lodari}}, \bibinfo {author} {\bibfnamefont
  {F.}~\bibnamefont {Luthi}}, \bibinfo {author} {\bibfnamefont {D.~J.}\
  \bibnamefont {Michalak}}, \bibinfo {author} {\bibfnamefont {B.~K.}\
  \bibnamefont {Mueller}}, \bibinfo {author} {\bibfnamefont {S.}~\bibnamefont
  {Neyens}}, \bibinfo {author} {\bibfnamefont {J.}~\bibnamefont {Roberts}},
  \bibinfo {author} {\bibfnamefont {N.}~\bibnamefont {Samkharadze}}, \bibinfo
  {author} {\bibfnamefont {G.}~\bibnamefont {Zheng}}, \bibinfo {author}
  {\bibfnamefont {O.~K.}\ \bibnamefont {Zietz}}, \bibinfo {author}
  {\bibfnamefont {G.}~\bibnamefont {Scappucci}}, \bibinfo {author}
  {\bibfnamefont {M.}~\bibnamefont {Veldhorst}}, \bibinfo {author}
  {\bibfnamefont {L.~M.~K.}\ \bibnamefont {Vandersypen}},\ and\ \bibinfo
  {author} {\bibfnamefont {J.~S.}\ \bibnamefont {Clarke}},\ }\bibfield  {title}
  {\bibinfo {title} {Qubits made by advanced semiconductor manufacturing},\
  }\href {http://arxiv.org/abs/2101.12650} {\bibfield  {journal} {\bibinfo
  {journal} {arXiv:2101.12650}\ } (\bibinfo {year} {2021})}\BibitemShut
  {NoStop}%
\bibitem [{\citenamefont {McArdle}\ \emph {et~al.}(2020)\citenamefont
  {McArdle}, \citenamefont {Endo}, \citenamefont {Aspuru-Guzik}, \citenamefont
  {Benjamin},\ and\ \citenamefont {Yuan}}]{mcardle2020quantum}%
  \BibitemOpen
  \bibfield  {author} {\bibinfo {author} {\bibfnamefont {S.}~\bibnamefont
  {McArdle}}, \bibinfo {author} {\bibfnamefont {S.}~\bibnamefont {Endo}},
  \bibinfo {author} {\bibfnamefont {A.}~\bibnamefont {Aspuru-Guzik}}, \bibinfo
  {author} {\bibfnamefont {S.~C.}\ \bibnamefont {Benjamin}},\ and\ \bibinfo
  {author} {\bibfnamefont {X.}~\bibnamefont {Yuan}},\ }\bibfield  {title}
  {\bibinfo {title} {Quantum computational chemistry},\ }\href
  {https://doi.org/10.1103/RevModPhys.92.015003} {\bibfield  {journal}
  {\bibinfo  {journal} {Rev. Mod. Phys.}\ }\textbf {\bibinfo {volume} {92}},\
  \bibinfo {pages} {015003} (\bibinfo {year} {2020})}\BibitemShut {NoStop}%
\bibitem [{\citenamefont {Nielsen}\ and\ \citenamefont
  {Chuang}(2002)}]{nielsen2002quantum}%
  \BibitemOpen
  \bibfield  {author} {\bibinfo {author} {\bibfnamefont {M.~A.}\ \bibnamefont
  {Nielsen}}\ and\ \bibinfo {author} {\bibfnamefont {I.}~\bibnamefont
  {Chuang}},\ }\href@noop {} {\emph {\bibinfo {title} {Quantum computation and
  quantum information}}}\ (\bibinfo  {publisher} {American Association of
  Physics Teachers},\ \bibinfo {year} {2002})\BibitemShut {NoStop}%
\bibitem [{\citenamefont {Preskill}(2018)}]{preskill2018quantum}%
  \BibitemOpen
  \bibfield  {author} {\bibinfo {author} {\bibfnamefont {J.}~\bibnamefont
  {Preskill}},\ }\bibfield  {title} {\bibinfo {title} {Quantum {Computing} in
  the {NISQ} era and beyond},\ }\href
  {https://doi.org/10.22331/q-2018-08-06-79} {\bibfield  {journal} {\bibinfo
  {journal} {Quantum}\ }\textbf {\bibinfo {volume} {2}},\ \bibinfo {pages} {79}
  (\bibinfo {year} {2018})}\BibitemShut {NoStop}%
\bibitem [{\citenamefont {Veldhorst}\ \emph {et~al.}(2015)\citenamefont
  {Veldhorst}, \citenamefont {Yang}, \citenamefont {Hwang}, \citenamefont
  {Huang}, \citenamefont {Dehollain}, \citenamefont {Muhonen}, \citenamefont
  {Simmons}, \citenamefont {Laucht}, \citenamefont {Hudson}, \citenamefont
  {Itoh}, \citenamefont {Morello},\ and\ \citenamefont
  {Dzurak}}]{veldhorst2015two}%
  \BibitemOpen
  \bibfield  {author} {\bibinfo {author} {\bibfnamefont {M.}~\bibnamefont
  {Veldhorst}}, \bibinfo {author} {\bibfnamefont {C.~H.}\ \bibnamefont {Yang}},
  \bibinfo {author} {\bibfnamefont {J.~C.~C.}\ \bibnamefont {Hwang}}, \bibinfo
  {author} {\bibfnamefont {W.}~\bibnamefont {Huang}}, \bibinfo {author}
  {\bibfnamefont {J.~P.}\ \bibnamefont {Dehollain}}, \bibinfo {author}
  {\bibfnamefont {J.~T.}\ \bibnamefont {Muhonen}}, \bibinfo {author}
  {\bibfnamefont {S.}~\bibnamefont {Simmons}}, \bibinfo {author} {\bibfnamefont
  {A.}~\bibnamefont {Laucht}}, \bibinfo {author} {\bibfnamefont {F.~E.}\
  \bibnamefont {Hudson}}, \bibinfo {author} {\bibfnamefont {K.~M.}\
  \bibnamefont {Itoh}}, \bibinfo {author} {\bibfnamefont {A.}~\bibnamefont
  {Morello}},\ and\ \bibinfo {author} {\bibfnamefont {A.~S.}\ \bibnamefont
  {Dzurak}},\ }\bibfield  {title} {\bibinfo {title} {A two-qubit logic gate in
  silicon},\ }\href {https://doi.org/10.1038/nature15263} {\bibfield  {journal}
  {\bibinfo  {journal} {Nature}\ }\textbf {\bibinfo {volume} {526}},\ \bibinfo
  {pages} {410} (\bibinfo {year} {2015})}\BibitemShut {NoStop}%
\bibitem [{\citenamefont {Zajac}\ \emph {et~al.}(2016)\citenamefont {Zajac},
  \citenamefont {Hazard}, \citenamefont {Mi}, \citenamefont {Nielsen},\ and\
  \citenamefont {Petta}}]{zajac2016scalable}%
  \BibitemOpen
  \bibfield  {author} {\bibinfo {author} {\bibfnamefont {D.~M.}\ \bibnamefont
  {Zajac}}, \bibinfo {author} {\bibfnamefont {T.~M.}\ \bibnamefont {Hazard}},
  \bibinfo {author} {\bibfnamefont {X.}~\bibnamefont {Mi}}, \bibinfo {author}
  {\bibfnamefont {E.}~\bibnamefont {Nielsen}},\ and\ \bibinfo {author}
  {\bibfnamefont {J.~R.}\ \bibnamefont {Petta}},\ }\bibfield  {title} {\bibinfo
  {title} {Scalable {Gate} {Architecture} for a {One}-{Dimensional} {Array} of
  {Semiconductor} {Spin} {Qubits}},\ }\href
  {https://doi.org/10.1103/PhysRevApplied.6.054013} {\bibfield  {journal}
  {\bibinfo  {journal} {Phys. Rev. Applied}\ }\textbf {\bibinfo {volume} {6}},\
  \bibinfo {pages} {054013} (\bibinfo {year} {2016})}\BibitemShut {NoStop}%
\bibitem [{\citenamefont {Vandersypen}\ \emph {et~al.}(2017)\citenamefont
  {Vandersypen}, \citenamefont {Bluhm}, \citenamefont {Clarke}, \citenamefont
  {Dzurak}, \citenamefont {Ishihara}, \citenamefont {Morello}, \citenamefont
  {Reilly}, \citenamefont {Schreiber},\ and\ \citenamefont
  {Veldhorst}}]{vandersypen2017interfacing}%
  \BibitemOpen
  \bibfield  {author} {\bibinfo {author} {\bibfnamefont {L.~M.~K.}\
  \bibnamefont {Vandersypen}}, \bibinfo {author} {\bibfnamefont
  {H.}~\bibnamefont {Bluhm}}, \bibinfo {author} {\bibfnamefont {J.~S.}\
  \bibnamefont {Clarke}}, \bibinfo {author} {\bibfnamefont {A.~S.}\
  \bibnamefont {Dzurak}}, \bibinfo {author} {\bibfnamefont {R.}~\bibnamefont
  {Ishihara}}, \bibinfo {author} {\bibfnamefont {A.}~\bibnamefont {Morello}},
  \bibinfo {author} {\bibfnamefont {D.~J.}\ \bibnamefont {Reilly}}, \bibinfo
  {author} {\bibfnamefont {L.~R.}\ \bibnamefont {Schreiber}},\ and\ \bibinfo
  {author} {\bibfnamefont {M.}~\bibnamefont {Veldhorst}},\ }\bibfield  {title}
  {\bibinfo {title} {Interfacing spin qubits in quantum dots and donors--hot,
  dense, and coherent},\ }\href {https://doi.org/10.1038/s41534-017-0038-y}
  {\bibfield  {journal} {\bibinfo  {journal} {npj Quantum Inf.}\ }\textbf
  {\bibinfo {volume} {3}},\ \bibinfo {pages} {34} (\bibinfo {year}
  {2017})}\BibitemShut {NoStop}%
\bibitem [{\citenamefont {Li}\ \emph {et~al.}(2018)\citenamefont {Li},
  \citenamefont {Petit}, \citenamefont {Franke}, \citenamefont {Dehollain},
  \citenamefont {Helsen}, \citenamefont {Steudtner}, \citenamefont {Thomas},
  \citenamefont {Yoscovits}, \citenamefont {Singh}, \citenamefont {Wehner},
  \citenamefont {Vandersypen}, \citenamefont {Clarke},\ and\ \citenamefont
  {Veldhorst}}]{li2018crossbar}%
  \BibitemOpen
  \bibfield  {author} {\bibinfo {author} {\bibfnamefont {R.}~\bibnamefont
  {Li}}, \bibinfo {author} {\bibfnamefont {L.}~\bibnamefont {Petit}}, \bibinfo
  {author} {\bibfnamefont {D.~P.}\ \bibnamefont {Franke}}, \bibinfo {author}
  {\bibfnamefont {J.~P.}\ \bibnamefont {Dehollain}}, \bibinfo {author}
  {\bibfnamefont {J.}~\bibnamefont {Helsen}}, \bibinfo {author} {\bibfnamefont
  {M.}~\bibnamefont {Steudtner}}, \bibinfo {author} {\bibfnamefont {N.~K.}\
  \bibnamefont {Thomas}}, \bibinfo {author} {\bibfnamefont {Z.~R.}\
  \bibnamefont {Yoscovits}}, \bibinfo {author} {\bibfnamefont {K.~J.}\
  \bibnamefont {Singh}}, \bibinfo {author} {\bibfnamefont {S.}~\bibnamefont
  {Wehner}}, \bibinfo {author} {\bibfnamefont {L.~M.~K.}\ \bibnamefont
  {Vandersypen}}, \bibinfo {author} {\bibfnamefont {J.~S.}\ \bibnamefont
  {Clarke}},\ and\ \bibinfo {author} {\bibfnamefont {M.}~\bibnamefont
  {Veldhorst}},\ }\bibfield  {title} {\bibinfo {title} {A crossbar network for
  silicon quantum dot qubits},\ }\href {https://doi.org/10.1126/sciadv.aar3960}
  {\bibfield  {journal} {\bibinfo  {journal} {Sci. Adv.}\ }\textbf {\bibinfo
  {volume} {4}},\ \bibinfo {pages} {eaar3960} (\bibinfo {year}
  {2018})}\BibitemShut {NoStop}%
\bibitem [{\citenamefont {Yoneda}\ \emph {et~al.}(2018)\citenamefont {Yoneda},
  \citenamefont {Takeda}, \citenamefont {Otsuka}, \citenamefont {Nakajima},
  \citenamefont {Delbecq}, \citenamefont {Allison}, \citenamefont {Honda},
  \citenamefont {Kodera}, \citenamefont {Oda}, \citenamefont {Hoshi},
  \citenamefont {Usami}, \citenamefont {Itoh},\ and\ \citenamefont
  {Tarucha}}]{yoneda2018quantum}%
  \BibitemOpen
  \bibfield  {author} {\bibinfo {author} {\bibfnamefont {J.}~\bibnamefont
  {Yoneda}}, \bibinfo {author} {\bibfnamefont {K.}~\bibnamefont {Takeda}},
  \bibinfo {author} {\bibfnamefont {T.}~\bibnamefont {Otsuka}}, \bibinfo
  {author} {\bibfnamefont {T.}~\bibnamefont {Nakajima}}, \bibinfo {author}
  {\bibfnamefont {M.~R.}\ \bibnamefont {Delbecq}}, \bibinfo {author}
  {\bibfnamefont {G.}~\bibnamefont {Allison}}, \bibinfo {author} {\bibfnamefont
  {T.}~\bibnamefont {Honda}}, \bibinfo {author} {\bibfnamefont
  {T.}~\bibnamefont {Kodera}}, \bibinfo {author} {\bibfnamefont
  {S.}~\bibnamefont {Oda}}, \bibinfo {author} {\bibfnamefont {Y.}~\bibnamefont
  {Hoshi}}, \bibinfo {author} {\bibfnamefont {N.}~\bibnamefont {Usami}},
  \bibinfo {author} {\bibfnamefont {K.~M.}\ \bibnamefont {Itoh}},\ and\
  \bibinfo {author} {\bibfnamefont {S.}~\bibnamefont {Tarucha}},\ }\bibfield
  {title} {\bibinfo {title} {A quantum-dot spin qubit with coherence limited by
  charge noise and fidelity higher than 99.9\%},\ }\href
  {https://doi.org/10.1038/s41565-017-0014-x} {\bibfield  {journal} {\bibinfo
  {journal} {Nat. Nanotechnol.}\ }\textbf {\bibinfo {volume} {13}},\ \bibinfo
  {pages} {102} (\bibinfo {year} {2018})}\BibitemShut {NoStop}%
\bibitem [{\citenamefont {Yang}\ \emph
  {et~al.}(2019{\natexlab{a}})\citenamefont {Yang}, \citenamefont {Chan},
  \citenamefont {Harper}, \citenamefont {Huang}, \citenamefont {Evans},
  \citenamefont {Hwang}, \citenamefont {Hensen}, \citenamefont {Laucht},
  \citenamefont {Tanttu}, \citenamefont {Hudson}, \citenamefont {Flammia},
  \citenamefont {Itoh}, \citenamefont {Morello}, \citenamefont {Bartlett},\
  and\ \citenamefont {Dzurak}}]{yang2019silicon}%
  \BibitemOpen
  \bibfield  {author} {\bibinfo {author} {\bibfnamefont {C.~H.}\ \bibnamefont
  {Yang}}, \bibinfo {author} {\bibfnamefont {K.~W.}\ \bibnamefont {Chan}},
  \bibinfo {author} {\bibfnamefont {R.}~\bibnamefont {Harper}}, \bibinfo
  {author} {\bibfnamefont {W.}~\bibnamefont {Huang}}, \bibinfo {author}
  {\bibfnamefont {T.}~\bibnamefont {Evans}}, \bibinfo {author} {\bibfnamefont
  {J.~C.~C.}\ \bibnamefont {Hwang}}, \bibinfo {author} {\bibfnamefont
  {B.}~\bibnamefont {Hensen}}, \bibinfo {author} {\bibfnamefont
  {A.}~\bibnamefont {Laucht}}, \bibinfo {author} {\bibfnamefont
  {T.}~\bibnamefont {Tanttu}}, \bibinfo {author} {\bibfnamefont {F.~E.}\
  \bibnamefont {Hudson}}, \bibinfo {author} {\bibfnamefont {S.~T.}\
  \bibnamefont {Flammia}}, \bibinfo {author} {\bibfnamefont {K.~M.}\
  \bibnamefont {Itoh}}, \bibinfo {author} {\bibfnamefont {A.}~\bibnamefont
  {Morello}}, \bibinfo {author} {\bibfnamefont {S.~D.}\ \bibnamefont
  {Bartlett}},\ and\ \bibinfo {author} {\bibfnamefont {A.~S.}\ \bibnamefont
  {Dzurak}},\ }\bibfield  {title} {\bibinfo {title} {Silicon qubit fidelities
  approaching incoherent noise limits via pulse engineering},\ }\href
  {https://doi.org/10.1038/s41928-019-0234-1} {\bibfield  {journal} {\bibinfo
  {journal} {Nat. Electron.}\ }\textbf {\bibinfo {volume} {2}},\ \bibinfo
  {pages} {151} (\bibinfo {year} {2019}{\natexlab{a}})}\BibitemShut {NoStop}%
\bibitem [{\citenamefont {Hendrickx}\ \emph {et~al.}(2021)\citenamefont
  {Hendrickx}, \citenamefont {Lawrie}, \citenamefont {Russ}, \citenamefont {van
  Riggelen}, \citenamefont {de~Snoo}, \citenamefont {Schouten}, \citenamefont
  {Sammak}, \citenamefont {Scappucci},\ and\ \citenamefont
  {Veldhorst}}]{hendrickx_four-qubit_2021}%
  \BibitemOpen
  \bibfield  {author} {\bibinfo {author} {\bibfnamefont {N.~W.}\ \bibnamefont
  {Hendrickx}}, \bibinfo {author} {\bibfnamefont {W.~I.~L.}\ \bibnamefont
  {Lawrie}}, \bibinfo {author} {\bibfnamefont {M.}~\bibnamefont {Russ}},
  \bibinfo {author} {\bibfnamefont {F.}~\bibnamefont {van Riggelen}}, \bibinfo
  {author} {\bibfnamefont {S.~L.}\ \bibnamefont {de~Snoo}}, \bibinfo {author}
  {\bibfnamefont {R.~N.}\ \bibnamefont {Schouten}}, \bibinfo {author}
  {\bibfnamefont {A.}~\bibnamefont {Sammak}}, \bibinfo {author} {\bibfnamefont
  {G.}~\bibnamefont {Scappucci}},\ and\ \bibinfo {author} {\bibfnamefont
  {M.}~\bibnamefont {Veldhorst}},\ }\bibfield  {title} {\bibinfo {title} {A
  four-qubit germanium quantum processor},\ }\href
  {https://doi.org/10.1038/s41586-021-03332-6} {\bibfield  {journal} {\bibinfo
  {journal} {Nature}\ }\textbf {\bibinfo {volume} {591}},\ \bibinfo {pages}
  {580} (\bibinfo {year} {2021})}\BibitemShut {NoStop}%
\bibitem [{\citenamefont {Xue}\ \emph {et~al.}(2019)\citenamefont {Xue},
  \citenamefont {Watson}, \citenamefont {Helsen}, \citenamefont {Ward},
  \citenamefont {Savage}, \citenamefont {Lagally}, \citenamefont {Coppersmith},
  \citenamefont {Eriksson}, \citenamefont {Wehner},\ and\ \citenamefont
  {Vandersypen}}]{xue2019benchmarking}%
  \BibitemOpen
  \bibfield  {author} {\bibinfo {author} {\bibfnamefont {X.}~\bibnamefont
  {Xue}}, \bibinfo {author} {\bibfnamefont {T.~F.}\ \bibnamefont {Watson}},
  \bibinfo {author} {\bibfnamefont {J.}~\bibnamefont {Helsen}}, \bibinfo
  {author} {\bibfnamefont {D.~R.}\ \bibnamefont {Ward}}, \bibinfo {author}
  {\bibfnamefont {D.~E.}\ \bibnamefont {Savage}}, \bibinfo {author}
  {\bibfnamefont {M.~G.}\ \bibnamefont {Lagally}}, \bibinfo {author}
  {\bibfnamefont {S.~N.}\ \bibnamefont {Coppersmith}}, \bibinfo {author}
  {\bibfnamefont {M.~A.}\ \bibnamefont {Eriksson}}, \bibinfo {author}
  {\bibfnamefont {S.}~\bibnamefont {Wehner}},\ and\ \bibinfo {author}
  {\bibfnamefont {L.~M.~K.}\ \bibnamefont {Vandersypen}},\ }\bibfield  {title}
  {\bibinfo {title} {Benchmarking {Gate} {Fidelities} in a
  $\mathrm{Si}/\mathrm{SiGe}$ {Two-Qubit} {Device}},\ }\href
  {https://doi.org/10.1103/PhysRevX.9.021011} {\bibfield  {journal} {\bibinfo
  {journal} {Phys. Rev. X}\ }\textbf {\bibinfo {volume} {9}},\ \bibinfo {pages}
  {021011} (\bibinfo {year} {2019})}\BibitemShut {NoStop}%
\bibitem [{\citenamefont {Huang}\ \emph {et~al.}(2019)\citenamefont {Huang},
  \citenamefont {Yang}, \citenamefont {Chan}, \citenamefont {Tanttu},
  \citenamefont {Hensen}, \citenamefont {Leon}, \citenamefont {Fogarty},
  \citenamefont {Hwang}, \citenamefont {Hudson}, \citenamefont {Itoh},
  \citenamefont {Morello}, \citenamefont {Laucht},\ and\ \citenamefont
  {Dzurak}}]{huang2019fidelity}%
  \BibitemOpen
  \bibfield  {author} {\bibinfo {author} {\bibfnamefont {W.}~\bibnamefont
  {Huang}}, \bibinfo {author} {\bibfnamefont {C.~H.}\ \bibnamefont {Yang}},
  \bibinfo {author} {\bibfnamefont {K.~W.}\ \bibnamefont {Chan}}, \bibinfo
  {author} {\bibfnamefont {T.}~\bibnamefont {Tanttu}}, \bibinfo {author}
  {\bibfnamefont {B.}~\bibnamefont {Hensen}}, \bibinfo {author} {\bibfnamefont
  {R.~C.~C.}\ \bibnamefont {Leon}}, \bibinfo {author} {\bibfnamefont {M.~A.}\
  \bibnamefont {Fogarty}}, \bibinfo {author} {\bibfnamefont {J.~C.~C.}\
  \bibnamefont {Hwang}}, \bibinfo {author} {\bibfnamefont {F.~E.}\ \bibnamefont
  {Hudson}}, \bibinfo {author} {\bibfnamefont {K.~M.}\ \bibnamefont {Itoh}},
  \bibinfo {author} {\bibfnamefont {A.}~\bibnamefont {Morello}}, \bibinfo
  {author} {\bibfnamefont {A.}~\bibnamefont {Laucht}},\ and\ \bibinfo {author}
  {\bibfnamefont {A.~S.}\ \bibnamefont {Dzurak}},\ }\bibfield  {title}
  {\bibinfo {title} {Fidelity benchmarks for two-qubit gates in silicon},\
  }\href {https://doi.org/10.1038/s41586-019-1197-0} {\bibfield  {journal}
  {\bibinfo  {journal} {Nature}\ }\textbf {\bibinfo {volume} {569}},\ \bibinfo
  {pages} {532} (\bibinfo {year} {2019})}\BibitemShut {NoStop}%
\bibitem [{\citenamefont {Takeda}\ \emph {et~al.}(2021)\citenamefont {Takeda},
  \citenamefont {Noiri}, \citenamefont {Nakajima}, \citenamefont {Yoneda},
  \citenamefont {Kobayashi},\ and\ \citenamefont
  {Tarucha}}]{takeda2020quantum}%
  \BibitemOpen
  \bibfield  {author} {\bibinfo {author} {\bibfnamefont {K.}~\bibnamefont
  {Takeda}}, \bibinfo {author} {\bibfnamefont {A.}~\bibnamefont {Noiri}},
  \bibinfo {author} {\bibfnamefont {T.}~\bibnamefont {Nakajima}}, \bibinfo
  {author} {\bibfnamefont {J.}~\bibnamefont {Yoneda}}, \bibinfo {author}
  {\bibfnamefont {T.}~\bibnamefont {Kobayashi}},\ and\ \bibinfo {author}
  {\bibfnamefont {S.}~\bibnamefont {Tarucha}},\ }\bibfield  {title} {\bibinfo
  {title} {Quantum tomography of an entangled three-qubit state in silicon},\
  }\bibfield  {journal} {\bibinfo  {journal} {Nat. Nanotechnol.}\ }\href
  {https://doi.org/10.1038/s41565-021-00925-0} {10.1038/s41565-021-00925-0}
  (\bibinfo {year} {2021})\BibitemShut {NoStop}%
\bibitem [{\citenamefont {Watson}\ \emph {et~al.}(2018)\citenamefont {Watson},
  \citenamefont {Philips}, \citenamefont {Kawakami}, \citenamefont {Ward},
  \citenamefont {Scarlino}, \citenamefont {Veldhorst}, \citenamefont {Savage},
  \citenamefont {Lagally}, \citenamefont {Friesen}, \citenamefont
  {Coppersmith}, \citenamefont {Eriksson},\ and\ \citenamefont
  {Vandersypen}}]{watson2018programmable}%
  \BibitemOpen
  \bibfield  {author} {\bibinfo {author} {\bibfnamefont {T.~F.}\ \bibnamefont
  {Watson}}, \bibinfo {author} {\bibfnamefont {S.~G.~J.}\ \bibnamefont
  {Philips}}, \bibinfo {author} {\bibfnamefont {E.}~\bibnamefont {Kawakami}},
  \bibinfo {author} {\bibfnamefont {D.~R.}\ \bibnamefont {Ward}}, \bibinfo
  {author} {\bibfnamefont {P.}~\bibnamefont {Scarlino}}, \bibinfo {author}
  {\bibfnamefont {M.}~\bibnamefont {Veldhorst}}, \bibinfo {author}
  {\bibfnamefont {D.~E.}\ \bibnamefont {Savage}}, \bibinfo {author}
  {\bibfnamefont {M.~G.}\ \bibnamefont {Lagally}}, \bibinfo {author}
  {\bibfnamefont {M.}~\bibnamefont {Friesen}}, \bibinfo {author} {\bibfnamefont
  {S.~N.}\ \bibnamefont {Coppersmith}}, \bibinfo {author} {\bibfnamefont
  {M.~A.}\ \bibnamefont {Eriksson}},\ and\ \bibinfo {author} {\bibfnamefont
  {L.~M.~K.}\ \bibnamefont {Vandersypen}},\ }\bibfield  {title} {\bibinfo
  {title} {A programmable two-qubit quantum processor in silicon},\ }\href
  {https://doi.org/10.1038/nature25766} {\bibfield  {journal} {\bibinfo
  {journal} {Nature}\ }\textbf {\bibinfo {volume} {555}},\ \bibinfo {pages}
  {633} (\bibinfo {year} {2018})}\BibitemShut {NoStop}%
\bibitem [{\citenamefont {Xue}\ \emph {et~al.}(2021)\citenamefont {Xue},
  \citenamefont {Patra}, \citenamefont {van Dijk}, \citenamefont {Samkharadze},
  \citenamefont {Subramanian}, \citenamefont {Corna}, \citenamefont
  {Paquelet~Wuetz}, \citenamefont {Jeon}, \citenamefont {Sheikh}, \citenamefont
  {Juarez-Hernandez}, \citenamefont {Esparza}, \citenamefont {Rampurawala},
  \citenamefont {Carlton}, \citenamefont {Ravikumar}, \citenamefont {Nieva},
  \citenamefont {Kim}, \citenamefont {Lee}, \citenamefont {Sammak},
  \citenamefont {Scappucci}, \citenamefont {Veldhorst}, \citenamefont
  {Sebastiano}, \citenamefont {Babaie}, \citenamefont {Pellerano},
  \citenamefont {Charbon},\ and\ \citenamefont {Vandersypen}}]{xue2020cmos}%
  \BibitemOpen
  \bibfield  {author} {\bibinfo {author} {\bibfnamefont {X.}~\bibnamefont
  {Xue}}, \bibinfo {author} {\bibfnamefont {B.}~\bibnamefont {Patra}}, \bibinfo
  {author} {\bibfnamefont {J.~P.~G.}\ \bibnamefont {van Dijk}}, \bibinfo
  {author} {\bibfnamefont {N.}~\bibnamefont {Samkharadze}}, \bibinfo {author}
  {\bibfnamefont {S.}~\bibnamefont {Subramanian}}, \bibinfo {author}
  {\bibfnamefont {A.}~\bibnamefont {Corna}}, \bibinfo {author} {\bibfnamefont
  {B.}~\bibnamefont {Paquelet~Wuetz}}, \bibinfo {author} {\bibfnamefont
  {C.}~\bibnamefont {Jeon}}, \bibinfo {author} {\bibfnamefont {F.}~\bibnamefont
  {Sheikh}}, \bibinfo {author} {\bibfnamefont {E.}~\bibnamefont
  {Juarez-Hernandez}}, \bibinfo {author} {\bibfnamefont {B.~P.}\ \bibnamefont
  {Esparza}}, \bibinfo {author} {\bibfnamefont {H.}~\bibnamefont
  {Rampurawala}}, \bibinfo {author} {\bibfnamefont {B.}~\bibnamefont
  {Carlton}}, \bibinfo {author} {\bibfnamefont {S.}~\bibnamefont {Ravikumar}},
  \bibinfo {author} {\bibfnamefont {C.}~\bibnamefont {Nieva}}, \bibinfo
  {author} {\bibfnamefont {S.}~\bibnamefont {Kim}}, \bibinfo {author}
  {\bibfnamefont {H.-J.}\ \bibnamefont {Lee}}, \bibinfo {author} {\bibfnamefont
  {A.}~\bibnamefont {Sammak}}, \bibinfo {author} {\bibfnamefont
  {G.}~\bibnamefont {Scappucci}}, \bibinfo {author} {\bibfnamefont
  {M.}~\bibnamefont {Veldhorst}}, \bibinfo {author} {\bibfnamefont
  {F.}~\bibnamefont {Sebastiano}}, \bibinfo {author} {\bibfnamefont
  {M.}~\bibnamefont {Babaie}}, \bibinfo {author} {\bibfnamefont
  {S.}~\bibnamefont {Pellerano}}, \bibinfo {author} {\bibfnamefont
  {E.}~\bibnamefont {Charbon}},\ and\ \bibinfo {author} {\bibfnamefont
  {L.~M.~K.}\ \bibnamefont {Vandersypen}},\ }\bibfield  {title} {\bibinfo
  {title} {{CMOS}-based cryogenic control of silicon quantum circuits},\ }\href
  {https://doi.org/10.1038/s41586-021-03469-4} {\bibfield  {journal} {\bibinfo
  {journal} {Nature}\ }\textbf {\bibinfo {volume} {593}},\ \bibinfo {pages}
  {205} (\bibinfo {year} {2021})}\BibitemShut {NoStop}%
\bibitem [{\citenamefont {Elzerman}\ \emph {et~al.}(2004)\citenamefont
  {Elzerman}, \citenamefont {Hanson}, \citenamefont {Willems~van Beveren},
  \citenamefont {Witkamp}, \citenamefont {Vandersypen},\ and\ \citenamefont
  {Kouwenhoven}}]{elzerman2004single}%
  \BibitemOpen
  \bibfield  {author} {\bibinfo {author} {\bibfnamefont {J.~M.}\ \bibnamefont
  {Elzerman}}, \bibinfo {author} {\bibfnamefont {R.}~\bibnamefont {Hanson}},
  \bibinfo {author} {\bibfnamefont {L.~H.}\ \bibnamefont {Willems~van
  Beveren}}, \bibinfo {author} {\bibfnamefont {B.}~\bibnamefont {Witkamp}},
  \bibinfo {author} {\bibfnamefont {L.~M.~K.}\ \bibnamefont {Vandersypen}},\
  and\ \bibinfo {author} {\bibfnamefont {L.~P.}\ \bibnamefont {Kouwenhoven}},\
  }\bibfield  {title} {\bibinfo {title} {Single-shot read-out of an individual
  electron spin in a quantum dot},\ }\href
  {https://doi.org/10.1038/nature02693} {\bibfield  {journal} {\bibinfo
  {journal} {Nature}\ }\textbf {\bibinfo {volume} {430}},\ \bibinfo {pages}
  {431} (\bibinfo {year} {2004})}\BibitemShut {NoStop}%
\bibitem [{\citenamefont {Pioro-Ladri{\`e}re}\ \emph
  {et~al.}(2008)\citenamefont {Pioro-Ladri{\`e}re}, \citenamefont {Obata},
  \citenamefont {Tokura}, \citenamefont {Shin}, \citenamefont {Kubo},
  \citenamefont {Yoshida}, \citenamefont {Taniyama},\ and\ \citenamefont
  {Tarucha}}]{pioro2008electrically}%
  \BibitemOpen
  \bibfield  {author} {\bibinfo {author} {\bibfnamefont {M.}~\bibnamefont
  {Pioro-Ladri{\`e}re}}, \bibinfo {author} {\bibfnamefont {T.}~\bibnamefont
  {Obata}}, \bibinfo {author} {\bibfnamefont {Y.}~\bibnamefont {Tokura}},
  \bibinfo {author} {\bibfnamefont {Y.-S.}\ \bibnamefont {Shin}}, \bibinfo
  {author} {\bibfnamefont {T.}~\bibnamefont {Kubo}}, \bibinfo {author}
  {\bibfnamefont {K.}~\bibnamefont {Yoshida}}, \bibinfo {author} {\bibfnamefont
  {T.}~\bibnamefont {Taniyama}},\ and\ \bibinfo {author} {\bibfnamefont
  {S.}~\bibnamefont {Tarucha}},\ }\bibfield  {title} {\bibinfo {title}
  {Electrically driven single-electron spin resonance in a slanting zeeman
  field},\ }\href {https://doi.org/10.1038/nphys1053} {\bibfield  {journal}
  {\bibinfo  {journal} {Nat. Phys.}\ }\textbf {\bibinfo {volume} {4}},\
  \bibinfo {pages} {776} (\bibinfo {year} {2008})}\BibitemShut {NoStop}%
\bibitem [{\citenamefont {Vandersypen}\ and\ \citenamefont
  {Chuang}(2005)}]{vandersypen2005nmr}%
  \BibitemOpen
  \bibfield  {author} {\bibinfo {author} {\bibfnamefont {L.~M.~K.}\
  \bibnamefont {Vandersypen}}\ and\ \bibinfo {author} {\bibfnamefont {I.~L.}\
  \bibnamefont {Chuang}},\ }\bibfield  {title} {\bibinfo {title} {{NMR}
  techniques for quantum control and computation},\ }\href
  {https://doi.org/10.1103/RevModPhys.76.1037} {\bibfield  {journal} {\bibinfo
  {journal} {Rev. Mod. Phys.}\ }\textbf {\bibinfo {volume} {76}},\ \bibinfo
  {pages} {1037} (\bibinfo {year} {2005})}\BibitemShut {NoStop}%
\bibitem [{\citenamefont {Loss}\ and\ \citenamefont
  {DiVincenzo}(1998)}]{loss1998quantum}%
  \BibitemOpen
  \bibfield  {author} {\bibinfo {author} {\bibfnamefont {D.}~\bibnamefont
  {Loss}}\ and\ \bibinfo {author} {\bibfnamefont {D.~P.}\ \bibnamefont
  {DiVincenzo}},\ }\bibfield  {title} {\bibinfo {title} {Quantum computation
  with quantum dots},\ }\href {https://doi.org/10.1103/PhysRevA.57.120}
  {\bibfield  {journal} {\bibinfo  {journal} {Phys. Rev. A}\ }\textbf {\bibinfo
  {volume} {57}},\ \bibinfo {pages} {120} (\bibinfo {year} {1998})}\BibitemShut
  {NoStop}%
\bibitem [{\citenamefont {Petta}\ \emph {et~al.}(2005)\citenamefont {Petta},
  \citenamefont {Johnson}, \citenamefont {Taylor}, \citenamefont {Laird},
  \citenamefont {Yacoby}, \citenamefont {Lukin}, \citenamefont {Marcus},
  \citenamefont {Hanson},\ and\ \citenamefont {Gossard}}]{petta2005coherent}%
  \BibitemOpen
  \bibfield  {author} {\bibinfo {author} {\bibfnamefont {J.~R.}\ \bibnamefont
  {Petta}}, \bibinfo {author} {\bibfnamefont {A.~C.}\ \bibnamefont {Johnson}},
  \bibinfo {author} {\bibfnamefont {J.~M.}\ \bibnamefont {Taylor}}, \bibinfo
  {author} {\bibfnamefont {E.~A.}\ \bibnamefont {Laird}}, \bibinfo {author}
  {\bibfnamefont {A.}~\bibnamefont {Yacoby}}, \bibinfo {author} {\bibfnamefont
  {M.~D.}\ \bibnamefont {Lukin}}, \bibinfo {author} {\bibfnamefont {C.~M.}\
  \bibnamefont {Marcus}}, \bibinfo {author} {\bibfnamefont {M.~P.}\
  \bibnamefont {Hanson}},\ and\ \bibinfo {author} {\bibfnamefont {A.~C.}\
  \bibnamefont {Gossard}},\ }\bibfield  {title} {\bibinfo {title} {Coherent
  {Manipulation} of {Coupled} {Electron} {Spins} in {Semiconductor} {Quantum}
  {Dots}},\ }\href {https://doi.org/10.1126/science.1116955} {\bibfield
  {journal} {\bibinfo  {journal} {Science}\ }\textbf {\bibinfo {volume}
  {309}},\ \bibinfo {pages} {2180} (\bibinfo {year} {2005})}\BibitemShut
  {NoStop}%
\bibitem [{\citenamefont {Meunier}\ \emph {et~al.}(2011)\citenamefont
  {Meunier}, \citenamefont {Calado},\ and\ \citenamefont
  {Vandersypen}}]{meunier2011efficient}%
  \BibitemOpen
  \bibfield  {author} {\bibinfo {author} {\bibfnamefont {T.}~\bibnamefont
  {Meunier}}, \bibinfo {author} {\bibfnamefont {V.~E.}\ \bibnamefont
  {Calado}},\ and\ \bibinfo {author} {\bibfnamefont {L.~M.~K.}\ \bibnamefont
  {Vandersypen}},\ }\bibfield  {title} {\bibinfo {title} {Efficient
  controlled-phase gate for single-spin qubits in quantum dots},\ }\href
  {https://doi.org/10.1103/PhysRevB.83.121403} {\bibfield  {journal} {\bibinfo
  {journal} {Phys. Rev. B}\ }\textbf {\bibinfo {volume} {83}},\ \bibinfo
  {pages} {121403} (\bibinfo {year} {2011})}\BibitemShut {NoStop}%
\bibitem [{\citenamefont {Martins}\ \emph {et~al.}(2016)\citenamefont
  {Martins}, \citenamefont {Malinowski}, \citenamefont {Nissen}, \citenamefont
  {Barnes}, \citenamefont {Fallahi}, \citenamefont {Gardner}, \citenamefont
  {Manfra}, \citenamefont {Marcus},\ and\ \citenamefont
  {Kuemmeth}}]{martins2016noise}%
  \BibitemOpen
  \bibfield  {author} {\bibinfo {author} {\bibfnamefont {F.}~\bibnamefont
  {Martins}}, \bibinfo {author} {\bibfnamefont {F.~K.}\ \bibnamefont
  {Malinowski}}, \bibinfo {author} {\bibfnamefont {P.~D.}\ \bibnamefont
  {Nissen}}, \bibinfo {author} {\bibfnamefont {E.}~\bibnamefont {Barnes}},
  \bibinfo {author} {\bibfnamefont {S.}~\bibnamefont {Fallahi}}, \bibinfo
  {author} {\bibfnamefont {G.~C.}\ \bibnamefont {Gardner}}, \bibinfo {author}
  {\bibfnamefont {M.~J.}\ \bibnamefont {Manfra}}, \bibinfo {author}
  {\bibfnamefont {C.~M.}\ \bibnamefont {Marcus}},\ and\ \bibinfo {author}
  {\bibfnamefont {F.}~\bibnamefont {Kuemmeth}},\ }\bibfield  {title} {\bibinfo
  {title} {Noise {Suppression} {Using} {Symmetric} {Exchange} {Gates} in {Spin}
  {Qubits}},\ }\href {https://doi.org/10.1103/PhysRevLett.116.116801}
  {\bibfield  {journal} {\bibinfo  {journal} {Phys. Rev. Lett.}\ }\textbf
  {\bibinfo {volume} {116}},\ \bibinfo {pages} {116801} (\bibinfo {year}
  {2016})}\BibitemShut {NoStop}%
\bibitem [{\citenamefont {Reed}\ \emph {et~al.}(2016)\citenamefont {Reed},
  \citenamefont {Maune}, \citenamefont {Andrews}, \citenamefont {Borselli},
  \citenamefont {Eng}, \citenamefont {Jura}, \citenamefont {Kiselev},
  \citenamefont {Ladd}, \citenamefont {Merkel}, \citenamefont {Milosavljevic},
  \citenamefont {Pritchett}, \citenamefont {Rakher}, \citenamefont {Ross},
  \citenamefont {Schmitz}, \citenamefont {Smith}, \citenamefont {Wright},
  \citenamefont {Gyure},\ and\ \citenamefont {Hunter}}]{reed2016reduced}%
  \BibitemOpen
  \bibfield  {author} {\bibinfo {author} {\bibfnamefont {M.~D.}\ \bibnamefont
  {Reed}}, \bibinfo {author} {\bibfnamefont {B.~M.}\ \bibnamefont {Maune}},
  \bibinfo {author} {\bibfnamefont {R.~W.}\ \bibnamefont {Andrews}}, \bibinfo
  {author} {\bibfnamefont {M.~G.}\ \bibnamefont {Borselli}}, \bibinfo {author}
  {\bibfnamefont {K.}~\bibnamefont {Eng}}, \bibinfo {author} {\bibfnamefont
  {M.~P.}\ \bibnamefont {Jura}}, \bibinfo {author} {\bibfnamefont {A.~A.}\
  \bibnamefont {Kiselev}}, \bibinfo {author} {\bibfnamefont {T.~D.}\
  \bibnamefont {Ladd}}, \bibinfo {author} {\bibfnamefont {S.~T.}\ \bibnamefont
  {Merkel}}, \bibinfo {author} {\bibfnamefont {I.}~\bibnamefont
  {Milosavljevic}}, \bibinfo {author} {\bibfnamefont {E.~J.}\ \bibnamefont
  {Pritchett}}, \bibinfo {author} {\bibfnamefont {M.~T.}\ \bibnamefont
  {Rakher}}, \bibinfo {author} {\bibfnamefont {R.~S.}\ \bibnamefont {Ross}},
  \bibinfo {author} {\bibfnamefont {A.~E.}\ \bibnamefont {Schmitz}}, \bibinfo
  {author} {\bibfnamefont {A.}~\bibnamefont {Smith}}, \bibinfo {author}
  {\bibfnamefont {J.~A.}\ \bibnamefont {Wright}}, \bibinfo {author}
  {\bibfnamefont {M.~F.}\ \bibnamefont {Gyure}},\ and\ \bibinfo {author}
  {\bibfnamefont {A.~T.}\ \bibnamefont {Hunter}},\ }\bibfield  {title}
  {\bibinfo {title} {Reduced {Sensitivity} to {Charge} {Noise} in
  {Semiconductor} {Spin} {Qubits} via {Symmetric} {Operation}},\ }\href
  {https://doi.org/10.1103/PhysRevLett.116.110402} {\bibfield  {journal}
  {\bibinfo  {journal} {Phys. Rev. Lett.}\ }\textbf {\bibinfo {volume} {116}},\
  \bibinfo {pages} {110402} (\bibinfo {year} {2016})}\BibitemShut {NoStop}%
\bibitem [{\citenamefont {Zhang}\ \emph {et~al.}(2017)\citenamefont {Zhang},
  \citenamefont {Throckmorton}, \citenamefont {Yang}, \citenamefont {Wang},
  \citenamefont {Barnes},\ and\ \citenamefont {Sarma}}]{zhang_2017_tilt}%
  \BibitemOpen
  \bibfield  {author} {\bibinfo {author} {\bibfnamefont {C.}~\bibnamefont
  {Zhang}}, \bibinfo {author} {\bibfnamefont {R.~E.}\ \bibnamefont
  {Throckmorton}}, \bibinfo {author} {\bibfnamefont {X.-C.}\ \bibnamefont
  {Yang}}, \bibinfo {author} {\bibfnamefont {X.}~\bibnamefont {Wang}}, \bibinfo
  {author} {\bibfnamefont {E.}~\bibnamefont {Barnes}},\ and\ \bibinfo {author}
  {\bibfnamefont {S.~D.}\ \bibnamefont {Sarma}},\ }\bibfield  {title} {\bibinfo
  {title} {Randomized {Benchmarking} of {Barrier} versus {Tilt} {Control} of a
  {Singlet-Triplet} {Qubit}},\ }\href
  {https://doi.org/10.1103/PhysRevLett.118.216802} {\bibfield  {journal}
  {\bibinfo  {journal} {Phys. Rev. Lett.}\ }\textbf {\bibinfo {volume} {118}},\
  \bibinfo {pages} {216802} (\bibinfo {year} {2017})}\BibitemShut {NoStop}%
\bibitem [{\citenamefont {Blume-Kohout}\ \emph {et~al.}(2017)\citenamefont
  {Blume-Kohout}, \citenamefont {Gamble}, \citenamefont {Nielsen},
  \citenamefont {Rudinger}, \citenamefont {Mizrahi}, \citenamefont {Fortier},\
  and\ \citenamefont {Maunz}}]{blume2017demonstration}%
  \BibitemOpen
  \bibfield  {author} {\bibinfo {author} {\bibfnamefont {R.}~\bibnamefont
  {Blume-Kohout}}, \bibinfo {author} {\bibfnamefont {J.~K.}\ \bibnamefont
  {Gamble}}, \bibinfo {author} {\bibfnamefont {E.}~\bibnamefont {Nielsen}},
  \bibinfo {author} {\bibfnamefont {K.}~\bibnamefont {Rudinger}}, \bibinfo
  {author} {\bibfnamefont {J.}~\bibnamefont {Mizrahi}}, \bibinfo {author}
  {\bibfnamefont {K.}~\bibnamefont {Fortier}},\ and\ \bibinfo {author}
  {\bibfnamefont {P.}~\bibnamefont {Maunz}},\ }\bibfield  {title} {\bibinfo
  {title} {Demonstration of qubit operations below a rigorous fault tolerance
  threshold with gate set tomography},\ }\href
  {https://doi.org/10.1038/ncomms14485} {\bibfield  {journal} {\bibinfo
  {journal} {Nat. Commun.}\ }\textbf {\bibinfo {volume} {8}},\ \bibinfo {pages}
  {14485} (\bibinfo {year} {2017})}\BibitemShut {NoStop}%
\bibitem [{\citenamefont {Magesan}\ \emph {et~al.}(2012)\citenamefont
  {Magesan}, \citenamefont {Gambetta},\ and\ \citenamefont
  {Emerson}}]{magesan2012characterizing}%
  \BibitemOpen
  \bibfield  {author} {\bibinfo {author} {\bibfnamefont {E.}~\bibnamefont
  {Magesan}}, \bibinfo {author} {\bibfnamefont {J.~M.}\ \bibnamefont
  {Gambetta}},\ and\ \bibinfo {author} {\bibfnamefont {J.}~\bibnamefont
  {Emerson}},\ }\bibfield  {title} {\bibinfo {title} {Characterizing quantum
  gates via randomized benchmarking},\ }\href
  {https://doi.org/10.1103/PhysRevA.85.042311} {\bibfield  {journal} {\bibinfo
  {journal} {Phys. Rev. A}\ }\textbf {\bibinfo {volume} {85}},\ \bibinfo
  {pages} {042311} (\bibinfo {year} {2012})}\BibitemShut {NoStop}%
\bibitem [{\citenamefont {Dehollain}\ \emph {et~al.}(2016)\citenamefont
  {Dehollain}, \citenamefont {Muhonen}, \citenamefont {Blume-Kohout},
  \citenamefont {Rudinger}, \citenamefont {Gamble}, \citenamefont {Nielsen},
  \citenamefont {Laucht}, \citenamefont {Simmons}, \citenamefont {Kalra},
  \citenamefont {Dzurak},\ and\ \citenamefont
  {Morello}}]{dehollain2016optimization}%
  \BibitemOpen
  \bibfield  {author} {\bibinfo {author} {\bibfnamefont {J.~P.}\ \bibnamefont
  {Dehollain}}, \bibinfo {author} {\bibfnamefont {J.~T.}\ \bibnamefont
  {Muhonen}}, \bibinfo {author} {\bibfnamefont {R.}~\bibnamefont
  {Blume-Kohout}}, \bibinfo {author} {\bibfnamefont {K.~M.}\ \bibnamefont
  {Rudinger}}, \bibinfo {author} {\bibfnamefont {J.~K.}\ \bibnamefont
  {Gamble}}, \bibinfo {author} {\bibfnamefont {E.}~\bibnamefont {Nielsen}},
  \bibinfo {author} {\bibfnamefont {A.}~\bibnamefont {Laucht}}, \bibinfo
  {author} {\bibfnamefont {S.}~\bibnamefont {Simmons}}, \bibinfo {author}
  {\bibfnamefont {R.}~\bibnamefont {Kalra}}, \bibinfo {author} {\bibfnamefont
  {A.~S.}\ \bibnamefont {Dzurak}},\ and\ \bibinfo {author} {\bibfnamefont
  {A.}~\bibnamefont {Morello}},\ }\bibfield  {title} {\bibinfo {title}
  {Optimization of a solid-state electron spin qubit using gate set
  tomography},\ }\href {http://stacks.iop.org/1367-2630/18/i=10/a=103018}
  {\bibfield  {journal} {\bibinfo  {journal} {New J. Phys.}\ }\textbf {\bibinfo
  {volume} {18}},\ \bibinfo {pages} {103018} (\bibinfo {year}
  {2016})}\BibitemShut {NoStop}%
\bibitem [{\citenamefont {Greenbaum}(2015)}]{greenbaum2015introduction}%
  \BibitemOpen
  \bibfield  {author} {\bibinfo {author} {\bibfnamefont {D.}~\bibnamefont
  {Greenbaum}},\ }\bibfield  {title} {\bibinfo {title} {Introduction to
  {Quantum} {Gate} {Set} {Tomography}},\ }\href
  {http://arxiv.org/abs/1509.02921} {\bibfield  {journal} {\bibinfo  {journal}
  {arXiv:1509.02921}\ } (\bibinfo {year} {2015})}\BibitemShut {NoStop}%
\bibitem [{\citenamefont {Kelly}\ \emph {et~al.}(2014)\citenamefont {Kelly},
  \citenamefont {Barends}, \citenamefont {Campbell}, \citenamefont {Chen},
  \citenamefont {Chen}, \citenamefont {Chiaro}, \citenamefont {Dunsworth},
  \citenamefont {Fowler}, \citenamefont {Hoi}, \citenamefont {Jeffrey},
  \citenamefont {Megrant}, \citenamefont {Mutus}, \citenamefont {Neill},
  \citenamefont {O'Malley}, \citenamefont {Quintana}, \citenamefont {Roushan},
  \citenamefont {Sank}, \citenamefont {Vainsencher}, \citenamefont {Wenner},
  \citenamefont {White}, \citenamefont {Cleland},\ and\ \citenamefont
  {Martinis}}]{kelly2014optimal}%
  \BibitemOpen
  \bibfield  {author} {\bibinfo {author} {\bibfnamefont {J.}~\bibnamefont
  {Kelly}}, \bibinfo {author} {\bibfnamefont {R.}~\bibnamefont {Barends}},
  \bibinfo {author} {\bibfnamefont {B.}~\bibnamefont {Campbell}}, \bibinfo
  {author} {\bibfnamefont {Y.}~\bibnamefont {Chen}}, \bibinfo {author}
  {\bibfnamefont {Z.}~\bibnamefont {Chen}}, \bibinfo {author} {\bibfnamefont
  {B.}~\bibnamefont {Chiaro}}, \bibinfo {author} {\bibfnamefont
  {A.}~\bibnamefont {Dunsworth}}, \bibinfo {author} {\bibfnamefont {A.~G.}\
  \bibnamefont {Fowler}}, \bibinfo {author} {\bibfnamefont {I.-C.}\
  \bibnamefont {Hoi}}, \bibinfo {author} {\bibfnamefont {E.}~\bibnamefont
  {Jeffrey}}, \bibinfo {author} {\bibfnamefont {A.}~\bibnamefont {Megrant}},
  \bibinfo {author} {\bibfnamefont {J.}~\bibnamefont {Mutus}}, \bibinfo
  {author} {\bibfnamefont {C.}~\bibnamefont {Neill}}, \bibinfo {author}
  {\bibfnamefont {P.~J.~J.}\ \bibnamefont {O'Malley}}, \bibinfo {author}
  {\bibfnamefont {C.}~\bibnamefont {Quintana}}, \bibinfo {author}
  {\bibfnamefont {P.}~\bibnamefont {Roushan}}, \bibinfo {author} {\bibfnamefont
  {D.}~\bibnamefont {Sank}}, \bibinfo {author} {\bibfnamefont {A.}~\bibnamefont
  {Vainsencher}}, \bibinfo {author} {\bibfnamefont {J.}~\bibnamefont {Wenner}},
  \bibinfo {author} {\bibfnamefont {T.~C.}\ \bibnamefont {White}}, \bibinfo
  {author} {\bibfnamefont {A.~N.}\ \bibnamefont {Cleland}},\ and\ \bibinfo
  {author} {\bibfnamefont {J.~M.}\ \bibnamefont {Martinis}},\ }\bibfield
  {title} {\bibinfo {title} {Optimal quantum control using randomized
  benchmarking},\ }\href {https://doi.org/10.1103/PhysRevLett.112.240504}
  {\bibfield  {journal} {\bibinfo  {journal} {Phys. Rev. Lett.}\ }\textbf
  {\bibinfo {volume} {112}},\ \bibinfo {pages} {240504} (\bibinfo {year}
  {2014})}\BibitemShut {NoStop}%
\bibitem [{\citenamefont {Blume-Kohout}\ \emph {et~al.}(2021)\citenamefont
  {Blume-Kohout}, \citenamefont {da~Silva}, \citenamefont {Nielsen},
  \citenamefont {Proctor}, \citenamefont {Rudinger}, \citenamefont {Sarovar},\
  and\ \citenamefont {Young}}]{blume2021taxonomy}%
  \BibitemOpen
  \bibfield  {author} {\bibinfo {author} {\bibfnamefont {R.}~\bibnamefont
  {Blume-Kohout}}, \bibinfo {author} {\bibfnamefont {M.~P.}\ \bibnamefont
  {da~Silva}}, \bibinfo {author} {\bibfnamefont {E.}~\bibnamefont {Nielsen}},
  \bibinfo {author} {\bibfnamefont {T.}~\bibnamefont {Proctor}}, \bibinfo
  {author} {\bibfnamefont {K.}~\bibnamefont {Rudinger}}, \bibinfo {author}
  {\bibfnamefont {M.}~\bibnamefont {Sarovar}},\ and\ \bibinfo {author}
  {\bibfnamefont {K.}~\bibnamefont {Young}},\ }\bibfield  {title} {\bibinfo
  {title} {A taxonomy of small {Markovian} errors},\ }\href
  {http://arxiv.org/abs/2103.01928} {\bibfield  {journal} {\bibinfo  {journal}
  {arXiv:2103.01928}\ } (\bibinfo {year} {2021})}\BibitemShut {NoStop}%
\bibitem [{\citenamefont {Cerfontaine}\ \emph {et~al.}(2020)\citenamefont
  {Cerfontaine}, \citenamefont {Otten}, \citenamefont {Wolfe}, \citenamefont
  {Bethke},\ and\ \citenamefont {Bluhm}}]{cerfontaine2020high}%
  \BibitemOpen
  \bibfield  {author} {\bibinfo {author} {\bibfnamefont {P.}~\bibnamefont
  {Cerfontaine}}, \bibinfo {author} {\bibfnamefont {R.}~\bibnamefont {Otten}},
  \bibinfo {author} {\bibfnamefont {M.~A.}\ \bibnamefont {Wolfe}}, \bibinfo
  {author} {\bibfnamefont {P.}~\bibnamefont {Bethke}},\ and\ \bibinfo {author}
  {\bibfnamefont {H.}~\bibnamefont {Bluhm}},\ }\bibfield  {title} {\bibinfo
  {title} {High-fidelity gate set for exchange-coupled singlet-triplet
  qubits},\ }\href {https://doi.org/10.1103/PhysRevB.101.155311} {\bibfield
  {journal} {\bibinfo  {journal} {Phys. Rev. B}\ }\textbf {\bibinfo {volume}
  {101}},\ \bibinfo {pages} {155311} (\bibinfo {year} {2020})}\BibitemShut
  {NoStop}%
\bibitem [{\citenamefont {Pan}\ \emph {et~al.}(2020)\citenamefont {Pan},
  \citenamefont {Keating}, \citenamefont {Gyure}, \citenamefont {Pritchett},
  \citenamefont {Quinn}, \citenamefont {Ross}, \citenamefont {Ladd},\ and\
  \citenamefont {Kerckhoff}}]{pan_resonant_2020}%
  \BibitemOpen
  \bibfield  {author} {\bibinfo {author} {\bibfnamefont {A.}~\bibnamefont
  {Pan}}, \bibinfo {author} {\bibfnamefont {T.~E.}\ \bibnamefont {Keating}},
  \bibinfo {author} {\bibfnamefont {M.~F.}\ \bibnamefont {Gyure}}, \bibinfo
  {author} {\bibfnamefont {E.~J.}\ \bibnamefont {Pritchett}}, \bibinfo {author}
  {\bibfnamefont {S.}~\bibnamefont {Quinn}}, \bibinfo {author} {\bibfnamefont
  {R.~S.}\ \bibnamefont {Ross}}, \bibinfo {author} {\bibfnamefont {T.~D.}\
  \bibnamefont {Ladd}},\ and\ \bibinfo {author} {\bibfnamefont
  {J.}~\bibnamefont {Kerckhoff}},\ }\bibfield  {title} {\bibinfo {title}
  {Resonant exchange operation in triple-quantum-dot qubits for spin–photon
  transduction},\ }\href {https://doi.org/10.1088/2058-9565/ab86c9} {\bibfield
  {journal} {\bibinfo  {journal} {Quantum Sci. Technol.}\ }\textbf {\bibinfo
  {volume} {5}},\ \bibinfo {pages} {034005} (\bibinfo {year}
  {2020})}\BibitemShut {NoStop}%
\bibitem [{\citenamefont {Zajac}\ \emph {et~al.}(2017)\citenamefont {Zajac},
  \citenamefont {Sigillito}, \citenamefont {Russ}, \citenamefont {Borjans},
  \citenamefont {Taylor}, \citenamefont {Burkard},\ and\ \citenamefont
  {Petta}}]{zajac2018resonantly}%
  \BibitemOpen
  \bibfield  {author} {\bibinfo {author} {\bibfnamefont {D.~M.}\ \bibnamefont
  {Zajac}}, \bibinfo {author} {\bibfnamefont {A.~J.}\ \bibnamefont
  {Sigillito}}, \bibinfo {author} {\bibfnamefont {M.}~\bibnamefont {Russ}},
  \bibinfo {author} {\bibfnamefont {F.}~\bibnamefont {Borjans}}, \bibinfo
  {author} {\bibfnamefont {J.~M.}\ \bibnamefont {Taylor}}, \bibinfo {author}
  {\bibfnamefont {G.}~\bibnamefont {Burkard}},\ and\ \bibinfo {author}
  {\bibfnamefont {J.~R.}\ \bibnamefont {Petta}},\ }\bibfield  {title} {\bibinfo
  {title} {Resonantly driven cnot gate for electron spins},\ }\href
  {https://doi.org/10.1126/science.aao5965} {\bibfield  {journal} {\bibinfo
  {journal} {Science}\ }\textbf {\bibinfo {volume} {359}},\ \bibinfo {pages}
  {439} (\bibinfo {year} {2017})}\BibitemShut {NoStop}%
\bibitem [{\citenamefont {Martinis}\ and\ \citenamefont
  {Geller}(2014)}]{martinis_fast_2014}%
  \BibitemOpen
  \bibfield  {author} {\bibinfo {author} {\bibfnamefont {J.~M.}\ \bibnamefont
  {Martinis}}\ and\ \bibinfo {author} {\bibfnamefont {M.~R.}\ \bibnamefont
  {Geller}},\ }\bibfield  {title} {\bibinfo {title} {Fast adiabatic qubit gates
  using only $\sigma_z$ control},\ }\href
  {https://doi.org/10.1103/PhysRevA.90.022307} {\bibfield  {journal} {\bibinfo
  {journal} {Phys. Rev. A}\ }\textbf {\bibinfo {volume} {90}},\ \bibinfo
  {pages} {022307} (\bibinfo {year} {2014})}\BibitemShut {NoStop}%
\bibitem [{\citenamefont {Hempel}\ \emph {et~al.}(2018)\citenamefont {Hempel},
  \citenamefont {Maier}, \citenamefont {Romero}, \citenamefont {McClean},
  \citenamefont {Monz}, \citenamefont {Shen}, \citenamefont {Jurcevic},
  \citenamefont {Lanyon}, \citenamefont {Love}, \citenamefont {Babbush},
  \citenamefont {Aspuru-Guzik}, \citenamefont {Blatt},\ and\ \citenamefont
  {Roos}}]{hempel2018quantum}%
  \BibitemOpen
  \bibfield  {author} {\bibinfo {author} {\bibfnamefont {C.}~\bibnamefont
  {Hempel}}, \bibinfo {author} {\bibfnamefont {C.}~\bibnamefont {Maier}},
  \bibinfo {author} {\bibfnamefont {J.}~\bibnamefont {Romero}}, \bibinfo
  {author} {\bibfnamefont {J.}~\bibnamefont {McClean}}, \bibinfo {author}
  {\bibfnamefont {T.}~\bibnamefont {Monz}}, \bibinfo {author} {\bibfnamefont
  {H.}~\bibnamefont {Shen}}, \bibinfo {author} {\bibfnamefont {P.}~\bibnamefont
  {Jurcevic}}, \bibinfo {author} {\bibfnamefont {B.~P.}\ \bibnamefont
  {Lanyon}}, \bibinfo {author} {\bibfnamefont {P.}~\bibnamefont {Love}},
  \bibinfo {author} {\bibfnamefont {R.}~\bibnamefont {Babbush}}, \bibinfo
  {author} {\bibfnamefont {A.}~\bibnamefont {Aspuru-Guzik}}, \bibinfo {author}
  {\bibfnamefont {R.}~\bibnamefont {Blatt}},\ and\ \bibinfo {author}
  {\bibfnamefont {C.~F.}\ \bibnamefont {Roos}},\ }\bibfield  {title} {\bibinfo
  {title} {Quantum chemistry calculations on a trapped-ion quantum simulator},\
  }\href {https://doi.org/10.1103/PhysRevX.8.031022} {\bibfield  {journal}
  {\bibinfo  {journal} {Phys. Rev. X}\ }\textbf {\bibinfo {volume} {8}},\
  \bibinfo {pages} {031022} (\bibinfo {year} {2018})}\BibitemShut {NoStop}%
\bibitem [{\citenamefont {Chow}\ \emph {et~al.}(2010)\citenamefont {Chow},
  \citenamefont {DiCarlo}, \citenamefont {Gambetta}, \citenamefont
  {Nunnenkamp}, \citenamefont {Bishop}, \citenamefont {Frunzio}, \citenamefont
  {Devoret}, \citenamefont {Girvin},\ and\ \citenamefont
  {Schoelkopf}}]{chow2010detecting}%
  \BibitemOpen
  \bibfield  {author} {\bibinfo {author} {\bibfnamefont {J.~M.}\ \bibnamefont
  {Chow}}, \bibinfo {author} {\bibfnamefont {L.}~\bibnamefont {DiCarlo}},
  \bibinfo {author} {\bibfnamefont {J.~M.}\ \bibnamefont {Gambetta}}, \bibinfo
  {author} {\bibfnamefont {A.}~\bibnamefont {Nunnenkamp}}, \bibinfo {author}
  {\bibfnamefont {L.~S.}\ \bibnamefont {Bishop}}, \bibinfo {author}
  {\bibfnamefont {L.}~\bibnamefont {Frunzio}}, \bibinfo {author} {\bibfnamefont
  {M.~H.}\ \bibnamefont {Devoret}}, \bibinfo {author} {\bibfnamefont {S.~M.}\
  \bibnamefont {Girvin}},\ and\ \bibinfo {author} {\bibfnamefont {R.~J.}\
  \bibnamefont {Schoelkopf}},\ }\bibfield  {title} {\bibinfo {title} {Detecting
  highly entangled states with a joint qubit readout},\ }\href
  {https://doi.org/10.1103/PhysRevA.81.062325} {\bibfield  {journal} {\bibinfo
  {journal} {Phys. Rev. A}\ }\textbf {\bibinfo {volume} {81}},\ \bibinfo
  {pages} {062325} (\bibinfo {year} {2010})}\BibitemShut {NoStop}%
\bibitem [{\citenamefont {McClean}\ \emph {et~al.}(2020)\citenamefont
  {McClean}, \citenamefont {Rubin}, \citenamefont {Sung}, \citenamefont
  {Kivlichan}, \citenamefont {Bonet-Monroig}, \citenamefont {Cao},
  \citenamefont {Dai}, \citenamefont {Fried}, \citenamefont {Gidney},
  \citenamefont {Gimby}, \citenamefont {Gokhale}, \citenamefont {H{\"a}ner},
  \citenamefont {Hardikar}, \citenamefont {Havlicek}, \citenamefont {Higgott},
  \citenamefont {Huang}, \citenamefont {Izaac}, \citenamefont {Jiang},
  \citenamefont {Liu}, \citenamefont {McArdle}, \citenamefont {Neeley},
  \citenamefont {O'Brien}, \citenamefont {O'Gorman}, \citenamefont {Ozfidan},
  \citenamefont {Radin}, \citenamefont {Romero}, \citenamefont {Sawaya},
  \citenamefont {Senjean}, \citenamefont {Setia}, \citenamefont {Sim},
  \citenamefont {Steiger}, \citenamefont {Steudtner}, \citenamefont {Sun},
  \citenamefont {Sun}, \citenamefont {Wang}, \citenamefont {Zhang},\ and\
  \citenamefont {Babbush}}]{mcclean2020openfermion}%
  \BibitemOpen
  \bibfield  {author} {\bibinfo {author} {\bibfnamefont {J.~R.}\ \bibnamefont
  {McClean}}, \bibinfo {author} {\bibfnamefont {N.~C.}\ \bibnamefont {Rubin}},
  \bibinfo {author} {\bibfnamefont {K.~J.}\ \bibnamefont {Sung}}, \bibinfo
  {author} {\bibfnamefont {I.~D.}\ \bibnamefont {Kivlichan}}, \bibinfo {author}
  {\bibfnamefont {X.}~\bibnamefont {Bonet-Monroig}}, \bibinfo {author}
  {\bibfnamefont {Y.}~\bibnamefont {Cao}}, \bibinfo {author} {\bibfnamefont
  {C.}~\bibnamefont {Dai}}, \bibinfo {author} {\bibfnamefont {E.~S.}\
  \bibnamefont {Fried}}, \bibinfo {author} {\bibfnamefont {C.}~\bibnamefont
  {Gidney}}, \bibinfo {author} {\bibfnamefont {B.}~\bibnamefont {Gimby}},
  \bibinfo {author} {\bibfnamefont {P.}~\bibnamefont {Gokhale}}, \bibinfo
  {author} {\bibfnamefont {T.}~\bibnamefont {H{\"a}ner}}, \bibinfo {author}
  {\bibfnamefont {T.}~\bibnamefont {Hardikar}}, \bibinfo {author}
  {\bibfnamefont {V.}~\bibnamefont {Havlicek}}, \bibinfo {author}
  {\bibfnamefont {O.}~\bibnamefont {Higgott}}, \bibinfo {author} {\bibfnamefont
  {C.}~\bibnamefont {Huang}}, \bibinfo {author} {\bibfnamefont
  {J.}~\bibnamefont {Izaac}}, \bibinfo {author} {\bibfnamefont
  {Z.}~\bibnamefont {Jiang}}, \bibinfo {author} {\bibfnamefont
  {X.}~\bibnamefont {Liu}}, \bibinfo {author} {\bibfnamefont {S.}~\bibnamefont
  {McArdle}}, \bibinfo {author} {\bibfnamefont {M.}~\bibnamefont {Neeley}},
  \bibinfo {author} {\bibfnamefont {T.}~\bibnamefont {O'Brien}}, \bibinfo
  {author} {\bibfnamefont {B.}~\bibnamefont {O'Gorman}}, \bibinfo {author}
  {\bibfnamefont {I.}~\bibnamefont {Ozfidan}}, \bibinfo {author} {\bibfnamefont
  {M.~D.}\ \bibnamefont {Radin}}, \bibinfo {author} {\bibfnamefont
  {J.}~\bibnamefont {Romero}}, \bibinfo {author} {\bibfnamefont {N.~P.~D.}\
  \bibnamefont {Sawaya}}, \bibinfo {author} {\bibfnamefont {B.}~\bibnamefont
  {Senjean}}, \bibinfo {author} {\bibfnamefont {K.}~\bibnamefont {Setia}},
  \bibinfo {author} {\bibfnamefont {S.}~\bibnamefont {Sim}}, \bibinfo {author}
  {\bibfnamefont {D.~S.}\ \bibnamefont {Steiger}}, \bibinfo {author}
  {\bibfnamefont {M.}~\bibnamefont {Steudtner}}, \bibinfo {author}
  {\bibfnamefont {Q.}~\bibnamefont {Sun}}, \bibinfo {author} {\bibfnamefont
  {W.}~\bibnamefont {Sun}}, \bibinfo {author} {\bibfnamefont {D.}~\bibnamefont
  {Wang}}, \bibinfo {author} {\bibfnamefont {F.}~\bibnamefont {Zhang}},\ and\
  \bibinfo {author} {\bibfnamefont {R.}~\bibnamefont {Babbush}},\ }\bibfield
  {title} {\bibinfo {title} {{OpenFermion}: the electronic structure package
  for quantum computers},\ }\href {https://doi.org/10.1088/2058-9565/ab8ebc}
  {\bibfield  {journal} {\bibinfo  {journal} {Quantum Sci. Technol.}\ }\textbf
  {\bibinfo {volume} {5}},\ \bibinfo {pages} {034014} (\bibinfo {year}
  {2020})}\BibitemShut {NoStop}%
\bibitem [{\citenamefont {Ganzhorn}\ \emph {et~al.}(2019)\citenamefont
  {Ganzhorn}, \citenamefont {Egger}, \citenamefont {Barkoutsos}, \citenamefont
  {Ollitrault}, \citenamefont {Salis}, \citenamefont {Moll}, \citenamefont
  {Roth}, \citenamefont {Fuhrer}, \citenamefont {Mueller}, \citenamefont
  {Woerner}, \citenamefont {Tavernelli},\ and\ \citenamefont
  {Filipp}}]{ganzhorn2019gate}%
  \BibitemOpen
  \bibfield  {author} {\bibinfo {author} {\bibfnamefont {M.}~\bibnamefont
  {Ganzhorn}}, \bibinfo {author} {\bibfnamefont {D.}~\bibnamefont {Egger}},
  \bibinfo {author} {\bibfnamefont {P.}~\bibnamefont {Barkoutsos}}, \bibinfo
  {author} {\bibfnamefont {P.}~\bibnamefont {Ollitrault}}, \bibinfo {author}
  {\bibfnamefont {G.}~\bibnamefont {Salis}}, \bibinfo {author} {\bibfnamefont
  {N.}~\bibnamefont {Moll}}, \bibinfo {author} {\bibfnamefont {M.}~\bibnamefont
  {Roth}}, \bibinfo {author} {\bibfnamefont {A.}~\bibnamefont {Fuhrer}},
  \bibinfo {author} {\bibfnamefont {P.}~\bibnamefont {Mueller}}, \bibinfo
  {author} {\bibfnamefont {S.}~\bibnamefont {Woerner}}, \bibinfo {author}
  {\bibfnamefont {I.}~\bibnamefont {Tavernelli}},\ and\ \bibinfo {author}
  {\bibfnamefont {S.}~\bibnamefont {Filipp}},\ }\bibfield  {title} {\bibinfo
  {title} {Gate-efficient simulation of molecular eigenstates on a quantum
  computer},\ }\href {https://doi.org/10.1103/PhysRevApplied.11.044092}
  {\bibfield  {journal} {\bibinfo  {journal} {Phys. Rev. Applied}\ }\textbf
  {\bibinfo {volume} {11}},\ \bibinfo {pages} {044092} (\bibinfo {year}
  {2019})}\BibitemShut {NoStop}%
\bibitem [{\citenamefont {Madzik}\ \emph {et~al.}(2021)\citenamefont {Madzik},
  \citenamefont {Asaad}, \citenamefont {Youssry}, \citenamefont {Joecker},
  \citenamefont {Rudinger}, \citenamefont {Nielsen}, \citenamefont {Young},
  \citenamefont {Proctor}, \citenamefont {Baczewski}, \citenamefont {Laucht},
  \citenamefont {Schmitt}, \citenamefont {Hudson}, \citenamefont {Itoh},
  \citenamefont {Jakob}, \citenamefont {Johnson}, \citenamefont {Jamieson},
  \citenamefont {Dzurak}, \citenamefont {Ferrie}, \citenamefont
  {Blume-Kohout},\ and\ \citenamefont {Morello}}]{madzik2021precision}%
  \BibitemOpen
  \bibfield  {author} {\bibinfo {author} {\bibfnamefont {M.~T.}\ \bibnamefont
  {Madzik}}, \bibinfo {author} {\bibfnamefont {S.}~\bibnamefont {Asaad}},
  \bibinfo {author} {\bibfnamefont {A.}~\bibnamefont {Youssry}}, \bibinfo
  {author} {\bibfnamefont {B.}~\bibnamefont {Joecker}}, \bibinfo {author}
  {\bibfnamefont {K.~M.}\ \bibnamefont {Rudinger}}, \bibinfo {author}
  {\bibfnamefont {E.}~\bibnamefont {Nielsen}}, \bibinfo {author} {\bibfnamefont
  {K.~C.}\ \bibnamefont {Young}}, \bibinfo {author} {\bibfnamefont {T.~J.}\
  \bibnamefont {Proctor}}, \bibinfo {author} {\bibfnamefont {A.~D.}\
  \bibnamefont {Baczewski}}, \bibinfo {author} {\bibfnamefont {A.}~\bibnamefont
  {Laucht}}, \bibinfo {author} {\bibfnamefont {V.}~\bibnamefont {Schmitt}},
  \bibinfo {author} {\bibfnamefont {F.~E.}\ \bibnamefont {Hudson}}, \bibinfo
  {author} {\bibfnamefont {K.~M.}\ \bibnamefont {Itoh}}, \bibinfo {author}
  {\bibfnamefont {A.~M.}\ \bibnamefont {Jakob}}, \bibinfo {author}
  {\bibfnamefont {B.~C.}\ \bibnamefont {Johnson}}, \bibinfo {author}
  {\bibfnamefont {D.~N.}\ \bibnamefont {Jamieson}}, \bibinfo {author}
  {\bibfnamefont {A.~S.}\ \bibnamefont {Dzurak}}, \bibinfo {author}
  {\bibfnamefont {C.}~\bibnamefont {Ferrie}}, \bibinfo {author} {\bibfnamefont
  {R.}~\bibnamefont {Blume-Kohout}},\ and\ \bibinfo {author} {\bibfnamefont
  {A.}~\bibnamefont {Morello}},\ }\bibfield  {title} {\bibinfo {title}
  {Precision tomography of a three-qubit electron-nuclear quantum processor in
  silicon},\ }\href {http://arxiv.org/abs/2106.03082} {\bibfield  {journal}
  {\bibinfo  {journal} {arXiv:2106.03082}\ } (\bibinfo {year}
  {2021})}\BibitemShut {NoStop}%
\bibitem [{\citenamefont {Zheng}\ \emph {et~al.}(2019)\citenamefont {Zheng},
  \citenamefont {Samkharadze}, \citenamefont {Noordam}, \citenamefont {Kalhor},
  \citenamefont {Brousse}, \citenamefont {Sammak}, \citenamefont {Scappucci},\
  and\ \citenamefont {Vandersypen}}]{zheng2019rapid}%
  \BibitemOpen
  \bibfield  {author} {\bibinfo {author} {\bibfnamefont {G.}~\bibnamefont
  {Zheng}}, \bibinfo {author} {\bibfnamefont {N.}~\bibnamefont {Samkharadze}},
  \bibinfo {author} {\bibfnamefont {M.~L.}\ \bibnamefont {Noordam}}, \bibinfo
  {author} {\bibfnamefont {N.}~\bibnamefont {Kalhor}}, \bibinfo {author}
  {\bibfnamefont {D.}~\bibnamefont {Brousse}}, \bibinfo {author} {\bibfnamefont
  {A.}~\bibnamefont {Sammak}}, \bibinfo {author} {\bibfnamefont
  {G.}~\bibnamefont {Scappucci}},\ and\ \bibinfo {author} {\bibfnamefont
  {L.~M.~K.}\ \bibnamefont {Vandersypen}},\ }\bibfield  {title} {\bibinfo
  {title} {Rapid gate-based spin read-out in silicon using an on-chip
  resonator},\ }\href {https://doi.org/10.1038/s41565-019-0488-9} {\bibfield
  {journal} {\bibinfo  {journal} {Nat. Nanotechnol.}\ }\textbf {\bibinfo
  {volume} {14}},\ \bibinfo {pages} {742} (\bibinfo {year} {2019})}\BibitemShut
  {NoStop}%
\bibitem [{\citenamefont {Schaal}\ \emph {et~al.}(2020)\citenamefont {Schaal},
  \citenamefont {Ahmed}, \citenamefont {Haigh}, \citenamefont {Hutin},
  \citenamefont {Bertrand}, \citenamefont {Barraud}, \citenamefont {Vinet},
  \citenamefont {Lee}, \citenamefont {Stelmashenko}, \citenamefont {Robinson},
  \citenamefont {Qiu}, \citenamefont {Hacohen-Gourgy}, \citenamefont {Siddiqi},
  \citenamefont {Gonzalez-Zalba},\ and\ \citenamefont
  {Morton}}]{schaal2020fast}%
  \BibitemOpen
  \bibfield  {author} {\bibinfo {author} {\bibfnamefont {S.}~\bibnamefont
  {Schaal}}, \bibinfo {author} {\bibfnamefont {I.}~\bibnamefont {Ahmed}},
  \bibinfo {author} {\bibfnamefont {J.~A.}\ \bibnamefont {Haigh}}, \bibinfo
  {author} {\bibfnamefont {L.}~\bibnamefont {Hutin}}, \bibinfo {author}
  {\bibfnamefont {B.}~\bibnamefont {Bertrand}}, \bibinfo {author}
  {\bibfnamefont {S.}~\bibnamefont {Barraud}}, \bibinfo {author} {\bibfnamefont
  {M.}~\bibnamefont {Vinet}}, \bibinfo {author} {\bibfnamefont {C.-M.}\
  \bibnamefont {Lee}}, \bibinfo {author} {\bibfnamefont {N.}~\bibnamefont
  {Stelmashenko}}, \bibinfo {author} {\bibfnamefont {J.~W.~A.}\ \bibnamefont
  {Robinson}}, \bibinfo {author} {\bibfnamefont {J.~Y.}\ \bibnamefont {Qiu}},
  \bibinfo {author} {\bibfnamefont {S.}~\bibnamefont {Hacohen-Gourgy}},
  \bibinfo {author} {\bibfnamefont {I.}~\bibnamefont {Siddiqi}}, \bibinfo
  {author} {\bibfnamefont {M.~F.}\ \bibnamefont {Gonzalez-Zalba}},\ and\
  \bibinfo {author} {\bibfnamefont {J.~J.~L.}\ \bibnamefont {Morton}},\
  }\bibfield  {title} {\bibinfo {title} {Fast gate-based readout of silicon
  quantum dots using josephson parametric amplification},\ }\href
  {https://doi.org/10.1103/PhysRevLett.124.067701} {\bibfield  {journal}
  {\bibinfo  {journal} {Phys. Rev. Lett.}\ }\textbf {\bibinfo {volume} {124}},\
  \bibinfo {pages} {067701} (\bibinfo {year} {2020})}\BibitemShut {NoStop}%
\bibitem [{\citenamefont {Reed}(2013)}]{reed2013entanglement}%
  \BibitemOpen
  \bibfield  {author} {\bibinfo {author} {\bibfnamefont {M.}~\bibnamefont
  {Reed}},\ }\bibfield  {title} {\bibinfo {title} {Entanglement and {Quantum}
  {Error} {Correction} with {Superconducting} {Qubits}},\ }\href
  {http://arxiv.org/abs/1311.6759} {\bibfield  {journal} {\bibinfo  {journal}
  {{PhD Thesis, Yale Univ.}}\ } (\bibinfo {year} {2013})}\BibitemShut {NoStop}%
\bibitem [{\citenamefont {Russ}\ \emph {et~al.}(2018)\citenamefont {Russ},
  \citenamefont {Zajac}, \citenamefont {Sigillito}, \citenamefont {Borjans},
  \citenamefont {Taylor}, \citenamefont {Petta},\ and\ \citenamefont
  {Burkard}}]{russ_high-fidelity_2018}%
  \BibitemOpen
  \bibfield  {author} {\bibinfo {author} {\bibfnamefont {M.}~\bibnamefont
  {Russ}}, \bibinfo {author} {\bibfnamefont {D.~M.}\ \bibnamefont {Zajac}},
  \bibinfo {author} {\bibfnamefont {A.~J.}\ \bibnamefont {Sigillito}}, \bibinfo
  {author} {\bibfnamefont {F.}~\bibnamefont {Borjans}}, \bibinfo {author}
  {\bibfnamefont {J.~M.}\ \bibnamefont {Taylor}}, \bibinfo {author}
  {\bibfnamefont {J.~R.}\ \bibnamefont {Petta}},\ and\ \bibinfo {author}
  {\bibfnamefont {G.}~\bibnamefont {Burkard}},\ }\bibfield  {title} {\bibinfo
  {title} {High-fidelity quantum gates in {Si}/{SiGe} double quantum dots},\
  }\href {https://doi.org/10.1103/PhysRevB.97.085421} {\bibfield  {journal}
  {\bibinfo  {journal} {Phys. Rev. B}\ }\textbf {\bibinfo {volume} {97}},\
  \bibinfo {pages} {085421} (\bibinfo {year} {2018})}\BibitemShut {NoStop}%
\bibitem [{\citenamefont {Yang}\ \emph
  {et~al.}(2019{\natexlab{b}})\citenamefont {Yang}, \citenamefont
  {Coppersmith},\ and\ \citenamefont {Friesen}}]{yang_achieving_2019}%
  \BibitemOpen
  \bibfield  {author} {\bibinfo {author} {\bibfnamefont {Y.-C.}\ \bibnamefont
  {Yang}}, \bibinfo {author} {\bibfnamefont {S.~N.}\ \bibnamefont
  {Coppersmith}},\ and\ \bibinfo {author} {\bibfnamefont {M.}~\bibnamefont
  {Friesen}},\ }\bibfield  {title} {\bibinfo {title} {Achieving high-fidelity
  single-qubit gates in a strongly driven charge qubit with 1/f charge noise},\
  }\href {https://doi.org/10.1038/s41534-019-0127-1} {\bibfield  {journal}
  {\bibinfo  {journal} {npj Quantum Information}\ }\textbf {\bibinfo {volume}
  {5}},\ \bibinfo {pages} {12} (\bibinfo {year}
  {2019}{\natexlab{b}})}\BibitemShut {NoStop}%
\bibitem [{\citenamefont {Koski}\ \emph {et~al.}(2020)\citenamefont {Koski},
  \citenamefont {Landig}, \citenamefont {Russ}, \citenamefont {Abadillo-Uriel},
  \citenamefont {Scarlino}, \citenamefont {Kratochwil}, \citenamefont {Reichl},
  \citenamefont {Wegscheider}, \citenamefont {Burkard}, \citenamefont
  {Friesen}, \citenamefont {Coppersmith}, \citenamefont {Wallraff},
  \citenamefont {Ensslin},\ and\ \citenamefont {Ihn}}]{koski_strong_2020}%
  \BibitemOpen
  \bibfield  {author} {\bibinfo {author} {\bibfnamefont {J.~V.}\ \bibnamefont
  {Koski}}, \bibinfo {author} {\bibfnamefont {A.~J.}\ \bibnamefont {Landig}},
  \bibinfo {author} {\bibfnamefont {M.}~\bibnamefont {Russ}}, \bibinfo {author}
  {\bibfnamefont {J.~C.}\ \bibnamefont {Abadillo-Uriel}}, \bibinfo {author}
  {\bibfnamefont {P.}~\bibnamefont {Scarlino}}, \bibinfo {author}
  {\bibfnamefont {B.}~\bibnamefont {Kratochwil}}, \bibinfo {author}
  {\bibfnamefont {C.}~\bibnamefont {Reichl}}, \bibinfo {author} {\bibfnamefont
  {W.}~\bibnamefont {Wegscheider}}, \bibinfo {author} {\bibfnamefont
  {G.}~\bibnamefont {Burkard}}, \bibinfo {author} {\bibfnamefont
  {M.}~\bibnamefont {Friesen}}, \bibinfo {author} {\bibfnamefont {S.~N.}\
  \bibnamefont {Coppersmith}}, \bibinfo {author} {\bibfnamefont
  {A.}~\bibnamefont {Wallraff}}, \bibinfo {author} {\bibfnamefont
  {K.}~\bibnamefont {Ensslin}},\ and\ \bibinfo {author} {\bibfnamefont
  {T.}~\bibnamefont {Ihn}},\ }\bibfield  {title} {\bibinfo {title} {Strong
  photon coupling to the quadrupole moment of an electron in a solid-state
  qubit},\ }\href {https://doi.org/10.1038/s41567-020-0862-4} {\bibfield
  {journal} {\bibinfo  {journal} {Nature Physics}\ }\textbf {\bibinfo {volume}
  {16}},\ \bibinfo {pages} {642} (\bibinfo {year} {2020})}\BibitemShut
  {NoStop}%
\bibitem [{\citenamefont {Russ}\ \emph {et~al.}(2021)\citenamefont {Russ},
  \citenamefont {Philips}, \citenamefont {Xue},\ and\ \citenamefont
  {Vandersypen}}]{russ_soon_to_be_appear}%
  \BibitemOpen
  \bibfield  {author} {\bibinfo {author} {\bibfnamefont {M.}~\bibnamefont
  {Russ}}, \bibinfo {author} {\bibfnamefont {S.}~\bibnamefont {Philips}},
  \bibinfo {author} {\bibfnamefont {X.}~\bibnamefont {Xue}},\ and\ \bibinfo
  {author} {\bibfnamefont {L.~M.~K.}\ \bibnamefont {Vandersypen}},\ }\bibfield
  {title} {\bibinfo {title} {The path to high fidelity multi-qubit gates for
  quantum dot spin qubits},\ }\href@noop {} {\bibfield  {journal} {\bibinfo
  {journal} {Unpublished}\ } (\bibinfo {year} {2021})}\BibitemShut {NoStop}%
\bibitem [{\citenamefont {Nielsen}\ \emph
  {et~al.}(2020{\natexlab{a}})\citenamefont {Nielsen}, \citenamefont {Gamble},
  \citenamefont {Rudinger}, \citenamefont {Scholten}, \citenamefont {Young},\
  and\ \citenamefont {Blume-Kohout}}]{nielsen2020gate}%
  \BibitemOpen
  \bibfield  {author} {\bibinfo {author} {\bibfnamefont {E.}~\bibnamefont
  {Nielsen}}, \bibinfo {author} {\bibfnamefont {J.~K.}\ \bibnamefont {Gamble}},
  \bibinfo {author} {\bibfnamefont {K.}~\bibnamefont {Rudinger}}, \bibinfo
  {author} {\bibfnamefont {T.}~\bibnamefont {Scholten}}, \bibinfo {author}
  {\bibfnamefont {K.}~\bibnamefont {Young}},\ and\ \bibinfo {author}
  {\bibfnamefont {R.}~\bibnamefont {Blume-Kohout}},\ }\bibfield  {title}
  {\bibinfo {title} {Gate set tomography},\ }\href
  {http://arxiv.org/abs/2009.07301} {\bibfield  {journal} {\bibinfo  {journal}
  {arXiv:2009.07301}\ } (\bibinfo {year} {2020}{\natexlab{a}})}\BibitemShut
  {NoStop}%
\bibitem [{\citenamefont {Nielsen}\ \emph {et~al.}(2019)\citenamefont
  {Nielsen}, \citenamefont {Blume-Kohout}, \citenamefont {Rudinger},
  \citenamefont {Proctor}, \citenamefont {Saldyt},\ and\ \citenamefont
  {USDOE}}]{nielsen_python_2019}%
  \BibitemOpen
  \bibfield  {author} {\bibinfo {author} {\bibfnamefont {E.}~\bibnamefont
  {Nielsen}}, \bibinfo {author} {\bibfnamefont {R.~J.}\ \bibnamefont
  {Blume-Kohout}}, \bibinfo {author} {\bibfnamefont {K.~M.}\ \bibnamefont
  {Rudinger}}, \bibinfo {author} {\bibfnamefont {T.~J.}\ \bibnamefont
  {Proctor}}, \bibinfo {author} {\bibfnamefont {L.}~\bibnamefont {Saldyt}},\
  and\ \bibinfo {author} {\bibnamefont {USDOE}},\ }\href
  {https://doi.org/10.11578/dc.20190722.2} {\emph {\bibinfo {title} {Python
  {GST} {Implementation} ({PyGSTi}) v. 0.9}}},\ \bibinfo {type} {Tech. Rep.}\
  \bibinfo {number} {PyGSTi}\ (\bibinfo {year} {2019})\BibitemShut {NoStop}%
\bibitem [{\citenamefont {Nielsen}\ \emph
  {et~al.}(2020{\natexlab{b}})\citenamefont {Nielsen}, \citenamefont
  {Rudinger}, \citenamefont {Proctor}, \citenamefont {Russo}, \citenamefont
  {Young},\ and\ \citenamefont {Blume-Kohout}}]{nielsen_probing_2020}%
  \BibitemOpen
  \bibfield  {author} {\bibinfo {author} {\bibfnamefont {E.}~\bibnamefont
  {Nielsen}}, \bibinfo {author} {\bibfnamefont {K.}~\bibnamefont {Rudinger}},
  \bibinfo {author} {\bibfnamefont {T.}~\bibnamefont {Proctor}}, \bibinfo
  {author} {\bibfnamefont {A.}~\bibnamefont {Russo}}, \bibinfo {author}
  {\bibfnamefont {K.}~\bibnamefont {Young}},\ and\ \bibinfo {author}
  {\bibfnamefont {R.}~\bibnamefont {Blume-Kohout}},\ }\bibfield  {title}
  {\bibinfo {title} {Probing quantum processor performance with {pyGSTi}},\
  }\href {https://doi.org/10.1088/2058-9565/ab8aa4} {\bibfield  {journal}
  {\bibinfo  {journal} {Quantum Sci. Technol.}\ }\textbf {\bibinfo {volume}
  {5}},\ \bibinfo {pages} {044002} (\bibinfo {year}
  {2020}{\natexlab{b}})}\BibitemShut {NoStop}%
\bibitem [{\citenamefont {White}\ \emph {et~al.}(2007)\citenamefont {White},
  \citenamefont {Gilchrist}, \citenamefont {Pryde}, \citenamefont {O'Brien},
  \citenamefont {Bremner},\ and\ \citenamefont
  {Langford}}]{white_measuring_2007}%
  \BibitemOpen
  \bibfield  {author} {\bibinfo {author} {\bibfnamefont {A.~G.}\ \bibnamefont
  {White}}, \bibinfo {author} {\bibfnamefont {A.}~\bibnamefont {Gilchrist}},
  \bibinfo {author} {\bibfnamefont {G.~J.}\ \bibnamefont {Pryde}}, \bibinfo
  {author} {\bibfnamefont {J.~L.}\ \bibnamefont {O'Brien}}, \bibinfo {author}
  {\bibfnamefont {M.~J.}\ \bibnamefont {Bremner}},\ and\ \bibinfo {author}
  {\bibfnamefont {N.~K.}\ \bibnamefont {Langford}},\ }\bibfield  {title}
  {\bibinfo {title} {Measuring two-qubit gates},\ }\href
  {https://doi.org/10.1364/JOSAB.24.000172} {\bibfield  {journal} {\bibinfo
  {journal} {J. Opt. Soc. Am. B}\ }\textbf {\bibinfo {volume} {24}},\ \bibinfo
  {pages} {172} (\bibinfo {year} {2007})}\BibitemShut {NoStop}%
\bibitem [{\citenamefont {Jamiolkowski}(1972)}]{jamiolkowski_linear_1972}%
  \BibitemOpen
  \bibfield  {author} {\bibinfo {author} {\bibfnamefont {A.}~\bibnamefont
  {Jamiolkowski}},\ }\bibfield  {title} {\bibinfo {title} {Linear
  transformations which preserve trace and positive semidefiniteness of
  operators},\ }\href
  {https://doi.org/https://doi.org/10.1016/0034-4877(72)90011-0} {\bibfield
  {journal} {\bibinfo  {journal} {Reports on Mathematical Physics}\ }\textbf
  {\bibinfo {volume} {3}},\ \bibinfo {pages} {275} (\bibinfo {year}
  {1972})}\BibitemShut {NoStop}%
\bibitem [{\citenamefont {Taube}\ and\ \citenamefont
  {Bartlett}(2006)}]{taube2006new}%
  \BibitemOpen
  \bibfield  {author} {\bibinfo {author} {\bibfnamefont {A.~G.}\ \bibnamefont
  {Taube}}\ and\ \bibinfo {author} {\bibfnamefont {R.~J.}\ \bibnamefont
  {Bartlett}},\ }\bibfield  {title} {\bibinfo {title} {New perspectives on
  unitary coupled-cluster theory},\ }\href {https://doi.org/10.1002/qua.21198}
  {\bibfield  {journal} {\bibinfo  {journal} {International Journal of Quantum
  Chemistry}\ }\textbf {\bibinfo {volume} {106}},\ \bibinfo {pages} {3393}
  (\bibinfo {year} {2006})}\BibitemShut {NoStop}%
\end{thebibliography}%
